\documentclass[twoside]{IEEEtran}
\ifCLASSINFOpdf
\else
\fi

\usepackage{graphics}
\usepackage{float}              
\usepackage{tikz}               
\usepackage{amssymb}
\usepackage{bm}                 
\usepackage{dsfont}
\usepackage{bbold}
\usepackage[cmex10]{amsmath}
\usepackage{algorithmic}
\usepackage{color}
\usepackage[tight,footnotesize]{subfigure}
\usepackage[utf8]{inputenc} 
\usepackage{algorithm}
\usepackage{setspace} 
\providecommand{\e}[1]{\mathrm{e}^{#1} }

\providecommand{\taumax}{\bar \tau_n}
\providecommand{\cumR}{{\overline R}}
\newcommand{\farg}[1]{\mathopen{}\left( #1 \right)}
\newcommand{\fargbig}[1]{\mathopen{}\big( #1 \big)}

\newcommand{\vect}[1]{\boldsymbol{\mathbf{#1}}}
\providecommand{\scseq}{{\bar k_n}}

\providecommand{\C}[1]{C\farg{#1}}

\providecommand{\tr}{^{\mathrm{T}}}
\providecommand{\e}[1]{\exp \left(#1\right)}
\providecommand{\indi}[1]{\mathds{1}\mathopen{}\left\{#1\right\}}

\providecommand{\len}[1]{\text{len}\farg{#1}}

\providecommand{\tr}{^{\mathrm{T}}}

\providecommand{\C}[1]{C\mathopen{}\left(#1\right)}
\providecommand{\vectornorm}[1]{\left\lVert#1\right\rVert}

\providecommand{\vectornormbig}[1]{\big\lVert#1\big\rVert}

\newcounter{MYtempeqncnt}

\usepackage{ifthen}
\newboolean{thesis}
\setboolean{thesis}{false}

\newcommand{\pos}[1]{\left[#1 \right]^+}

\newcommand{\E}[1]{\mathbb{E}\mathopen{}\left[#1\right]}
\newcommand{\Ebig}[1]{\mathbb{E}\mathopen{}\big[#1\big]}
\newcommand{\EBig}[1]{\mathbb{E}\mathopen{}\Big[#1\Big]}

\newcommand{\EBigg}[1]{\mathbb{E}\mathopen{}\Bigg[#1\Bigg]}

\newcommand{\range}[2]{\left[#1 {:} #2\right]}

\newcommand{\pr}[1]{\mathbb{P}\mathopen{}\left[#1\right]}
\newcommand{\prBigg}[1]{\mathbb{P}\mathopen{}\Bigg[#1\Bigg]}
\newcommand{\prBig}[1]{\mathbb{P}\mathopen{}\Big[#1\Big]}

\newcommand{\prbig}[1]{\mathbb{P}\mathopen{}\big[#1\big]}
\newcommand{\ppr}[1]{\mathbb{\overline P}\mathopen{}\left[#1\right]}
\newcommand{\pprr}{\mathbb{\overline P}}

\newcommand{\dd}{\mathop{}\!\mathrm{d}}

\newcommand{\unit}[1]{\ 
    \ifmmode & \left[\textup{#1}\right] \hspace{1cm}
    \else \hfill $\left[\textup{#1}\right]$ \hspace{1cm}
    \fi
}
\newcommand{\feedset}{\mathds{F}}

\newcommand*{\PathBRQ}{.}%

\newtheorem{definition}{Definition}
\newtheorem{remark}{Remark}
\newtheorem{theorem}{Theorem}
\newtheorem{lemma}[theorem]{Lemma}

\begin{document}
%
\title{Generalized HARQ Protocols with Delayed Channel State Information and Average Latency Constraints}

%
%
%

\author{\thanks{The work of K. F. Trillingsgaard and P. Popovski  was supported in part by the European Research Council (ERC Consolidator Grant Nr. 648382 WILLOW) within the Horizon 2020 Program. The material of this paper was presented in part at the 2014 IEEE International Conference on Communication \cite{backtrack}.}\thanks{K. F. Trillingsgaard and P. Popovski are with the Department of Eletronic Systems, Aalborg University, 9220, Aalborg Øst, Denmark (e-mail: \{kft,petarp\}@es.aau.dk).}\thanks{Copyright (c) 2014 IEEE. Personal use of this material is permitted.  However, permission to use this material for any other purposes must be obtained from the IEEE by sending a request to pubs-permissions@ieee.org.} Kasper Fløe Trillingsgaard,~\IEEEmembership{Student Member,~IEEE}, and Petar Popovski, \IEEEmembership{Fellow,~IEEE}}

\maketitle

\begin{abstract}
In many wireless systems, the signal-to-interference-and-noise ratio that is applicable to a certain transmission, referred to as channel state information (CSI), can only be learned after the transmission has taken place and is thereby delayed (outdated). In such systems, hybrid automatic repeat request (HARQ) protocols are often used to achieve high throughput with low latency. This paper put forth the family of expandable message space (EMS) protocols that generalize the HARQ protocol and allow for rate adaptation based on delayed CSI at the transmitter (CSIT). Assuming a block-fading channel, the proposed EMS protocols are analyzed using dynamic programming. When full delayed CSIT is available and there is a constraint on the average decoding time, it is shown that the optimal zero outage EMS protocol has a particularly simple operational interpretation and that the throughput is identical to that of the backtrack retransmission request (BRQ) protocol. We also devise EMS protocols for the case in which CSIT is only available through a finite number of feedback messages. The numerical results demonstrate that the throughput of BRQ approaches the ergodic capacity quickly compared to HARQ, while EMS protocols with only three and four feedback messages achieve throughputs that are only slightly worse than that of BRQ.
\end{abstract}


%
\IEEEpeerreviewmaketitle


\section{Introduction}
%
%
%
%

\ifthenelse{\boolean{thesis}}{Channel}{\IEEEPARstart{C}{hannel}} state information at the transmitter is important for achieving high throughput in wireless systems. Preferably, CSIT is known before a transmission takes place since, in that case, the transmitter is able to optimize the transmission parameters such as rate  and power. The transmitter may acquire an estimate of the CSI in advance in various ways; for example, by using channel reciprocity or via explicit feedback from the receiver. This is referred to as \emph{prior CSIT}. A wireless channel is, however, dynamic and in many cases the channel changes from the time the CSI has been acquired to the time at which the channel is actually used for transmission \cite[pp.~211--213]{Tse}. In addition, even if the channel is static, during the transmission there may be an unpredictable amount of interference at the receiver. In such cases, prior CSI is different from the actual conditions at the receiver when the data transmission takes place and thus of limited use for adapting the transmission parameters. On the other hand, it is viable to assume that the transmitter gets feedback about the CSI \emph{after} the data transmission has been made. We refer to this as \emph{delayed CSIT} as it carries information to the transmitter about the conditions at the receiver in the past. The simplest form of delayed CSIT is the $1-$bit feedback used in ARQ protocols: (ACK) the transmission was successful, i.e., the channel could support the chosen data rate and (NACK) the channel could not support the data rate. In the most elementary form of ARQ, a failed packet is retransmitted in the subsequent time slots until it is successfully decoded or until a strict decoding time constraint is violated. In order to increase throughput compared to ARQ, one can use chase combining (CC) or send incremental redundancy (IR)  instead of retransmissions that consist of pure packet repetition. Such extensions are referred to as HARQ-CC and HARQ-INR, respectively \cite{Caire2001}. In this paper, we focus on IR-based protocols.

The ergodic capacity represents an upper bound on the throughput for any communication protocol and can be approached by fixed-length coding across many time slots. HARQ-type protocols attempt to get as close as possible to this upper bound while keeping the average or maximum decoding time as small as possible. Specifically, as the rate $R$,  which is used in the first transmission opportunity, tends to infinity, the average decoding time of HARQ-INR also tends to infinity and the throughput of HARQ-INR approaches the ergodic capacity of the underlying channel provided that there is no strict constraint on the decoding time. If a strict or average decoding time constraint is present, the achievable throughput is strictly lower than the ergodic capacity. 

The purpose of this paper is to put forth and investigate a type of retransmission protocol which is fundamentally different from conventional HARQ protocols and uses rate adaptation based on delayed CSIT to achieve high throughput subject to an \emph{average decoding time constraint}. As with most prior work in the area of HARQ-INR, we assume the channel is modeled by a Gaussian block-fading channel, with each time slot consisting of $n$ channel uses. The channel gain is kept constant during a single time slot but varies independently from time slot to time slot. Feedback, such as delayed CSIT or acknowledgements (ACKs), can only be received by the transmitter at the end of each time slot. The main problem with an HARQ-INR protocol for a block-fading channel is that resources are wasted when the receiver sends NACK, while it only needs a small amount of additional information to be able to decode. This results in under-utilization of the last time slot and may significantly reduce the throughput when the average decoding time is small. Our key idea is to append \emph{new information bits} in each time slot such that the last time slot is rarely under-utilized and the throughput degradation is reduced. We achieve this by using delayed CSIT which allows the transmitter to estimate the amount of unresolved information at the beginning of each time slot.

\subsection{Prior work}
Caire and Tuninetti \cite{Caire2001} were among the first who analyzed HARQ from an information-theoretic perspective. Here, the throughput measure was defined through the renewal-reward theorem (see also \cite{wolff} and \cite{renewal_arq}) and achievability and converse results were proved for the HARQ-INR protocol.
Several lines of works has since improved the throughput of HARQ-INR by using available side information in combination with either power adaptation or rate adaptation.

One line of work uses power or rate adaptation to enhance the throughput of HARQ-INR with either prior or no CSIT. For example, \cite{Tuninetti2007} investigates HARQ-INR protocols that maximize the throughput over a block-fading channel with independent channel gains under both a strict decoding time constraint and a long-term power constraint. The long-term power constraint allows the use of slot-based power allocation. It is found that HARQ-INR in combination with slot-based power allocation increases the throughput. The key idea is that the probability of having to retransmit $m$ times is decreasing in $m$. This implies that the throughput is increased by using more power in the first slots. In addition, it is shown that if the single feedback bit is used to convey a one-bit quantization of the prior CSI rather than an ACK/NACK message, then this can result in significant throughput gains. The results from \cite{Tuninetti2007} are further extended to any number of feedback bits per slot in~\cite{Tuninetti11}. 
Under the same channel conditions, \cite{VariableRateRetransmission} considers rate adaptation for an HARQ-INR protocol without prior nor delayed CSIT. Dynamic programming is used to maximize the throughput under an outage constraint and it is found that rate adaptation provides significantly lower outage probabilities. The assumption of independent channel gains is relaxed in \cite{OptimalRateAdaptation}, where optimal rate adaptation policies are found for the cases in which the channel gains are correlated. 

Although prior CSIT improves the throughput of HARQ-INR remarkably, CSIT is often delayed when it is obtained by the transmitter. This has led to another line of work which studies the benefits of delayed CSIT in context of HARQ-INR protocols. Specifically, \cite{Szczecinski13} and \cite{MultipacketHybridARQ} considers a point-to-point channel with independent block-fading in a setting identical to ours. Apart from the statistics of the channel gain, the transmitter has no knowledge about the current CSI, but the transmitter is informed about the CSI of the previous slot. In their protocol, the channel uses of each slot are divided among a large number of parallel HARQ-INR instances transmitting separate messages in a time division multiplexing (TDM) fashion. In particular, for a specific HARQ-INR instance, the number of channel uses used for the $k$th retransmission is some percentage $0\leq\ell_k \leq 1$ of the number of channel uses spend in the first transmission. This implies that new HARQ-INR instances, with new data, can be initiated in each slot. The objective is to maximize the throughput under a constraint on the outage probability. It is found that delayed CSIT significantly decreases the outage probabilities.
A similar setting was considered in \cite{PowerAdapt}, where power adaptation was investigated. Here, the authors used a conventional HARQ-INR instance, but adapted the power in each slot according to the delayed CSIT. In contrast to \cite{Szczecinski13}, in which the authors design composite protocols based on a large number of HARQ-INR instances, the protocol proposed in \cite{PowerAdapt} only uses a single HARQ-INR instance with power adaptation which is optimized using dynamic programming. Rate adaptation can also be achieved using superposition coding. A multi-layer broadcast approach to fading channels without prior CSIT is proposed in \cite{broadcast_approach_isit}. Specifically, a transmission is initiated in large number of superposition coded layers and the number of decoded layers at the receiver depends on the actual CSI, which is assumed not to be known in advance. This approach provides an alternative to HARQ protocols in the sense that it provides variable-rate transmission with a fixed transmission length of one slot. The approach, however, has the disadvantage that the throughput in practical implementations suffer as the number of layers increases. A more practical approach is taken in \cite{multilayer_broadcast} which combines the approach in \cite{broadcast_approach_isit} for few layers with HARQ-INR. Specifically, the proposed protocols initiate an HARQ-INR instance in each layer. In a certain slot, the receiver feeds back the number of decoded layers and, in the subsequent slot, the transmitter only conveys IR for the layers not decoded. For the layers that are decoded, the transmitter initiates new HARQ-INR instances with new data. 
Finally, although not directly related to our work, it was shown in \cite{stale_csi} that delayed CSIT, which is possibly completely independent of the current channel state, increases the multiplexing gains in a multiple-input multiple-output (MIMO) broadcast channel with $K$ transmit antennas and $K$ receivers each with one receive antenna.

In contrast to previous works, this paper is motivated by the \emph{backtrack retransmission request (BRQ)} protocol proposed in \cite{backtrack}. BRQ is suited for systems in which the transmission opportunities come in slots of a predefined number of channel uses. 
 This prevents conventional HARQ-INR to optimize the throughput, as the number of channel uses cannot be adapted to the required amount of IR. BRQ overcomes this problem by appending additional new information bits before the information bits sent in previous slot have been decoded. The number of new information bits is adapted according to the reported delayed CSIT. 
Our approach in this paper combines the idea of appending new data during a transmission for HARQ in \cite{backtrack}, \cite{Szczecinski13}, and \cite{multilayer_broadcast} with streaming codes proposed in \cite{streaming_transmission}  and \cite{Khisti2014}. The streaming codes in \cite{streaming_transmission}  and \cite{Khisti2014} are a family of codes that allow the transmitter to append new information bits during a transmission in such a way that all information bits can be jointly decoded as one code. In \cite{streaming_transmission}, each message has the same absolute deadline at which all messages need to be decoded. In \cite{Khisti2014}, each message is required to be decoded within a certain number slots after arrival. Both \cite{streaming_transmission} and \cite{Khisti2014} use a transmission scheme that enlarges the message space in each slot. In coding theory, streaming codes, as those investigated in \cite{streaming_transmission} and \cite{Khisti2014}, are also known as cross-packet codes. Cross-packet codes based on Turbo codes and LDPC codes have previously been considered in the context of HARQ in \cite{CrossPacketConv} and \cite{CrossPacketLDPC}, respectively. 
The EMS protocols proposed in this paper extend streaming codes to an HARQ-INR setting in which the amount of new information bits that are appended within a retransmission is adaptive, as it depends on the delayed CSIT in manner similar to BRQ.

EMS protocols are thus variable-rate protocols in a sense similar to \cite{Szczecinski13} and \cite{multilayer_broadcast}. However, to the best of our knowledge, all previously proposed protocols that allow for rate adaptation are composite protocols based on a conventional HARQ-INR protocol as building block, where rate adaptation is achieved by using a large number of parallel HARQ-INR instances in a TDM fashion or in superposition coded layers. These approaches incur rate penalties in practical implementations because each HARQ-INR instance only uses a small fraction of the available resources (channel uses/power) in each slot.
In contrast, EMS protocols differ fundamentally from HARQ-INR in the way new information bits are appended in each slot. This implies that, in principle, one can use our scheme instead of HARQ-INR as a building block and devise protocols similar to \cite{Szczecinski13} and \cite{multilayer_broadcast}. Consequently, we consider HARQ-INR and HARQ-INR with power adaptation based on delayed CSIT as relevant baseline protocols for comparison. 

\subsection{The backtrack retransmission protocol}\label{sec:brq_desc}
Since our work is motivated by BRQ, we shall provide a brief description of the protocol below. Suppose the transmitter sends to the receiver in slots, where each slot is a fixed communication resource that consists of $n$ channel uses. The channel is modeled as a Gaussian block-fading channel with channel gains $\{H_t\}$ of the slots being independent and identically distributed. Assume also that the transmitter uses unit transmission power such that $H_t$ is the SNR in the $t$th slot. The channel gain $H_t$ is fed back to the transmitter by the end of the $t$th slot. The BRQ protocol uses a single channel code with blocklength $n$ and a fixed rate $R$ in each slot such that the receiver can decode if $C(H_t) > R$, where we have defined
\begin{IEEEeqnarray}{rCl}
    C(h) \triangleq \frac{1}{2}\log_2(1+h).\label{eq:cap_def}
\end{IEEEeqnarray}
In the first slot, the transmitter sends $n R$ bits of new information using the fixed channel code. If the realized channel gain $H_1$ satisfies $C(H_1) > R$, the receiver decodes the packet, extracts the $nR$ information bits, and the protocol terminates with a decoding time of one slot. On the other hand, if $C(H_1) \leq R$, the receiver cannot decode the packet, it feeds back the CSI of the first slot, and the protocol 
continues in slot $2$. Considering the $k$th slot, with $k\geq 2$ and assuming that $C(H_t) \leq R$ for all $t\in\{1,\cdots,k-1\}$, the transmitter forms the packet of $nR$ bits for the $k$th slot as follows:
\begin{enumerate}
  \item The first $n(R-C(H_{k-1}))$ bits are IR that allow the decoding of the packet in slot $k-1$. 
  \item The remaining $n C(H_{k-1})$ bits are new information bits.
\end{enumerate}
Note that $H_{k-1}$ is fed back to the transmitter by the end of slot $k-1$ and thereby known at the transmitter in slot $k$.
If $C(H_k) \leq R$, the receiver feeds back the CSI of the slot and the protocol continues in slot $k+1$. If $C(H_k) > R$, the receiver can decode the packet in slot $k$ and it can recover the $n C(H_{k-1})$ information bits. It also recovers the $n(R-C(H_{k-1}))$ bits of IR for the packet in slot $k-1$. At this time, the receiver can decode the packet conveyed in the $(k-1)$th slot using the side information from the IR bits in slot $k$. Next, the decoder sequentially decodes the packets $(k-2),(k-3), \cdots, 1$ in a similar fashion,  thereby recovering all the $n(R+C(H_1)+\cdots+C(H_{k-1}))$ bits. Over the same slots, one could have transmitted $n(C(H_1)+\cdots+C(H_{k}))$ information bits if the channel gains had been available a priori (and assuming that power adaptation was not used). The loss in throughput by BRQ is thus only due to the difference $C_k - R$. The throughput of BRQ, reported in \cite{backtrack}, is restated in Theorem~\ref{thm:brq_opt}.

We note that the IR bits and the new information bits are only separable in the digital domain, but not at the physical layer. Hence, the receiver needs to decode the whole packet, which is transmitted using the fixed  channel code with rate $R$, in order to extract the IR bits and the new information bits.

We observe that BRQ relies on appending information bits to the parity bits. The transmission rate used in BRQ is predefined to be $R$ in each slot. The number of appended information bits is computed based on delayed CSIT but chosen such that the a priori probability of decoding a certain slot is kept constant. Hence, the BRQ protocol ends a transmission as soon as the CSI is above a level that is sufficient for decoding the predefined rate $R$.

\subsection{Contribution}
In this paper, we generalize the BRQ protocol from \cite{backtrack}. First, we propose a family of EMS protocols that allow the transmitter to expand the message space in manner similar to BRQ. In contrast to BRQ, however, the EMS protocols are based on streaming codes and all information bits are decoded jointly. The notion of an EMS protocol introduced here is sufficiently general to include protocols like ARQ, HARQ-INR, and BRQ. Next, we prove a converse and an achievability result for the EMS protocols, and it is shown that the throughput of the optimal zero outage EMS protocol given a constraint on the average decoding time and full delayed CSIT is identical to the throughput of BRQ. Then, we address the same problem with only a finite number of feedback messages in each slot. In this case, we put forth heuristic EMS protocols which have a structure similar to BRQ, but are designed to work with a finite number of feedback messages. Finally, the throughput of BRQ and the proposed finite feedback EMS protocols are evaluated and compared to relevant baseline protocols. Specifically, we compute the throughput in terms of SNR and in terms of average decoding time. Our numerical results confirm that the throughput of BRQ converges to the ergodic capacity faster than the throughput of HARQ-INR. Moreover, the proposed finite feedback EMS protocol using only three feedback messages per slot achieves throughput which is only slightly worse than that of BRQ. We remark that EMS protocols have previously been introduced in \cite{ems}, where we used finite blocklength analysis to investigate a protocol similar to BRQ in a simplified setup. In a similar setting, optimal rate adaptation policies were optimized using error exponents in \cite{EMS2}.

\subsubsection*{Notation}
Vectors are denoted by boldface (e.g., $\vect{a}$), while their entries are denoted by roman letters (e.g., $a_i$). The transpose of a vector $\mathbf{a}$ is denoted by $\mathbf{a}\tr$, the length of a vector by $\len{\cdot}$, and the tuple $(a_i,\cdots,a_j)$, for $i\leq j$, is denoted by $a_i^j$.
Similarly, we denote a tuple of random variables $(X_i,\cdots,X_j)$, $j\geq i$, by $X_i^j$. We adopt the convention that $\sum_{i=j}^{j-1} a_i = 0$ and likewise we let $X_{i}^{i-1}$ be the empty tuple. Let $\mathbb{N}$ be the natural numbers, $\mathbb{R}$ be the reals, and $\mathbb{R}_+$ be the nonnegative reals. Moreover, the range of integers $\{i,\cdots,j\}$, $i\leq j$, is denoted by $\range{i}{j}$.  We also use the standard asymptotic notation $f(n) = \mathcal{O}(g(n))$ and $f(n) = o(g(n))$ which means that $\limsup_{n\rightarrow \infty} |f(n)/g(n)| < \infty$ and that $\limsup_{n\rightarrow \infty} |f(n)/g(n)| = 0$, respectively.
Finally, we let $\left[x\right]^- \triangleq \min\{x,0\}$.
 
\section{System Model}\label{sec:system_model}
We consider a single-user block-fading channel with Gaussian noise. The transmitter sends to the receiver in slots of $n$ channel uses, where $n$ is sufficiently large to offer reliable communication that is optimal in an information-theoretic sense. The received signal vector in slot $t\in\mathbb{N}$ is given by 
\begin{align}
  \mathbf{Y}_t = \sqrt{H_t} \mathbf{X}_t + \mathbf{Z}_t
\end{align}
where $\mathbf{Z}_t \sim \mathcal{N}(\mathbf{0}_n,\mathbf{I}_n)$ is an $n$-dimensional noise vector distributed according to the Gaussian distribution with zero mean and identity covariance matrix, $\mathbf{X}_t\in\mathbb{R}^n$ is the transmitted vector satisfying 
\begin{IEEEeqnarray}{rCl}
\frac{1}{n}\mathbf{X}_t\tr \mathbf{X}_t\leq 1\label{eq:power_const}
\end{IEEEeqnarray}
and $H_t\geq 0$ denotes the instantaneous channel gain, drawn independently from a smooth probability density $P_{H}(\cdot)$ with support on $\mathbb{R}_+$. The cumulative distribution function of $H_t$ is given by $F_H(\cdot)$. The instantaneous channel gain $H_t$ is unknown at the transmitter prior to the transmission of $\mathbf{X}_t$ but is known at the receiver after observing $\mathbf{Y}_t$. Moreover, the receiver is able to provide feedback based on the CSI. Specifically, we assume that feedback is given by a sequence of feedback functions $\mathbb{v}_t: \mathbb{R}_+^{t} \rightarrow \feedset$ that maps $H_1^t$ to a feedback alphabet $\feedset$ such that $V_t = \mathbb{v}_t(H_1^t)$ is observed at the transmitter before transmission in the $(t+1)$th slot. The \textit{feedback cost} is defined as the cardinality of the feedback alphabet $|\feedset|$ and may be finite, countably infinite, or uncountably infinite. The transmitter is said to have full delayed CSIT if $H_{t}$ can be recovered from $V_t$.

If a transmission is to be done over slot $t$ alone, the maximum supported rate is given by $\C{H_t}$, whereas the maximum achievable rate if a transmission is done over many slots approaches the ergodic capacity \cite{Goldsmith}
\begin{align}
  C_{\text{erg}} = \E{\C{H}}
\end{align}
as the number of slots tends to infinity. 
Here, $H$ denotes a random variable distributed according to $P_H$. 
If, however, a transmission is to be done over few slots, high throughput cannot be achieved without either layered transmissions as in \cite{multilayer_broadcast} or a HARQ technique. The latter approach is commonly applied in practical systems due to its relative simplicity compared to the layered transmissions.

A comment on the block-fading assumption is in order. The block-fading channel model is an abstraction of a practical system model. In particular, if slots are transmitted consecutively in time as this model suggests, the channel gains cannot be assumed to be independent.  In practical systems, however, the delay of ACK/NACK feedback can often spread over multiple slots in time. Therefore, in wireless systems such as LTE, multiple HARQ instances are interleaved in time \cite[Ch.~12]{book4G}; while the transmitter waits for feedback from one HARQ instance, it transmits to other users. In the uplink in LTE, a synchronous version of HARQ is employed \cite[Ch.~12]{book4G}. This ensures that the time between each retransmission is fixed and known by both the transmitter and the receiver. The fact that each transmission opportunity is spaced apart by a fixed number of slots implies that channel gains can be assumed to be independent for many scenarios. In addition to these considerations, one cannot expect that each transmission opportunity occurs in the same frequency slot; this further justifies the use of a block-fading model.

An EMS protocol is now defined by
\begin{itemize}
  \item A sequence of feedback functions $\mathbb{v}_t: \mathbb{R}_+^{t} \mapsto \feedset$ that maps $H_1^t$ to the feedback alphabet $\feedset$ such that
  \begin{align}
  V_t \triangleq \mathbb{v}_t(H_1^t).
  \end{align}
  \item A sequence of rate selection functions $\mathbb{r}^{(n)}_t : \feedset \mapsto \mathbb{R}_+$ that satisfy $R_t^{(n)} \triangleq \mathbb{r}_t^{(n)}(V_{t-1})$, $\mathbb{r}^{(n+1)}_t(\cdot) \geq \mathbb{r}^{(n)}_t(\cdot)$ for all $t\in\mathbb{N}$, and $R_t^{(n)} \leq \mathbb{r}_{\text{max}}$ for some positive constant $\mathbb{r}_{\text{max}}$. We also define the cumulative rates $\cumR^{(n)}_t \triangleq \sum_{k=1}^t R_k^{(n)}$.
  \item A sequence of encoding functions $\mathbb{f}_{t}^{(n)} : \mathfrak{B}\mapsto \mathbb{R}^n$ such that
  \begin{align}
    \mathbf{X}_t \triangleq \mathbb{f}_{t}^{(n)}\farg{B_1^{\lceil n \cumR_t^{(n)}\rceil}}.\label{eq:enc_func}
  \end{align}
  Here, $\mathfrak{B}$ denotes all binary vectors (of arbitrary length), i.e., $\mathfrak{B}\triangleq \{[]\}\cup\bigcup_{i=1}^\infty \{0,1\}^i$, where $[]$ denotes the vector of length $0$; $B_i$ are independent Bernoulli variables with parameter $1/2$; and the tuple $(B_i,\cdots,B_j)$ is denoted by $B_i^j$. 
  \item A sequence of decoding functions $\mathbb{g}_{t}^{(n)}: \mathbb{R}^{t n} \times \mathbb{R}_+^{t}\mapsto \mathfrak{B}$. 
  \item A sequence of nonnegative integer-valued random variables $\{\tau_n\}_{n=1}^\infty$, which are stopping times with respect to the filtration $\mathcal{F}_t \triangleq \sigma\{V^t\}$ (see e.g. \cite[p.~488]{billingsley}) and satisfy $\tau_{n+1}\geq \tau_n$ and $\sup_n \E{\tau_n}<\infty$.
  \end{itemize}
  The error event of an EMS protocol is given by
  \begin{IEEEeqnarray}{rCl}
\mathcal{E}_n &\triangleq& \Big\{\mathbb{g}^{(n)}_{\tau_n}(\mathbf{Y}_1^{\tau_n}, H^{\tau_n}) \not= B_1^{\lceil n \cumR_{\tau_n} \rceil}\Big\}.\label{eq:error_event}
\end{IEEEeqnarray}
  We also define the limiting rate selection functions and stopping time of an EMS protocol:
\begin{IEEEeqnarray}{rCl}
  \mathbb{r}_t &\triangleq& \lim_{n\rightarrow \infty}\mathbb{r}^{(n)}_t\\
  \tau &\triangleq& \sup_{n} \tau_n.
\end{IEEEeqnarray}
The limit of $\mathbb{r}^{(n)}_t$ exists because $\mathbb{r}_t^{(n)}$ is non-decreasing in $n$ and bounded above by $\mathbb{r}_{\text{max}}$. On the other hand, we define $\tau$ as the supremum over $\tau_n$ since the existence of the limit of $\tau_n$ cannot be guaranteed because only $\E{\tau_n}$ is bounded above for increasing $n$.

The random variables $B^{\infty}\in\{0,1\}^\infty$ correspond to the binary sequence of information bits, which size in bits is unbounded. We assume that all the information bits are available prior to the transmission in the first slot. This implies that the stopping time $\tau_n$ is also the decoding time and the transmission time in slots. In the remainder of this paper, we shall refer to $\tau_n$ as a decoding time. We note that our definition of decoding time deviates from some other works. For example, in \cite{VariableRateRetransmission} and \cite{Szczecinski13}, the decoding time is measured as the time from the information bits are appended to the time at which they are decoded. 

As an implication of the definition of an EMS protocol, $\mathbf{X}_t$ becomes a function of the information bits $B_{1}^{\lceil n \cumR_t^{(n)}\rceil}=(B_1,\cdots, B_{\lceil n \cumR_t^{(n)}\rceil})$.  This enables the encoder to combine IR and new information bits, i.e., in each slot the encoder fetches $n R_t^{(n)}$ information bits and encodes them jointly with the previously  encoded $n \cumR_{t-1}^{(n)}$ information bits. This message structure is different from other works on HARQ-INR protocols. In light of \cite{variable_rate}, HARQ-INR can be seen as fixed-to-variable coding because the number of transmitted information bits are prespecified while the number of channel observations at the receiver depends on channel realization. On the other hand, for an EMS protocol, both the number of information bits and the number of channel observations depend on the channel realization. This concept has previously been used in \cite{Szczecinski13} and \cite{multilayer_broadcast}; however, none of these works alter the conventional HARQ-INR protocol. They rather use it as a building block and initiate a large number of HARQ-INR instances which run in parallel consecutively in time or in multiple superposition coded layers.

Following other HARQ works \cite{Caire2001,renewal_arq,multilayer_broadcast}, we define the throughput $\eta$ of an EMS protocol in terms of a renewal-reward process. A renewal event occurs at time $\tau_n$ and the reward is the sum of all rates appended since time $1$. Likewise, the inter-renewal time corresponds to the decoding time $\tau_n$. Hence, we define the throughput of an EMS protocol as $\lim_{n\rightarrow \infty}\Ebig{\cumR^{(n)}_{\tau_n}}/ \E{\tau_n} $. 
This leads us to the definition of a zero outage EMS protocol.
\begin{definition}
\label{def:ems}
  An EMS protocol is called an $(\eta,T)$-zero outage EMS protocol if there exists a non-decreasing integer-valued sequence $\{\taumax\}$ such that  $\tau_n \leq \taumax$, $\E{\tau_n}\leq T$, $\lim_{n\rightarrow \infty}\Ebig{\cumR_{\tau_n}^{(n)}}/ \E{\tau_n} \geq \eta$, 
  \begin{IEEEeqnarray}{rCl}
  \lim_{n \rightarrow \infty}\pr{\mathcal{E}_n} = 0\label{eq:outage_error_cond}
  \end{IEEEeqnarray}
  and
  \begin{IEEEeqnarray}{rCl}
   \lim_{n\rightarrow \infty} \max_{\substack{t\in\range{1}{\taumax-1}:\\ \pr{\tau_n = t}>0}} \pr{\mathcal{E}_n | \tau_n = t} = 0.\label{eq:strong_error_cond}
  \end{IEEEeqnarray}
\end{definition}

Our focus is on the characterization of optimal zero outage EMS protocols:
\begin{IEEEeqnarray}{rCl}
  \eta_{\text{opt}}(T) \triangleq \sup \{\eta: \exists (\eta, T)\text{-zero outage EMS protocol}\}.\IEEEeqnarraynumspace\label{eq:eta_opt_def}
\end{IEEEeqnarray}
The condition in \eqref{eq:outage_error_cond} ensures that the outage probability of the EMS protocol is zero, while the condition in  \eqref{eq:strong_error_cond} ensures that the conditional probability of error given a decoding time vanishes uniformly for all decoding times except for $\taumax$. We note that our converse result does not hinge on the condition in \eqref{eq:strong_error_cond}; it is only introduced to strengthen the achievability result.

We note that most other HARQ works consider strict latency constraints which naturally arise in wireless communication systems having either a strict deadline or a limited buffer size. We consider average decoding time constraints and zero outage protocols for two reasons:
\begin{itemize}
\item A strict latency constraint does not naturally arise in systems without a strict deadline or limited buffer size, and hence, in such applications, there is no reason to choose a specific deadline  $T$ in the strict decoding time constraint $\pr{\tau_n \leq T} = 1$. For example, consider an application that requires high reliability. In this case, imposing a strict latency constraint for the HARQ protocol only implies that the receiver will request a retransmission of the data in outage. This is the case for LTE, which uses HARQ in the medium access control (MAC) layer, while it implements an ARQ protocol on a higher layer -- in the radio link control (RLC) layer --  that requests retransmissions for data in outage \cite[Ch.~12]{book4G}. In that sense, LTE attempts to achieve an outage probability close to zero, and an average decoding time constraint is therefore a natural constraint which attempts to keep the decoding time low on average but does not give any strict guarantees. As previously mentioned, LTE employs synchronous HARQ in the uplink which implies that the decoding time $\tau_n$ is indeed proportional to real decoding time in a system. We also point out that the customary metric for latency in queuing theory is the average waiting time.  
\item It turns out that the throughput of the optimal zero outage EMS protocol, under an average decoding time constraint, coincides with the throughput of the BRQ protocol proposed in \cite{backtrack}, i.e., the optimization problem in \eqref{eq:eta_opt_def} has a simple form.
\end{itemize}

\section{Achievability and Converse}
In this section, we state converse and achievability results that we shall apply in the subsequent sections. The achievability and converse results state conditions for when the probability of error tends to zero or one, respectively.
In order to state the results, it is convenient to introduce some notation. In particular, given rate selection functions and feedback functions, let
\begin{align}
 \mathbb{u}_{k,t}^{(n)}(h_1^{t})&\triangleq \sum_{i=k}^{t}\left(\mathbb{r}^{(n)}_{i}( \mathbb{v}_{i-1}(h_1^{i-1})) - \C{h_i}\right)\label{eq:bbm_u_def}
\end{align}
for $k\leq t$ and let $\mathbb{u}_{k,t}^{(n)}(\cdot) \triangleq 0$ for $t<k$.
Intuitively, $\mathbb{u}_{1,t}^{(n)}(h_1^{t})$ is the remaining amount of information needed to decode the information bits appended up to time $t$ given the channel gains $h_1^{t}=(h_1,\cdots,h_t)\in\mathbb{R}_+^t$. We also define 
\begin{IEEEeqnarray}{rCl}
  \mathbb{u}_{k,t}(h_1^{t}) &\triangleq& \lim_{n\rightarrow \infty}\mathbb{u}_{k,t}^{(n)}(h_1^{t}) \\
  &=& \sum_{i=k}^{t}\left(\mathbb{r}_{i}( \mathbb{v}_{i-1}(h_1^{i-1})) - \C{h_i}\right).
\end{IEEEeqnarray}
We prove the following results in Appendix~\ref{app:converse_proof} and Appendix~\ref{app:achiev_proof}.
\begin{lemma}[converse] 
  Given an EMS protocol, we have
    \begin{IEEEeqnarray}{rCl}
      \lim_{n\rightarrow \infty}\pr{ \mathcal{E}_n\Big| H^\infty =h^\infty} = 1
    \end{IEEEeqnarray}
    for every $h^\infty\in\mathbb{R}_+^\infty$ satisfying $\sup_{k\in\range{1}{\tau}}\mathbb{u}_{k,\tau}(h^\tau) > 0$ and $\tau < \infty$ given that $H^\infty = h^\infty$.
    \label{lem:strong_converse}
\end{lemma}
\begin{remark}
The conditions in Lemma~\ref{lem:strong_converse} are only given in terms of the asymptotic quantities $\tau$ and $\mathbb{r}_t$ and not $\tau_n$ and $\mathbb{r}_t^{(n)}$. Therefore, Lemma~\ref{lem:strong_converse} allows us to restrict the search for optimal zero outage EMS protocols to those EMS protocols for which $\sup_{k\in\range{1}{\tau}}\mathbb{u}_{k,\tau}(h^\tau) \leq 0$ almost surely.
\end{remark}
\begin{remark}
 The smallest limiting decoding time of a zero outage EMS protocol which is not ruled out by Lemma~\ref{lem:strong_converse} is given by
  \begin{align}
  \tau_{\text{opt}} &\triangleq \inf\mathopen{}\left\{t\geq 1: \mathbb{u}_{1,t}(H_1^t) \leq 0\right\}.\label{eq:opt_EMS_tau}
  \end{align}
To show that an EMS protocol with $\tau=\tau_{\text{opt}}$ is not ruled out by Lemma~\ref{lem:strong_converse}, note that by the definition of $\tau_{\text{opt}}$, we must have
\begin{align}
  \mathbb{u}_{1,1}(H_1^1)>0,\cdots,\mathbb{u}_{1,\tau_{\text{opt}}-1}(H_1^{\tau_{\text{opt}-1}}) > 0 
  \end{align}
   and
\begin{align}
    \mathbb{u}_{1,\tau_{\text{opt}}}(H_1^{\tau_{\text{opt}}}) \leq 0.
\end{align}
Thus, using the fact that $\mathbb{u}_{k,\tau_{\text{opt}}}(H_1^{\tau_{\text{opt}}}) = \mathbb{u}_{1,\tau_{\text{opt}}}(H_1^{\tau_{\text{opt}}})-\mathbb{u}_{1,k-1}(H_1^{k-1}) \leq 0$ for every $k\in\range{1}{\tau_{\text{opt}}}$, we find that the conditions in Lemma~1 cannot be simultaneously satisfied.
\end{remark}

    
\begin{lemma}[achievability]
    Let decoding times $\{\tau_n\}$, rate selection functions $\{\mathbb{r}_t^{(n)}\}$, and feedback functions $\{\mathbb{v}_t\}$ be given. Suppose that there exist positive sequences $c_n$, $g_n$, and $\taumax$ such that $\taumax\in\mathbb{N}$ is a nondecreasing sequence satistying $\tau_n\leq \taumax$ and  such that
    \begin{IEEEeqnarray}{rCl}
      \frac{\taumax^2}{n g_n c_n^2} &\rightarrow& 0\label{eq:achievability_conv_cond}
    \end{IEEEeqnarray}
    as $n\rightarrow \infty$. Moreover, define the event
    \begin{IEEEeqnarray}{rCl}
  \mathcal{\bar H}_n &\triangleq& \left\{ \max_{k\in\range{1}{\tau_n}}\mathbb{u}_{k,\tau_n}^{(n)}(H^{\tau_n}) \leq -c_n \right\}\label{eq:Hbar}
\end{IEEEeqnarray}
and assume for all sufficiently large $n$ that 
  \begin{IEEEeqnarray}{rCl}
  \min_{\substack{t\in\range{1}{\taumax}:\\ \pr{\tau_n = t|\mathcal{\bar H}_n} > 0 }} \pr{ \tau_n = t|\mathcal{\bar H}_n} &\geq& g_n.\label{eq:min_prob_cond}
  \end{IEEEeqnarray}
  Then, there exists an EMS protocol satisfying
  \begin{IEEEeqnarray}{rCl}
      \lim_{n\rightarrow \infty}\max_{\substack{t\in \range{1}{\taumax}:\\ \pr{\tau_n=t}>0}}\pr{\mathcal{E}_n\Big| \mathcal{\bar H}_n, \tau_n = t} = 0.\label{eq:lem_achievability_conclusion}
  \end{IEEEeqnarray}
  \label{lem:achievability}
\end{lemma}

\section{Full Delayed CSIT}
In this section, we consider the case in which the feedback alphabet is the positive reals, $\feedset = \mathbb{R}$, and the feedback functions are given by 
\begin{IEEEeqnarray}{rCl}
\mathbb{v}_t(h_1^t) \triangleq h_t.\label{eq:full_csi_vt}
\end{IEEEeqnarray}
This provides the transmitter with full delayed CSIT. In the following, we characterize the trade-off between throughput and the average decoding time for optimal zero outage EMS protocols. 
First, we specify an EMS protocol and we shall later show that it is an optimal zero outage EMS protocol. The EMS protocol is specified as follows
\begin{IEEEeqnarray}{rCl}
\mathbb{r}^{(n)}_t(v) &\triangleq&\left\{ \begin{array}{ll}
C(h_T) - \frac{c_1}{\log n}, & t = 1\\
\min\mathopen{}\left\{C(v),C(h_T)-\frac{c_1}{\log n}\right\}, & t \geq 2
\end{array}\right.\label{eq:full_csi_rt}
\end{IEEEeqnarray}
 for a positive constants $h_T$ and $c_1$. The decoding times are given by
 \begin{IEEEeqnarray}{rCl}
\tau_n &\triangleq& \min\{\taumax, \tau\}\label{eq:tau_n_def_full}
\end{IEEEeqnarray}
where 
\begin{IEEEeqnarray}{rCl}
\taumax &\triangleq& -\left\lfloor \frac{\log (c_2\sqrt{n})}{\log F_H(h_T)}\right\rfloor\label{eq:taumax_full}\\
\tau &\triangleq& \inf\mathopen{}\left\{t\geq 1: h_T < H_t \right\}\label{eq:achiev_au_full}
\end{IEEEeqnarray}
for an arbitrary constant $c_2>0$.
The particular choice of the rate selection functions has a simple operational interpretation when neglecting the vanishing term $c_1/\log n$. Consider a transmitter using a fixed-rate codebook with rate $\C{h_T}$ in each slot such that the minimal channel gain required to decode a slot is $h_T$. Based on the delayed CSI, in slot $t$, the transmitter sends the exact amount of IR that is required to decode the previous packet, i.e., $n(C(h_T)-\C{H_{t-1}})$ bits, along with $ n \C{H_{t-1}}$ bits of new information bits. This protocol resembles the BRQ protocol previously described in Section~\ref{sec:brq_desc} but formulated as an EMS protocol.  

\begin{figure}[!t]
\centering
\subfigure[HARQ-INR.]{
  \includegraphics[width=3.4in]{\PathBRQ/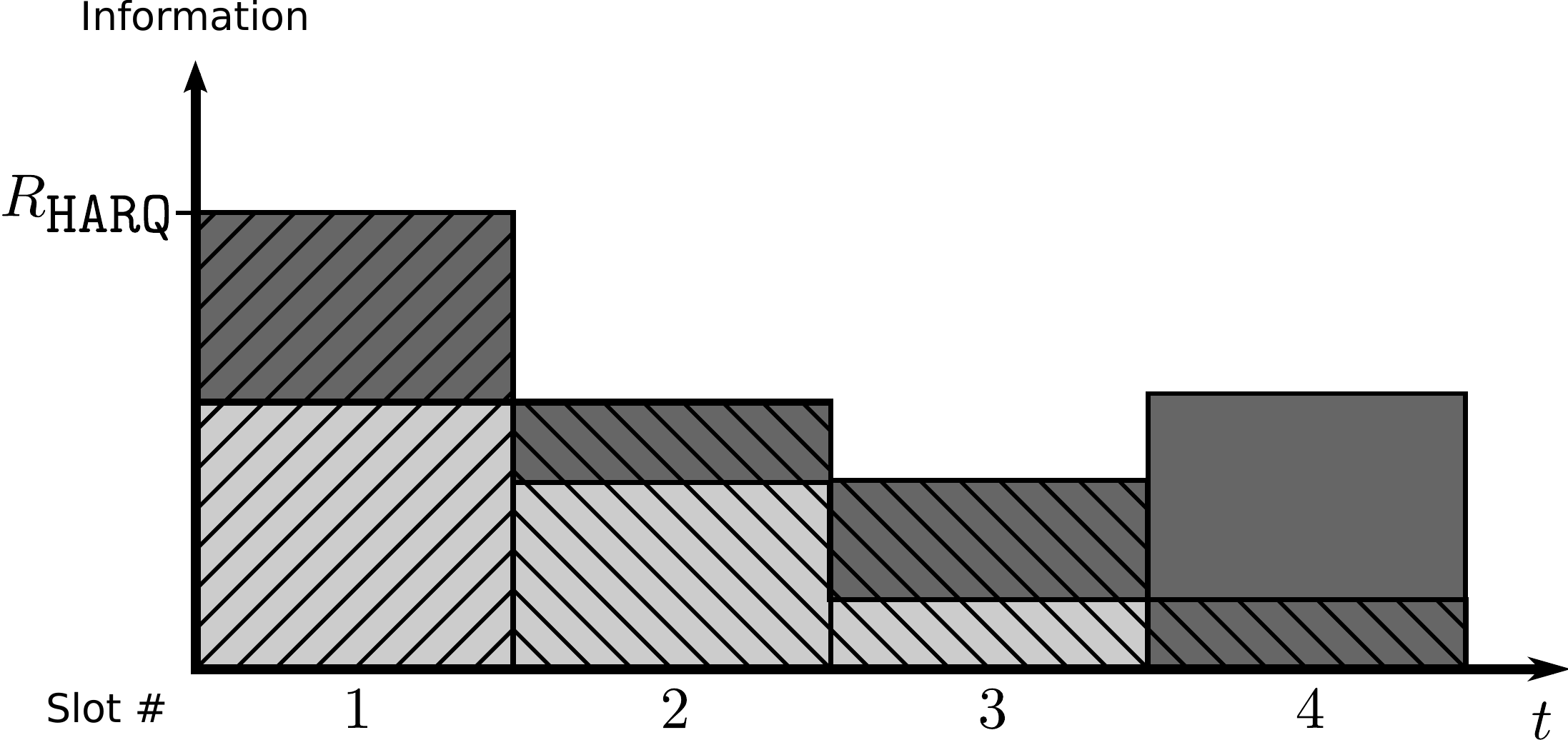}
  \label{fig:brq_comp}
}
\subfigure[BRQ.]{
  \includegraphics[width=3.4in]{\PathBRQ/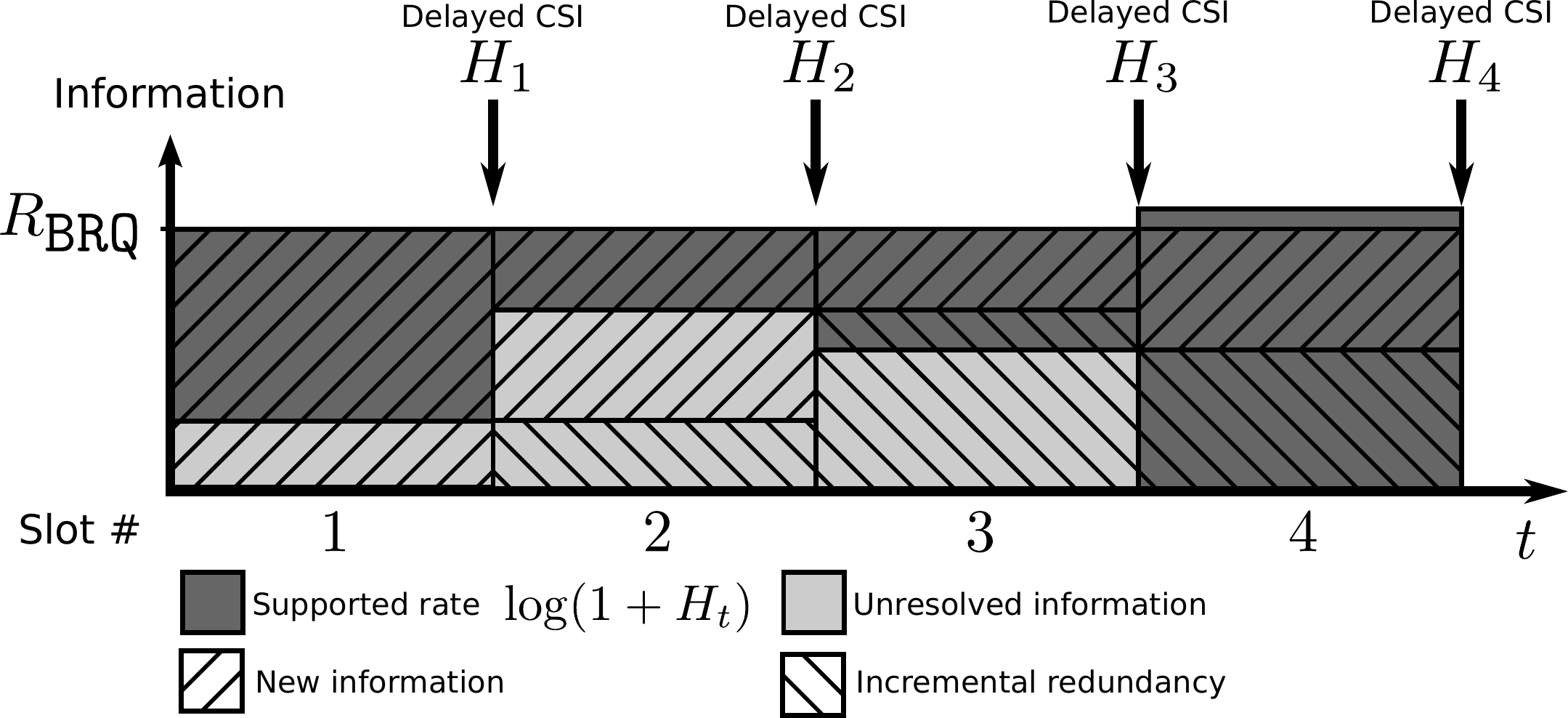}
  \label{fig:brq}
}
\caption{Comparison between HARQ-INR and BRQ. In slot $t$, the left and right striped areas correspond to the amount of unresolved information before receiving $\mathbf{Y}_t$. The dark grey areas designate the instantaneous supported rate and the light grey areas corresponds to the unresolved information after observing $\mathbf{Y}_t$. Note that for each time slot, the dark grey areas have the same size for both HARQ-INR and BRQ.}\label{fig:brq_ill}
\end{figure}

The operation of BRQ is illustrated and compared to HARQ-INR in Fig.~\ref{fig:brq_ill}. Initially, HARQ-INR transmits at a rate $R_{\text{HARQ}}$. The receiver accumulates information until the amount of unresolved information reaches zero. BRQ starts the transmission at a rate $R_{\text{BRQ}}$, but in contrast to HARQ-INR, it uses the delayed CSI to append new information bits in each slot to ensure that the amount of unresolved information, before the receiver observes $\mathbf{Y}_t$ and $H_t$, remains $R_{\text{BRQ}}$. Note that, in order to attain the same average decoding time for BRQ and HARQ-INR, $R_{\text{BRQ}}$ needs to be chosen smaller than $R_{\text{HARQ}}$ since no additional information bits are appended during transmission in the HARQ-INR protocol. This is why we have chosen $R_{\text{HARQ}} > R_{\text{BRQ}}$ in the figure. For the particular realization of channel gains depicted in Fig.~\ref{fig:brq_ill}, it is seen that HARQ-INR does not fully utilize the supported rate since the unresolved information, before $\mathbf{Y}_4$ and $H_4$ are observed, is significantly smaller than the supported rate in that slot. This phenomenon reduces the throughput at low average decoding times. The problem is partially circumvented in BRQ by ensuring that the amount of unresolved information, before $\mathbf{Y}_t$ and $H_t$ are observed, is kept constant. 
In contrast to BRQ, the EMS protocol specified by  \eqref{eq:full_csi_vt} and \eqref{eq:full_csi_rt} uses joint decoding over all slots. Since the EMS protocol specified by  \eqref{eq:full_csi_vt} and \eqref{eq:full_csi_rt} and BRQ are closely related, we shall refer to the proposed EMS protocol as ``BRQ-EMS'' to emphasize its relation to BRQ.

The following result analyzes the trade-off between throughput and average decoding time of BRQ-EMS. Specifically, we find that the throughput is identical to that of BRQ. Furthermore, we apply the converse result in Lemma~\ref{lem:strong_converse} and we demonstrate using dynamic programming that BRQ-EMS is optimal within the class of zero outage EMS protocols.
\begin{theorem}
For $T>1$, we have
  \begin{equation} \label{eq:AverageRateFullCSIT}\eta_{\text{opt}}(T) \geq \eta_{\text{BRQ}}(T) \triangleq\int_{0}^{h_T}
    P_{H}(h) \C{h} \mathrm{d}h + \frac{C(h_T)}{T}
  \end{equation}
  where
  \begin{align}
  h_T\triangleq F_H^{-1}\farg{1 - \frac{1}{T}}.\label{eq:expected_tau}
  \end{align}
  Moreover, we have that $\eta_{\text{BRQ}}(T) = \eta_{\text{opt}}(T)$  if
  \begin{IEEEeqnarray}{rCl}
  \frac{P_H(h)}{1-F_H(h)} + \frac{1}{(1+h) }+ \frac{P_H'(h)}{P_H(h)} \geq 0\label{eq:opt_cond}
\end{IEEEeqnarray}
for every $h \geq 0$. Here, $P'_H(\cdot)$ denotes the derivative of $P_H(\cdot)$.
  \label{thm:brq_opt}
\end{theorem}
\begin{remark}
  The throughput of BRQ, which is identical to \eqref{eq:AverageRateFullCSIT}, was first reported in \cite{backtrack}.
\end{remark}
\begin{remark}
  One can verify that \eqref{eq:opt_cond} is satisfied for the Rayleigh fading distribution $P_H(h) = \frac{1}{\Gamma} \mathrm{e}^{-h/\Gamma}$ for all $\Gamma>0$. Indeed, the LHS of \eqref{eq:opt_cond} yields $(1+h)^{-1}$ which is nonnegative for all $h\geq 0$.
\end{remark}
\begin{remark}
  It follows directly from \eqref{eq:AverageRateFullCSIT} that $\eta_{\text{BRQ}}(T) \rightarrow C_{\text{erg}}$ as $T\rightarrow \infty$. This is because $h_T\rightarrow \infty$ as $T\rightarrow \infty$,  and thus the first term in \eqref{eq:AverageRateFullCSIT} tends to $C_{\text{erg}}$ while the second term in \eqref{eq:AverageRateFullCSIT} tends to zero.
\end{remark}
\begin{remark}
  The second term on the RHS of \eqref{eq:AverageRateFullCSIT} is the throughput of the conventional ARQ protocol with a rate equal to $C(h_T)$. The first term on the RHS of \eqref{eq:AverageRateFullCSIT} thereby corresponds to the improvement of BRQ-EMS over ARQ.
\end{remark}
\begin{IEEEproof}
We shall first use Lemma~\ref{lem:achievability} to show that there exists an $(\eta_{\text{BRQ}}(T),T)$-zero outage EMS protocol with rate selection and feedback functions given by \eqref{eq:full_csi_vt} and \eqref{eq:full_csi_rt}, respectively. Then, we apply the converse result in Lemma~\ref{lem:strong_converse} to show that $\eta_{\text{opt}}(T) = \eta_{\text{BRQ}}(T)$ under the condition in \eqref{eq:opt_cond}. 

Fix positive constants $c_1$ and $c_2$, and $h_T$ as in \eqref{eq:expected_tau}. We first show that an EMS protocol specified by \eqref{eq:full_csi_vt}--\eqref{eq:tau_n_def_full} has throughput $\eta_{\text{BRQ}}(T)$ and average decoding time upper-bounded by $T$. Since $\{\tau_n\}$ is a non-decreasing sequence of random variables and since $\E{\tau_n}\leq \E{\tau}<\infty$, Lebesgue's monotone convergence theorem \cite[Th.~16.2]{billingsley} implies that
\begin{IEEEeqnarray}{rCl}
  \lim_{n\rightarrow \infty}\E{\tau_n}  &=& \E{\tau}\label{eq:monotone_conv_thm}\\
  &=& \sum_{i=1}^\infty i F_H\farg{h_T}^{i-1}\left(1-F_H\farg{h_T}\right)\label{eq:Emin_sum}\\
  &=& \frac{1}{1-F_H\farg{h_T}}\label{eq:Emin_sum2}\\
  &=& T\label{eq:Etau_eq_T}
\end{IEEEeqnarray}
Similarly, we also have that\ifthenelse{\boolean{thesis}}{
\begin{IEEEeqnarray}{rCl}
\lim_{n\rightarrow \infty}\E{\cumR_{\tau_n}^{(n)}} &=&\lim_{n\rightarrow \infty} \E{\sum_{t=1}^{\infty} \indi{\tau_n \geq t}\mathbb{r}^{(n)}_t\mathopen{}\left(\mathbb{v}_{t-1}\mathopen{}\left(H_1^{t-1}\right)\right)} \label{eq:inf_extension}\\
  &=& \E{\sum_{t=1}^{\infty}\lim_{n\rightarrow \infty} \indi{\tau_n \geq t}\mathbb{r}^{(n)}_t\mathopen{}\left(\mathbb{v}_{t-1}\mathopen{}\left(H_1^{t-1}\right)\right)}\label{eq:dominated_convergence_thm} \\
    &=& C(h_T) + \E{\sum_{t=2}^{\infty}\indi{\tau \geq t} \min\mathopen{}\left\{C(H_{t-1}),C(h_T)\right\}} \IEEEeqnarraynumspace\\
    &=& C(h_T) + \sum_{t=2}^{\infty}\EBig{ \indi{\tau \geq t}\min\mathopen{}\left\{C(H_{t-1}),C(h_T)\right\}} \label{eq:tonelli_brq}\\
    &=& C(h_T) + \E{ C(H)|H\leq h_T}(\E{\tau}-1)\\
 &=& C(h_T) + T\int_{0}^{h_T} P_H(h)C(h)\dd h.\label{eq:full_cumR}
\end{IEEEeqnarray}}{
\begin{IEEEeqnarray}{rCl}
\IEEEeqnarraymulticol{3}{l}{\lim_{n\rightarrow \infty}\E{\cumR_{\tau_n}^{(n)}}}\nonumber\\
 &=&\lim_{n\rightarrow \infty} \E{\sum_{t=1}^{\infty} \indi{\tau_n \geq t}\mathbb{r}^{(n)}_t\mathopen{}\left(\mathbb{v}_{t-1}\mathopen{}\left(H_1^{t-1}\right)\right)} \label{eq:inf_extension}\\
  &=& \E{\sum_{t=1}^{\infty}\lim_{n\rightarrow \infty} \indi{\tau_n \geq t}\mathbb{r}^{(n)}_t\mathopen{}\left(\mathbb{v}_{t-1}\mathopen{}\left(H_1^{t-1}\right)\right)}\label{eq:dominated_convergence_thm} \\
    &=& C(h_T) + \E{\sum_{t=2}^{\infty}\indi{\tau \geq t} \min\mathopen{}\left\{C(H_{t-1}),C(h_T)\right\}} \IEEEeqnarraynumspace\\
    &=& C(h_T) + \sum_{t=2}^{\infty}\EBig{ \indi{\tau \geq t}\min\mathopen{}\left\{C(H_{t-1}),C(h_T)\right\}} \label{eq:tonelli_brq}\\
    &=& C(h_T) + \E{ C(H)|H\leq h_T}(\E{\tau}-1)\\
 &=& C(h_T) + T\int_{0}^{h_T} P_H(h)C(h)\dd h.\label{eq:full_cumR}
\end{IEEEeqnarray}}
Here, \eqref{eq:dominated_convergence_thm} follows from Lebesgue's monotone convergence theorem \cite[Th.~16.2]{billingsley} because $\tau_n$ and $\mathbb{r}_t^{(n)}$ are non-decreasing in $n$.
%
 Moreover, \eqref{eq:tonelli_brq} follows from Tonelli's theorem \cite[Th.~18.3]{billingsley} and \eqref{eq:full_cumR} follows because 
 \begin{IEEEeqnarray}{rCl}
 \int_{0}^{h_T} P_H(h)C(h)\dd h = \E{C(H)|H\leq h_T} \pr{H\leq h_T}\IEEEeqnarraynumspace
 \end{IEEEeqnarray}
  and because $T=\E{\tau} = 1/\pr{H\geq h_T}$.
As a result of \eqref{eq:Etau_eq_T} and \eqref{eq:full_cumR}, we obtain the throughput
\begin{IEEEeqnarray}{rCl}
  \lim_{n\rightarrow \infty} \frac{\Ebig{\cumR_{\tau_n}^{(n)}}}{\E{\tau_n}}     &=& \eta_{\text{BRQ}}(T).\label{eq:throughput_ems}
\end{IEEEeqnarray}
To show the existence of an $(\eta_{\text{BRQ}}(T), T)$-zero outage EMS protocol, we need to demonstrate that BRQ-EMS satisfies \eqref{eq:outage_error_cond} and \eqref{eq:strong_error_cond}. Both of these conditions follow from \eqref{eq:lem_achievability_conclusion} if the conditions of Lemma~\ref{lem:achievability} can be verified.
 Let $c_n\triangleq c_1/\log n$. Then, we shall first show that $\tau \leq \taumax$ implies that $\max_{k\in\range{1}{\tau_n}}\mathbb{u}_{k,\tau_n}^{(n)}(H_1^{\tau_n}) \leq -c_n$, which in turn implies that
\begin{IEEEeqnarray}{rCl}
  \pr{\mathcal{\bar H}_n|\tau_n = t} = 1\label{eq:cond_prob_barH_full_csi_1}
\end{IEEEeqnarray}
for $t\in\range{1}{\taumax-1}$, where $\mathcal{\bar H}_n$ is defined in \eqref{eq:Hbar}. Because $\mathbb{u}_{1,t}^{(n)}(h^{t}) \leq \mathbb{u}_{1,t}(h^{t})-c_n$ for every $h^{t}\in\mathbb{R}_+^t$ and because $\mathbb{u}_{1,\tau_n}(H^{\tau_n})\leq 0$ when $\tau\leq \taumax$, this follows from
\begin{IEEEeqnarray}{rCl}
\mathbb{u}_{1,\tau_n}^{(n)}(H_1^{\tau_n}) \leq \mathbb{u}_{1,\tau_n}(H^{\tau_n}) - c_n \leq -c_n\label{eq:u1taun}
\end{IEEEeqnarray}
and from the following chain of inequalities\footnote{We use the convention that $\sum_{i=j}^{j-1}a_i = 0$ for all $a_i$ and for all integers $j$.}\ifthenelse{\boolean{thesis}}{
\begin{IEEEeqnarray}{rCl}
 \max_{k\in\range{2}{\tau_n}}\mathbb{u}_{k,\tau_n}^{(n)}(H_1^{\tau_n})
  &=& \max_{k\in\range{2}{\tau_n}}  \sum_{i=k}^{\tau_n}\mathopen{}\left(\min\mathopen{}\Big\{C(h_T)-c_n,C(H_{i-1})\right\}   -C(H_i)\Big)\label{eq:eqq1}\\
  &\leq& \max_{k\in\range{2}{\tau_n}}  \Bigg\{ C(H_{k-1}) - C(h_T)\nonumber\\
  && {}+ \sum_{i=k}^{\tau_n}\mathopen{}\Big(\min\mathopen{}\left\{C(h_T)-c_n,C(H_{i-1})\right\}  -C(H_{i-1}) \Big)\Bigg\}\label{eq:eqq2}\IEEEeqnarraynumspace\\
  &=& \max_{k\in\range{2}{\tau_n}}  \Bigg\{ \Big(C(H_{k-1}) + c_n- C(h_T)\Big) - c_n\nonumber\\
  &&\qquad\qquad{}+ \sum_{i=k}^{\tau_n}\Big[C(h_T)-c_n-C(H_{i-1})\Big]^{-}  \Bigg\}\label{eq:eqq3}\\
    &=& \max_{k\in\range{2}{\tau_n}}  \Bigg\{ \Big[C(H_{k-1}) + c_n- C(h_T) \Big]^{-} - c_n\nonumber\\
    &&\qquad\qquad{}+ \sum_{i=k+1}^{\tau_n}\Big[C(h_T)-c_n-C(H_{i-1})\Big]^{-}  \Bigg\}\label{eq:eqq4}\\
    &\leq& -c_n.\label{eq:eqq5}
\end{IEEEeqnarray}}{
\begin{IEEEeqnarray}{rCl}
 \IEEEeqnarraymulticol{3}{l}{\max_{k\in\range{2}{\tau_n}}\mathbb{u}_{k,\tau_n}^{(n)}(H_1^{\tau_n})}\nonumber\\
  &=& \max_{k\in\range{2}{\tau_n}}  \sum_{i=k}^{\tau_n}\mathopen{}\left[\min\mathopen{}\Big\{C(h_T)-c_n,C(H_{i-1})\right\}   -C(H_i)\Big]\label{eq:eqq1}\\
  &\leq& \max_{k\in\range{2}{\tau_n}}  \Bigg\{ C(H_{k-1}) - C(h_T)\nonumber\\
  && {}+ \sum_{i=k}^{\tau_n}\mathopen{}\Big[\min\mathopen{}\left\{C(h_T)-c_n,C(H_{i-1})\right\}  -C(H_{i-1}) \Big]\Bigg\}\label{eq:eqq2}\IEEEeqnarraynumspace\\
  &=& \max_{k\in\range{2}{\tau_n}}  \Bigg\{ (C(H_{k-1}) + c_n- C(h_T)) - c_n\nonumber\\
  &&\qquad\qquad{}+ \sum_{i=k}^{\tau_n}\Big[C(h_T)-c_n-C(H_{i-1})\Big]^{-}  \Bigg\}\label{eq:eqq3}\\
    &=& \max_{k\in\range{2}{\tau_n}}  \Bigg\{ \Big[C(H_{k-1}) + c_n- C(h_T) \Big]^{-} - c_n\nonumber\\
    &&\qquad\qquad{}+ \sum_{i=k+1}^{\tau_n}\Big[C(h_T)-c_n-C(H_{i-1})\Big]^{-}  \Bigg\}\label{eq:eqq4}\\
    &\leq& -c_n.\label{eq:eqq5}
\end{IEEEeqnarray}}
Here, \eqref{eq:eqq1} follows from \eqref{eq:bbm_u_def} and \eqref{eq:full_csi_rt}, \eqref{eq:eqq2} follows because  \eqref{eq:achiev_au_full} implies that $h_T < H_{\tau_n}$ when $\tau\leq\taumax$, \eqref{eq:eqq4} follows because $x + [-x]^{-} =  [x]^-$ for $x\in\mathbb{R}$, and \eqref{eq:eqq5} follows because $[\cdot]^- \leq 0$.
Next, we show that $g_n\triangleq\mathcal{O}(1/\sqrt{n})$ satisfies \eqref{eq:min_prob_cond}:\ifthenelse{\boolean{thesis}}{
\begin{IEEEeqnarray}{rCl}
  \min_{\substack{t\in\range{1}{\taumax}:\\ \pr{\tau_n = t|\mathcal{\bar H}_n} > 0 }}\pr{\tau_n = t|\mathcal{\bar H}_n}&\geq& \min_{\substack{t\in\range{1}{\taumax}:\\ \pr{\tau_n = t|\mathcal{\bar H}_n} > 0 }}  \pr{ \tau_n = t ,\mathcal{\bar H}_n}\\
  &\geq& \min_{\substack{t\in\range{1}{\taumax}:\\ \pr{\tau_n = t|\mathcal{\bar H}_n} > 0 }}  \pr{ \tau = t  }\label{eq:brq_11}\\
    &=& F_H(h_T)^{\taumax-1}(1-F_H(h_T))\\
    &=&\mathcal{O}(\e{\log(F_H(h_T)) \taumax})\\
     &=& \mathcal{O}\farg{\frac{1}{\sqrt{n}}} = g_n.\label{eq:achiev_full_csi_gn}
\end{IEEEeqnarray}}{
\begin{IEEEeqnarray}{rCl}
  \IEEEeqnarraymulticol{3}{l}{\min_{\substack{t\in\range{1}{\taumax}:\\ \pr{\tau_n = t|\mathcal{\bar H}_n} > 0 }}\pr{\tau_n = t|\mathcal{\bar H}_n}}\nonumber\\
    \qquad&\geq& \min_{\substack{t\in\range{1}{\taumax}:\\ \pr{\tau_n = t|\mathcal{\bar H}_n} > 0 }}  \pr{ \tau_n = t ,\mathcal{\bar H}_n}\\
  &\geq& \min_{\substack{t\in\range{1}{\taumax}:\\ \pr{\tau_n = t|\mathcal{\bar H}_n} > 0 }}  \pr{ \tau = t  }\label{eq:brq_11}\\
    &=& F_H(h_T)^{\taumax-1}(1-F_H(h_T))\\
    &=&\mathcal{O}(\e{\log(F_H(h_T)) \taumax})\\
     &=& \mathcal{O}\farg{\frac{1}{\sqrt{n}}} = g_n.\label{eq:achiev_full_csi_gn}
\end{IEEEeqnarray}}
Here, \eqref{eq:brq_11} follows because $\tau \leq \bar \tau_n$ implies that the event $\mathcal{\bar H}_n$ occurs.
It also follows that \eqref{eq:achievability_conv_cond} is satisfied:
\begin{IEEEeqnarray}{rCl}
  \frac{\taumax^2}{n g_n c_n^2} = \mathcal{O}\farg{\frac{\log^2(n)\log^2\farg{\sqrt{n}}}{\sqrt{n} }} = o(1)\label{eq:achiev_conv_full_csi}
\end{IEEEeqnarray}
as $n\rightarrow \infty$.
As a consequence of  \eqref{eq:achiev_full_csi_gn} and \eqref{eq:achiev_conv_full_csi}, Lemma~\ref{lem:achievability} implies that there exists an EMS protocol satisfying \eqref{eq:lem_achievability_conclusion}. In addition, the EMS protocol is also an $(\eta_{\text{BRQ}}(T),T)$-zero outage EMS protocol, which follows because the condition in \eqref{eq:strong_error_cond} is implied by \eqref{eq:lem_achievability_conclusion} and \eqref{eq:cond_prob_barH_full_csi_1}:\ifthenelse{\boolean{thesis}}{
\begin{IEEEeqnarray}{rCl}
  \max_{\substack{t\in\range{1}{\taumax-1}: \\ \pr{\tau_n = t}>0 }} \pr{\mathcal{E}_n|\tau_n = t}
   &=&   \max_{\substack{t\in\range{1}{\taumax-1}: \\ \pr{\tau_n = t}>0 }} \Big\{\pr{\mathcal{E}_n|\tau_n = t,\mathcal{\bar H}_n}\pr{\mathcal{\bar H}_n| \tau_n = t} \nonumber\\
  &&\qquad\qquad\quad{}+ \pr{\mathcal{E}_n|\tau_n = t,\mathcal{\bar H}_n^\complement}\pr{\mathcal{\bar H}_n^\complement | \tau_n = t} \Big\}\label{eq:Prob_error_bound1}\IEEEeqnarraynumspace\\
  &\leq&   \max_{\substack{t\in\range{1}{\taumax-1}: \\ \pr{\tau_n = t}>0 }} \pr{\mathcal{E}_n|\tau_n = t,\mathcal{\bar H}_n} \label{eq:Prob_error_bound2}\\
  &=& o(1)\label{eq:Prob_error_bound3}
\end{IEEEeqnarray}}{
\begin{IEEEeqnarray}{rCl}
  \IEEEeqnarraymulticol{3}{l}{\max_{\substack{t\in\range{1}{\taumax-1}: \\ \pr{\tau_n = t}>0 }} \pr{\mathcal{E}_n|\tau_n = t}}\nonumber\\
   &=&   \max_{\substack{t\in\range{1}{\taumax-1}: \\ \pr{\tau_n = t}>0 }} \Big\{\pr{\mathcal{E}_n|\tau_n = t,\mathcal{\bar H}_n}\pr{\mathcal{\bar H}_n| \tau_n = t} \nonumber\\
  &&\qquad\qquad\qquad{}+ \pr{\mathcal{E}_n|\tau_n = t,\mathcal{\bar H}_n^\complement}\pr{\mathcal{\bar H}_n^\complement | \tau_n = t} \Big\}\label{eq:Prob_error_bound1}\IEEEeqnarraynumspace\\
  &\leq&   \max_{\substack{t\in\range{1}{\taumax-1}: \\ \pr{\tau_n = t}>0 }} \pr{\mathcal{E}_n|\tau_n = t,\mathcal{\bar H}_n} \label{eq:Prob_error_bound2}\\
  &=& o(1)\label{eq:Prob_error_bound3}
\end{IEEEeqnarray}}
as $n\rightarrow \infty$. Here, $\mathcal{\bar H}_n^\complement$ denotes the complement of the event $\mathcal{\bar H}_n$ and \eqref{eq:Prob_error_bound2} follows \eqref{eq:cond_prob_barH_full_csi_1}. The condition in \eqref{eq:outage_error_cond} now follows from \eqref{eq:Prob_error_bound3} and because $\pr{\tau_n = \taumax}\rightarrow 0$ as $n\rightarrow \infty$. 

We prove in Appendix~\ref{app:converse_zero_outage} that no zero outage EMS protocol can achieve a throughput larger than that of the RHS of \eqref{eq:AverageRateFullCSIT}, i.e., we establish that $\eta_{\text{opt}}(T) = \eta_{\text{BRQ}}(T)$ for $T > 1$ under the condition in \eqref{eq:opt_cond}.
\end{IEEEproof}

\section{Finite Number of Feedback Messages}
Full delayed CSIT feedback is not always an viable assumption. This section addresses the case where the feedback cost is finite. While HARQ-INR does not allow for rate adaptations, EMS protocols with three or more feedback messages can be used to signal ACK/NACK, but also to instruct the transmitter to append additional information bits in the subsequent slot. The key difference from the case with full delayed CSIT is that the optimal amount of new information to be appended cannot be specified through the feedback. We provide a heuristic choice of the rate selection functions, feedback functions, and decoding times and demonstrate the existence of a zero outage EMS protocol. In Section~\ref{sec:num_res}, it is numerically shown that the throughput of the finite feedback cost EMS protocol is comparable with that of the BRQ protocol.

We shall construct an EMS code with feedback cost $f+1$, where $f\in\mathbb{N}$. Specifically, we define the rate selection and feedback functions as 
\begin{IEEEeqnarray}{rCl}
  \mathbb{v}_t(h_1^t) &\triangleq& \left\{\begin{array}{ll}
      \left\lfloor f- \frac{ \mathbb{u}_{1,t}(h_1^{t})}{\mathbb{r}}\right\rfloor, & \mathbb{u}_{1,t}(h_1^{t}) > 0 \\
      -1, & \mathbb{u}_{1,t}(h_1^{t}) \leq 0
  \end{array}\right.\label{eq:finite_rate_policy1} 
  \end{IEEEeqnarray}
  and\ifthenelse{\boolean{thesis}}{
  \begin{IEEEeqnarray}{rCl}
  \mathbb{r}_t^{(n)}(v_{t-1})&\triangleq&   \left\{\begin{array}{ll}
     \mathbb{r}f - c_n, & t=1 \\
   \min\{\mathbb{r}(f-1)-c_n,\mathbb{r}v_{t-1}\}\indi{v_{t-1}\not= -1}, & t\geq 2.
   \end{array}
   \right.\IEEEeqnarraynumspace
    \label{eq:finite_rate_policy2}
  \end{IEEEeqnarray}}{
  \begin{IEEEeqnarray}{rCl}
  \IEEEeqnarraymulticol{3}{l}{\mathbb{r}_t^{(n)}(v_{t-1})\triangleq}\nonumber\\
   &&  \left\{\begin{array}{ll}
     \mathbb{r}f - c_n, & t=1 \\
   \min\{\mathbb{r}(f-1)-c_n,\mathbb{r}v_{t-1}\}\indi{v_{t-1}\not= -1}, & t\geq 2.
   \end{array}
   \right.\IEEEeqnarraynumspace
    \label{eq:finite_rate_policy2}
  \end{IEEEeqnarray}}
  Here, $\mathbb{r}>0$ is a predefined constant, $\mathbb{F}=\range{-1}{f-1}$, and $c_n\triangleq c_1/\log(n)$ for an arbitrary positive constant $c_1$. The decoding time is given by
\begin{IEEEeqnarray}{rCl}
\tau_n &=& \min\{\taumax , \tau\}\label{eq:ems_stopping_time}
\end{IEEEeqnarray}
where
\begin{IEEEeqnarray}{rCl}
\tau &\triangleq& \inf\{t \geq 1: V_t = -1\}\\
\taumax &\triangleq& -\left\lfloor \frac{\log(c_2 \sqrt{n})}{\log F_C(\mathbb{r}(f-1))} \right\rfloor.
\end{IEEEeqnarray}
Here, $c_2$ is an arbitrary positive constant and the feedback ${-1}$ designates an ACK message. Since $\mathbb{v}_t^{(n)}$ can take at most $f+1=|\mathbb{F}|$ values, the corresponding EMS protocol has feedback cost $f+1$. 
We define the composite rate-feedback function as
\begin{align}
  \mathbb{\overline{r v}}(u) \triangleq   \mathbb{r} \min\mathopen{}\bigg\{ f-1, \bigg\lfloor f- \frac{\pos{u}}{\mathbb{r}} \bigg\rfloor \bigg\}. \label{eq:comp_rate_feedback1}
\end{align}
With this definition, we can write 
\begin{align}
  \mathbb{r}_t(\mathbb{v}_{t-1}(h_1^{t-1})) = \mathbb{\overline{rv}}( \mathbb{u}_{1,t-1}(h_1^{t-1}))
\end{align}
for all $t\geq 2$ and $h_1^{t-1}\in\mathbb{R}_+^{t-1}$ such that $\mathbb{u}_{1,t-1}(h_1^{t-1}) >0$.


The trade-off between throughput and average decoding time achievable by an EMS-$(f+1)$ protocol is characterized by the following theorem which provides a way to compute the throughput and average decoding time by solving two integral equations. Varying the parameter $\mathbb{r}$ determines the trade-off between throughput and average decoding time.  
\begin{theorem}\label{thm:finite_ems}
Define $W: [0,\mathbb{r} f]\mapsto \mathbb{R}_+$ and $M: [0,\mathbb{r}f]\mapsto \mathbb{R}_+$ through the integral equations
\begin{align}
  W(u) &\triangleq \mathbb{\overline{r v}}(u) + \int_{0}^{u + \mathbb{\overline{r v}}(u)} P_C(x) W(u + \mathbb{\overline{r v}}(u) - x)  \dd x\label{eq:finite_integral_eq}
\end{align}
and
\begin{align}
  M(u) &= 1 + \int_{0}^{u + \mathbb{\overline{r v}}(u)} P_C(x) M(u + \mathbb{\overline{r v}}(u) - x)  \dd x.\label{eq:finite_time_integral_eq}
\end{align}
Here, $P_C(\cdot)$ denotes the probability density function of $C(H)$.
Then, there exists an $(\eta,T)$-zero outage EMS protocol with
\begin{align}
 \eta &= \frac{\mathbb{r}f +  \EBig{\indi{ C(H) \leq  \mathbb{r}f}W( \mathbb{r}f - C(H))}}{1 +  \EBig{\indi{ C(H) \leq  \mathbb{r}f}M( \mathbb{r}f - C(H) ) }}\label{eq:brq_one_bit}
 \end{align}
 and
 \begin{IEEEeqnarray}{rCl}
 T &=& 1 +  \EBig{\indi{ C(H) \leq  \mathbb{r}f}M( \mathbb{r}f - C(H) ) }.\label{eq:brq_one_bit_T}
\end{IEEEeqnarray}
\end{theorem}
\begin{IEEEproof}
In order to show that \eqref{eq:finite_rate_policy1}--\eqref{eq:ems_stopping_time} define a zero outage EMS protocol, we need to verify the conditions of Lemma~\ref{lem:achievability}. We shall first show that \eqref{eq:min_prob_cond} is satisfied for $g_n = \mathcal{O}(1/\sqrt{n})$. The remaining conditions are verified using arguments similar to those in the proof of Theorem~\ref{thm:brq_opt}. Given that $\tau\leq \taumax$, we have for $k\in\range{2}{\tau_n}$\ifthenelse{\boolean{thesis}}{
\begin{IEEEeqnarray}{rCl}
  \IEEEeqnarraymulticol{3}{l}{\mathbb{u}_{k,\tau_n}^{(n)}(H^{\tau_n})}\nonumber\\
   &=& \sum_{i=k}^{\tau_n} \Bigg[ \min\mathopen{}\left\{\mathbb{r}(f-1)-c_n,\mathbb{r}\left\lfloor f- \frac{ \mathbb{u}_{1,i-1}(H_1^{i-1})}{\mathbb{r}}\right\rfloor\right\} - C(H_i) \Bigg]\label{eq:ems_eqq1}\\
     &\leq& \sum_{i=k}^{\tau_n} \Bigg[ \min\mathopen{}\left\{\mathbb{r}(f-1)-c_n,\mathbb{r}\left\lfloor f- \frac{ \mathbb{u}_{1,i-1}(H_1^{i-1})}{\mathbb{r}}\right\rfloor\right\} - C(H_i) \Bigg]\nonumber\\
     &&{} - \mathbb{u}_{1,\tau_n}(H^{\tau_n})\label{eq:ems_eqq2}\\
     &=& \sum_{i=k}^{\tau_n} \bigg[\mathbb{r}(f-1)-c_n-\mathbb{r}\bigg\lfloor f- \frac{ \mathbb{u}_{1,i-1}(H_1^{i-1})}{\mathbb{r}}\bigg\rfloor\bigg]^{-}  - \mathbb{u}_{1,k-1}(H^{k-1})\IEEEeqnarraynumspace \label{eq:ems_eqq3}\\
     &\leq& \sum_{i=k}^{\tau_n} \left[-c_n+\mathbb{u}_{1,i-1}(H_1^{i-1})\right]^{-}  -\mathbb{u}_{1,k-1}(H_1^{k-1})\label{eq:ems_eqq4}\\
       &\leq& \min\{-c_n, -\mathbb{u}_{1,k-1}(H_1^{k-1})\}+ \sum_{i=k+1}^{\tau_n} \left[-c_n+\mathbb{u}_{1,i-1}(H_1^{i-1})\right]^{-}  \label{eq:ems_eqq5}\\
       &\leq& -c_n.\label{eq:ems_eqq6}
\end{IEEEeqnarray}}{
\begin{IEEEeqnarray}{rCl}
  \IEEEeqnarraymulticol{3}{l}{\mathbb{u}_{k,\tau_n}^{(n)}(H^{\tau_n})}\nonumber\\
   &=& \sum_{i=k}^{\tau_n} \Bigg[ \min\mathopen{}\left\{\mathbb{r}(f-1)-c_n,\mathbb{r}\left\lfloor f- \frac{ \mathbb{u}_{1,i-1}(H_1^{i-1})}{\mathbb{r}}\right\rfloor\right\} \nonumber\\
   &&\qquad\qquad\qquad\qquad\qquad\qquad\qquad\qquad{}- C(H_i) \Bigg]\label{eq:ems_eqq1}\\
     &\leq& \sum_{i=k}^{\tau_n} \Bigg[ \min\mathopen{}\left\{\mathbb{r}(f-1)-c_n,\mathbb{r}\left\lfloor f- \frac{ \mathbb{u}_{1,i-1}(H_1^{i-1})}{\mathbb{r}}\right\rfloor\right\} \nonumber\\
     &&{}\qquad\qquad\qquad\qquad\qquad- C(H_i) \Bigg] - \mathbb{u}_{1,\tau_n}(H^{\tau_n})\label{eq:ems_eqq2}\\
     &=& \sum_{i=k}^{\tau_n} \bigg[\mathbb{r}(f-1)-c_n-\mathbb{r}\bigg\lfloor f- \frac{ \mathbb{u}_{1,i-1}(H_1^{i-1})}{\mathbb{r}}\bigg\rfloor\bigg]^{-} \nonumber\\
     &&{} - \mathbb{u}_{1,k-1}(H^{k-1})\IEEEeqnarraynumspace \label{eq:ems_eqq3}\\
     &\leq& \sum_{i=k}^{\tau_n} \left[-c_n+\mathbb{u}_{1,i-1}(H_1^{i-1})\right]^{-}  -\mathbb{u}_{1,k-1}(H_1^{k-1})\label{eq:ems_eqq4}\\
       &\leq& \min\{-c_n, -\mathbb{u}_{1,k-1}(H_1^{k-1})\} \nonumber\\
       &&{}+ \sum_{i=k+1}^{\tau_n} \left[-c_n+\mathbb{u}_{1,i-1}(H_1^{i-1})\right]^{-}  \label{eq:ems_eqq5}\\
       &\leq& -c_n.\label{eq:ems_eqq6}
\end{IEEEeqnarray}}
Here, \eqref{eq:ems_eqq1} follows from \eqref{eq:bbm_u_def} and \eqref{eq:finite_rate_policy1}--\eqref{eq:finite_rate_policy2}, \eqref{eq:ems_eqq2} follows because $\mathbb{u}_{1,\tau_n}(H^{\tau_n}) \leq 0$ when $\tau\leq\taumax$, \eqref{eq:ems_eqq4} follows from $\lfloor x\rfloor \in(x-1,x]$, \eqref{eq:ems_eqq6} follows because $[x]^{-}\leq 0$.  Using the same arguments as in \eqref{eq:u1taun}, it can also be shown that $\mathbb{u}^{(n)}_{1,\tau_n}(H^{\tau_n}) \leq -c_n$ when $\tau\leq \taumax$. Hence, we conclude that $\max_{k\in\range{1}{\tau_n}} \mathbb{u}^{(n)}_{k,\tau_n}(H^{\tau_n}) \leq -c_n$ when $\tau\leq \taumax$. 
An immediate implication of this is that
\begin{IEEEeqnarray}{rCl}
  \pr{\tau_n=t\Big| \mathcal{\bar H}_n} =\frac{\pr{\tau_n=t, \mathcal{\bar H}_n}}{\pr{\mathcal{\bar H}_n}} = \frac{\pr{\tau=t}}{\pr{\mathcal{\bar H}_n}} \geq \pr{\tau = t}\label{eq:prtau_barH_lower_bound}\IEEEeqnarraynumspace
\end{IEEEeqnarray}
for all $t\in\range{1}{\taumax}$. Note that $\tau$ is not necessarily Geometrically distributed as for the case with full delayed CSIT.
Instead, since $\lfloor x \rfloor \in (x-1,x]$ for any constant $x$, we have that\ifthenelse{\boolean{thesis}}{
\begin{IEEEeqnarray}{rCl}
  \mathbb{u}_{1,t}(h^{t}) + \mathbb{r}_{t+1}(\mathbb{v}_{t}(h^{t}))
  &=&\mathbb{u}_{1,t}(h^{t})  + \mathbb{r}\left\lfloor f- \frac{\mathbb{u}_{1,t}(h_1^{t})}{\mathbb{r}}\right\rfloor \in (\mathbb{r}(f-1),\mathbb{r}f]
\end{IEEEeqnarray}}{
\begin{IEEEeqnarray}{rCl}
  \IEEEeqnarraymulticol{3}{l}{\mathbb{u}_{1,t}(h^{t}) + \mathbb{r}_{t+1}(\mathbb{v}_{t}(h^{t}))}\nonumber\\
  &=&\mathbb{u}_{1,t}(h^{t})  + \mathbb{r}\left\lfloor f- \frac{\mathbb{u}_{1,t}(h_1^{t})}{\mathbb{r}}\right\rfloor \in (\mathbb{r}(f-1),\mathbb{r}f]
\end{IEEEeqnarray}}
for all $t\in\mathbb{N}$ and $h_1^{t}\in\mathbb{R}_+^{t}$ such that $\mathbb{u}_{1,t}(h_1^{t}) >0$.
Therefore, for all $t\in\mathbb{N}$, we also have that\ifthenelse{\boolean{thesis}}{
\begin{IEEEeqnarray}{rCl}
 \pr{\tau \geq t + 1 | \tau \geq t} &=& \prbig{ \mathbb{u}_{1,t}(H^{t}) > 0 \big| \tau \geq t}\in [F_C(\mathbb{r} (f-1)),F_C(\mathbb{r} f)].
\end{IEEEeqnarray}}{
\begin{IEEEeqnarray}{rCl}
 \pr{\tau \geq t + 1 | \tau \geq t} &=& \prbig{ \mathbb{u}_{1,t}(H^{t}) > 0 \big| \tau \geq t}\nonumber\\
 &&\in [F_C(\mathbb{r} (f-1)),F_C(\mathbb{r} f)].
\end{IEEEeqnarray}}
Thus,
\begin{IEEEeqnarray}{rCl}
 \pr{\tau = t} &=&  \pr{\tau = t|\tau\geq t}\prod_{i=1}^{t-1} \pr{\tau \geq i+1|\tau\geq i} \\
 &\geq& F_C(\mathbb{r}(f-1))^{t-1}(1-F_C(\mathbb{r}f)).\label{eq:prtau_lower_bound}
\end{IEEEeqnarray}
It follows from \eqref{eq:prtau_barH_lower_bound} and \eqref{eq:prtau_lower_bound} that \eqref{eq:min_prob_cond} is satisfied for $g_n = \mathcal{O}(1/\sqrt{n})$. The conditions in \eqref{eq:outage_error_cond}, \eqref{eq:strong_error_cond}, and \eqref{eq:achievability_conv_cond} follows using the same arguments as in the proof of Theorem~\ref{thm:brq_opt}. Similarly, we can also show that $\lim_{n\rightarrow \infty} \E{\tau_n} = \E{\tau}$ and that $\lim_{n\rightarrow \infty} \E{\cumR_{\tau_n}^{(n)}} = \E{\cumR_{\tau}}$. Hence, it only remains to compute the throughput given by $\E{\cumR_\tau}/\E{\tau}$ and the limiting average decoding time $\E{\tau}$.

We compute the throughput $\E{\cumR_\tau}/\E{\tau}$ via the rate selection functions, feedback functions, and the decoding time in \eqref{eq:finite_rate_policy1}--\eqref{eq:ems_stopping_time}. Using the following recursive relation
\begin{align}
\mathbb{u}_{1,t}(h^t) = \mathbb{u}_{1,t-1}(h^{t-1}) + \mathbb{\overline {r v}} ( \mathbb{u}_{1,t-1}(h^{t-1})) - C(h_t)
\end{align}
for $t\geq 2$, we observe that, if $t\geq k\geq 2$, then $\mathbb{u}_{1,t}(h^{k-1},H_k^{t})$ only depends on $h^{k-1}$ through $\mathbb{u}_{1,k-1}(h^{k-1})$. Therefore, we can define $\mathbb{\bar u}(u, h_k^{t})$ such that $\mathbb{\bar u}(\mathbb{u}_{1,k-1}(h^{k-1}) , h_k^{t}) = \mathbb{u}_{1,t}(h^{k-1},h_k^{t})$. In order to compute $\E{\cumR_\tau}$, define
\begin{align}
     W_t(u) &\triangleq  \EBigg{ \sum_{i=t}^{\tau_t(u)}\mathbb{\overline {r v}} ( \mathbb{\bar u}(u,H_t^{i-1}))}\label{eq:W_def}
  \end{align}
for $u\in[0,\mathbb{r}f]$, where 
\begin{align}
   \tau_t(u) \triangleq \inf\mathopen{}\left\{ \bar t \geq t: \mathbb{\bar u}(u,H_t^{\bar t}) < 0 \right\}.
\end{align}
Observe that 
\begin{IEEEeqnarray}{rCl}
\E{\cumR_\tau} = \mathbb{r}f + \E{\indi{ C(H_1) \leq  \mathbb{r}f  }W_1(\mathbb{u}_{1,1}(H_1)) }.
\end{IEEEeqnarray}
Rewriting the RHS of \eqref{eq:W_def} in terms of $W_{t+1}(\cdot)$, we obtain\ifthenelse{\boolean{thesis}}{
\begin{IEEEeqnarray}{rCl}
   W_t(u)
   &=& \mathbb{\overline {r v}} (u)+ \EBigg{\indi{u+\mathbb{\overline {r v}} (u) \geq C(H_t)}\nonumber\\
   &&{}\qquad\qquad\qquad\times\EBigg{\sum_{i=t+1}^{\tau_{t+1}(\mathbb{\bar u}(u,H_t))}\mathbb{\overline {r v}} ( \mathbb{\bar u}(\mathbb{\bar u}(u,H_t),H_{t+1}^{i-1}))\Bigg| H_t} }\\
    &=& \mathbb{\overline {r v}} (u)+ \E{\indi{u+\mathbb{\overline {r v}} (u)\geq C(H_t)} W_{t+1}(\mathbb{\bar u}(u,H_t))} \\
   &=& \mathbb{\overline{r v}}(u)+ \int_{0}^{u + \mathbb{\overline{r v}}(u)} P_C(x) W_{t+1}(u + \mathbb{\overline{r v}}(u) - x)  \dd x.\label{eq:Wt_integral}\IEEEeqnarraynumspace
\end{IEEEeqnarray}}{
\begin{IEEEeqnarray}{rCl}
   \IEEEeqnarraymulticol{3}{l}{W_t(u)}\\
   &=& \mathbb{\overline {r v}} (u)+ \EBigg{\indi{u+\mathbb{\overline {r v}} (u) \geq C(H_t)} \nonumber\\
   &&{}\times\EBigg{\sum_{i=t+1}^{\tau_{t+1}(\mathbb{\bar u}(u,H_t))}\mathbb{\overline {r v}} ( \mathbb{\bar u}(\mathbb{\bar u}(u,H_t),H_{t+1}^{i-1}))\Bigg| H_t} }\\
    &=& \mathbb{\overline {r v}} (u)\nonumber\\
    &&{}+ \E{\indi{u+\mathbb{\overline {r v}} (u)\geq C(H_t)} W_{t+1}(\mathbb{\bar u}(u,H_t))} \\
   &=& \mathbb{\overline{r v}}(u) \nonumber\\
   &&{}+ \int_{0}^{u + \mathbb{\overline{r v}}(u)} P_C(x) W_{t+1}(u + \mathbb{\overline{r v}}(u) - x)  \dd x.\label{eq:Wt_integral}\IEEEeqnarraynumspace
\end{IEEEeqnarray}}
By defining $W(\cdot)\triangleq W_1(\cdot)$ and by noting that $W_t(u) = W_{t+1}(u)$ for $u\in[0,\mathbb{r}f]$, we have the integral equation in \eqref{eq:finite_integral_eq}. The expected reward is thereby given by
\begin{align}
 \E{\cumR_\tau} = \mathbb{r} f+  \E{\indi{ C(H) \leq   \mathbb{r}f }W( \mathbb{r}f - C(H) ) }.
\end{align}
\begin{sloppypar}\noindent Using derivations similar to \eqref{eq:W_def}--\eqref{eq:Wt_integral}, we obtain $\E{\tau}=1 +  \E{\indi{ C(H) \leq  \mathbb{r}f }M(  \mathbb{r}f - C(H) ) }$. 
\end{sloppypar}\end{IEEEproof}
We remark that the integral equations in Theorem~\ref{thm:finite_ems} can be written as Fredholm equations of the second kind. These are readily solved as a system of linear equations when discretized or by using a quadrature method specifically for Fredholm equations \cite{fredholm_computation}.

\section{Numerical Results}
\label{sec:num_res}
In this section, the throughput of the described protocols are assessed and compared to the HARQ-INR protocol with and without power adaptation.

\subsection*{HARQ-INR}
In the HARQ-INR protocol, the transmitter uses a rate $R$ in the first slot and continues to send additional IR in the subsequent slots. By the end of each slot, the receiver attempts to decode and feeds back an ACK/NACK signal depending on whether the decoding was successful or not. The receiver is thereby able to accumulate mutual information until decoding is possible. The average decoding time of the HARQ-INR protocol is given by \cite{multilayer_broadcast}
\begin{align}
  \E{\tau} 
  &= \sum_{m=1}^\infty m (p_{\text{out}}^{m-1}(R) - p_{\text{out}}^{m}(R))\\
  &= 1 + \sum_{m=1}^\infty  p_{\text{out}}^m(R)
\end{align}
where $p_{\text{out}}^m(\cdot)$ is the outage probability after the $m$th retransmission and is given by 
\begin{align}
  p_{\text{out}}^m(r) = \pr{ \sum_{k=1}^m \C{H_k} < r }.\label{eq:po_m}
\end{align}
The maximal throughput of HARQ-INR subject to the average decoding time constraint is given by \cite{multilayer_broadcast}
\begin{subequations}
\begin{IEEEeqnarray}{rCl}
  \eta_{\text{HARQ-INR}}(T) = & \max_R &  \frac{R}{1+\sum_{m=1}^{\infty} p_{\text{out}}^m(R)}\\
  & \text{s.t.} & 1 + \sum_{m=1}^\infty  p_{\text{out}}^m(R) \leq T.
\end{IEEEeqnarray}
\label{eq:eta_HARQ}
\end{subequations}
We remark that $\sup_{T\in(1,\infty)} \eta_{\text{HARQ-INR}}(T) = C_{\text{erg}}$.

\subsection*{HARQ-INR with power adaptation}
A comparison between BRQ and HARQ-INR is not fair in the sense that HARQ-INR does not use the available delayed CSIT. It has been shown in literature that delayed CSIT can provide significant throughput benefits if the short-term power constraint in \eqref{eq:power_const} is relaxed. Power adaptation based on delayed CSIT has previously been proposed in a slightly different setting in \cite{PowerAdapt}. In this section, we optimize HARQ-INR with power adaptation under a constraint on the average decoding time. We follow \cite{Tuninetti2007} and redefine the power constraint in \eqref{eq:power_const} such that $\frac{1}{n}\vect{X}_t\tr \vect{X}_t \leq \rho_t$, where we require that the random variables $\{\rho_t\}$ depend only on $\{H_t\}_{t=1}^{t-1}$ and that $\{\rho_t\}_{t=1}^\infty$ satisfies
\begin{IEEEeqnarray}{rCl}
  \frac{\E{\sum_{i=1}^\tau \rho_i}}{\E{\tau}} &\leq& 1.\label{eq:long_term_power_const}
\end{IEEEeqnarray}
The constraint in \eqref{eq:long_term_power_const} ensures that the average power per slot over many runs of the protocol does not exceed one. Under this relaxation, we can design an HARQ-INR-type protocol that benefits from full delayed CSIT using power adaptation. In particular, full delayed CSIT provides the transmitter with knowledge about the amount of unresolved information at the receiver and is allowed to use this knowledge to optimize the power spend in the following slot. The transmitter sends in the first slot at a rate $R$ using power $\rho_1$. At the end of the slot, the transmitter receives the delayed CSIT which can be used to compute the unresolved information $I_1$ at the receiver. In the $t$th slot, the transmitter sends IR with power $\rho_t(I_{t-1})$, where $I_{t-1}$ is the amount of unresolved information at the receiver by the end of slot $t-1$ and $\rho_t(\cdot)$ denotes the power adaptation policy in the $t$th slot.
It follows that the unresolved information in slot $t$ satisfies
\begin{IEEEeqnarray}{rCl}
  I_t = I_{t-1} - C(H_{t} \rho_t(I_{t-1}))
\end{IEEEeqnarray}
where $I_0 \triangleq R$. We shall solve the following optimization problem using dynamic programming:
\begin{subequations}
\begin{IEEEeqnarray}{rCl}
  &\min_{\{\rho_t(\cdot)\}_{t=1}^\infty }& \E{\tau} \\
  & \text{s.t.} & \E{\sum_{t = 1}^\tau \rho_i(I_{t-1}) }\leq \E{\tau}.
\end{IEEEeqnarray}\label{eq:power_opt}\end{subequations}
Here, $\tau\triangleq \inf\{t: I_t < 0\}$.
First, we rewrite \eqref{eq:power_opt} as an unconstrained optimization problem using duality:
\begin{IEEEeqnarray}{rCl}
\IEEEeqnarraymulticol{3}{l}{ \max_{\lambda > 0} \min_{\{\rho_t(\cdot)\}_{t=1}^\infty}\mathopen{}\left\{ \E{\tau}(1-\lambda) +\lambda \E{\sum_{t = 1}^\tau \rho_t(I_{t-1}) }\right\}}.\label{eq:power_dual_opt}\IEEEeqnarraynumspace
\end{IEEEeqnarray}
Then, we rewrite the inner minimization in \eqref{eq:power_dual_opt} as an infinite-horizon dynamic programming problem. Specifically, we find that
\begin{IEEEeqnarray}{rCl}
\min_{\{\rho_t(\cdot)\}_{t=1}^\infty}\mathopen{}\left\{ \E{\tau}(1-\lambda) +\lambda \E{\sum_{t = 1}^\tau \rho_t(I_{t-1}) }\right\} = J_\lambda(R)\IEEEeqnarraynumspace
\end{IEEEeqnarray}
where the function $J_\lambda(\cdot)$ is defined by $J_{\lambda}(u) = 0$ for $u\leq 0$ and\ifthenelse{\boolean{thesis}}{
\begin{IEEEeqnarray}{rCl}
 J_{\lambda}(u)
 &=& \min_{\rho} \mathopen{}\Bigg\{1 + \lambda(\rho-1)+ \int_{0}^{ \frac{2^{2u}-1}{\rho} } P_H(h) J_{\lambda}(u - \C{h \rho}) \dd h\Bigg\}\IEEEeqnarraynumspace
\end{IEEEeqnarray}}{
\begin{IEEEeqnarray}{rCl}
 J_{\lambda}(u)
 &=& \min_{\rho} \mathopen{}\Bigg\{1 + \lambda(\rho-1)\nonumber\\
 &&\quad\qquad{} + \int_{0}^{ \frac{2^{u}-1}{\rho} } P_H(h) J_{\lambda}(u - \C{h \rho}) \dd h\Bigg\}\IEEEeqnarraynumspace
\end{IEEEeqnarray}}
for $u > 0$.
Consequently, we find that the solution to the optimization problem in \eqref{eq:power_opt} is given by $\max_{\lambda > 0} J_\lambda(R)$. 
The throughput of HARQ-INR with power adaptation under an average decoding time constraint is thereby given by
\begin{subequations}
\begin{IEEEeqnarray}{rCl}
 \eta_{\text{HARQ-INR-P}}(T) = &\max_{R >0 } &\frac{R}{\max_{\lambda > 0} J_\lambda(R)} \\
  & \text{s.t.} & \max_{\lambda > 0} J_\lambda(R) \leq T.
\end{IEEEeqnarray}
\label{eq:harq_power}
\end{subequations}

\subsection*{Assessment}
\begin{figure}[!t]
\centering
\subfigure[$\text{Average SNR} = 10\ \text{dB}$]{
  \includegraphics[width=5.7cm]{\PathBRQ/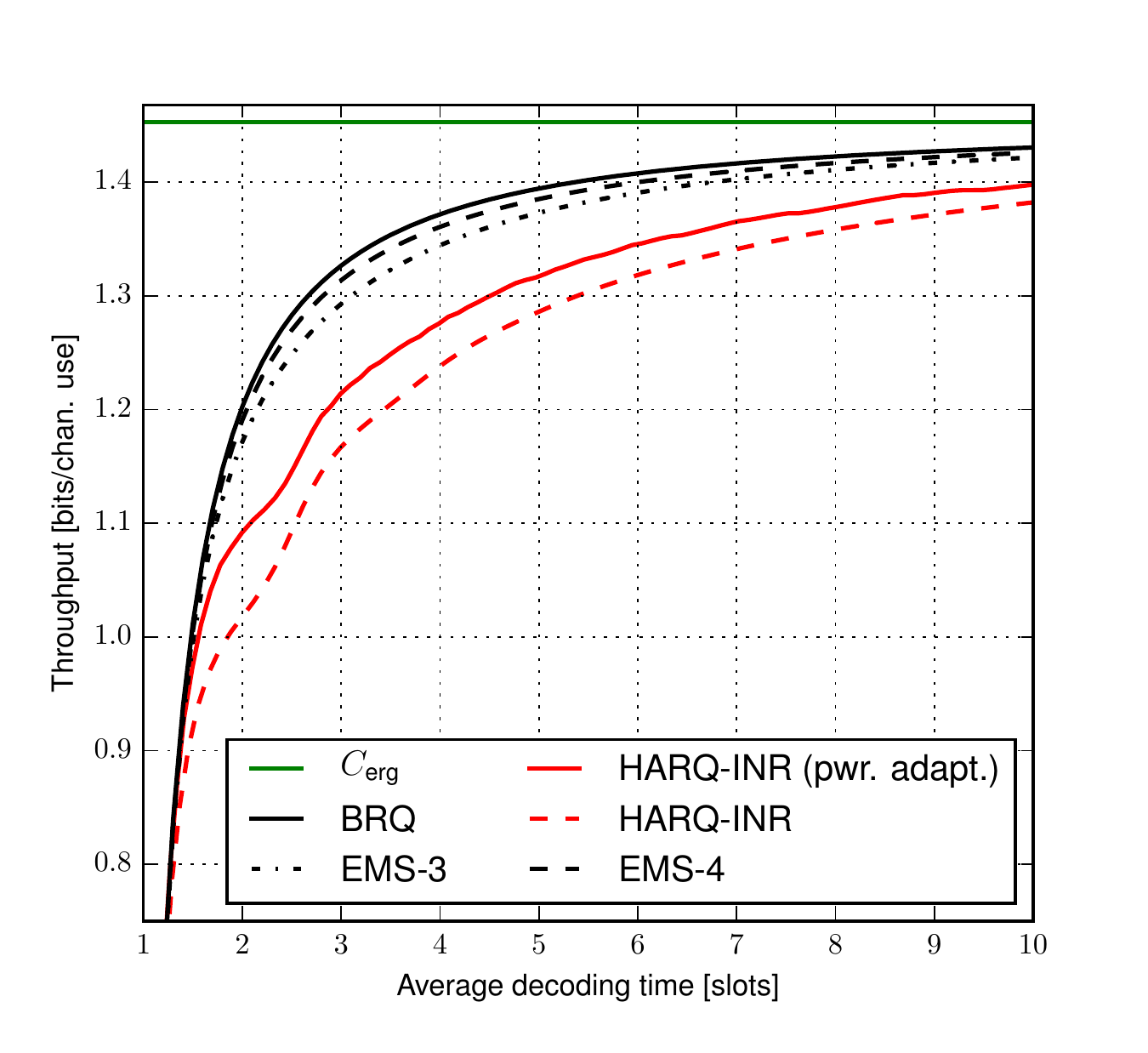}
  \label{fig:results_delay_throughput_snr_10}
}
\subfigure[$\text{Average SNR} = 30\ \text{dB}$]{  
\includegraphics[width=5.7cm]{\PathBRQ/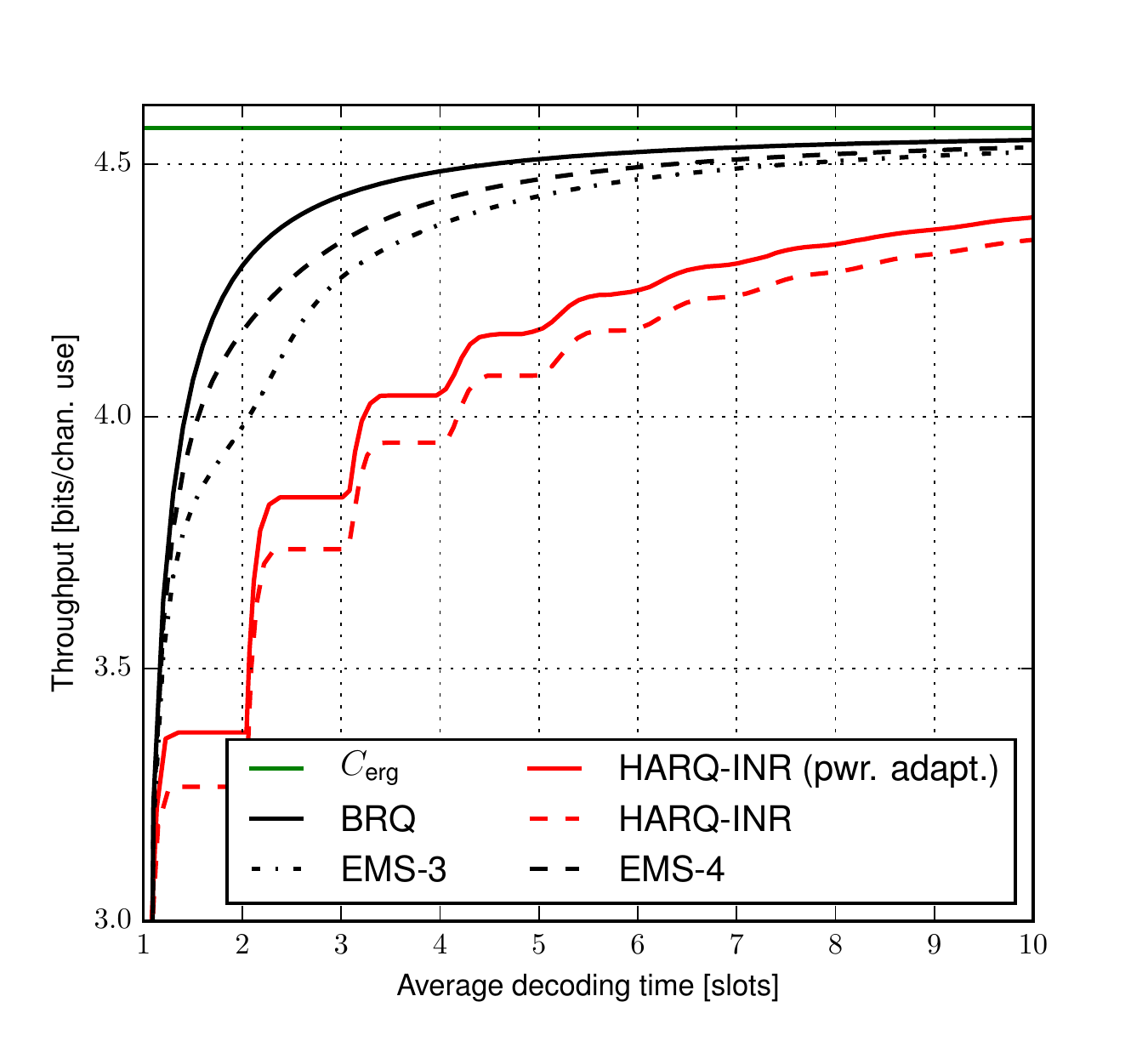}
  \label{fig:results_delay_throughput_snr_30}
}
\caption{Throughput versus average decoding time $\E{\tau}$ for the investigated protocols. The throughputs of HARQ-INR and HARQ-INR with power adaptation are computed using \eqref{eq:eta_HARQ} and \eqref{eq:harq_power}, respectively. The throughput of BRQ is computed using \eqref{eq:AverageRateFullCSIT} and for the EMS protocols we use \eqref{eq:brq_one_bit} and \eqref{eq:brq_one_bit_T}.}
\label{fig:results_delay_throughput}
\end{figure}
We evaluate the proposed protocols by assuming Rayleigh block-fading, independent from slot to slot, i.e., the probability density of $H$ is given by
\begin{align}
  P_H(h) = \frac{1}{\Gamma} e^{-h/\Gamma}.
\end{align}
  Fig.~\ref{fig:results_delay_throughput} depicts the throughput of various protocols as a function of average decoding time for SNR equal to $10\text{ dB}$ and $30\text{ dB}$. We remark that the stair-step behavior of the throughput of HARQ-INR at $\text{SNR}=30\text{ dB}$ origins because the probability distribution of $C(H)$ becomes increasingly concentrated around $C_{\text{erg}}$ as the SNR increases. For high SNR, this implies that the average decoding time, and therefore also the throughput, has a stair-step behavior when $R$ grows linearly. It is seen that the throughput of all protocols tend to the ergodic capacity as the allowed average decoding times are increased. We observe that BRQ and the EMS protocols with finite feedback cost significantly outperforms both HARQ-INR and HARQ-INR with power adaptation in terms of throughput. A particular interesting observation is that the proposed EMS protocols for finite feedback cost achieves throughputs that are very close to that of BRQ, even for the case $f=2$. Our interpretation of this is that the precise amount of additional information bits appended in each slot does not affect the throughput significantly. 
\begin{figure}[!t]
\centering
\subfigure[$\E{\tau}= 2.5$]{ 
\includegraphics[width=5.7cm]{\PathBRQ/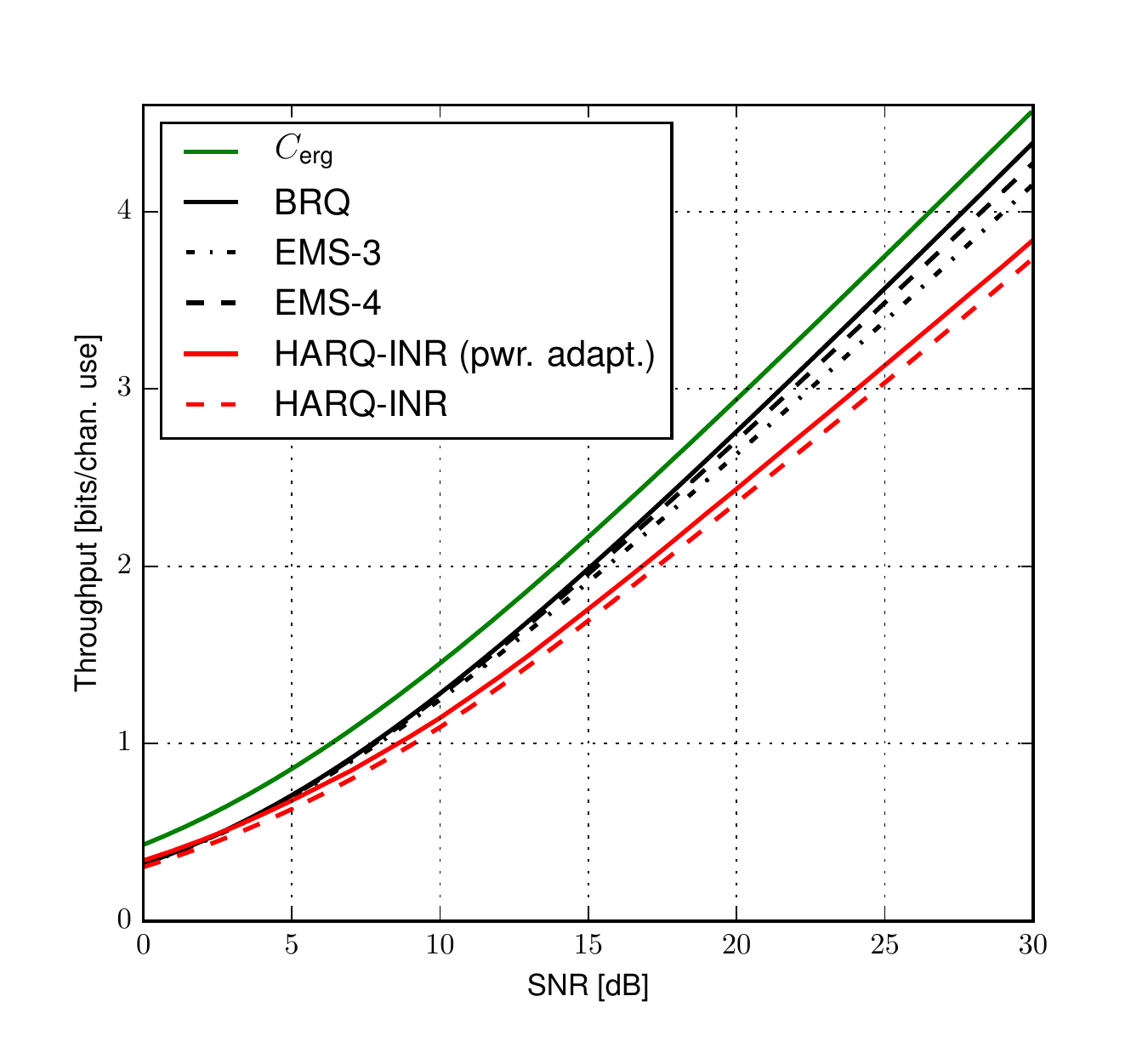}
  \label{fig:results_throughput_snr_10} 
} 
\subfigure[$\E{\tau} = 4.5$]{
\includegraphics[width=5.7cm]{\PathBRQ/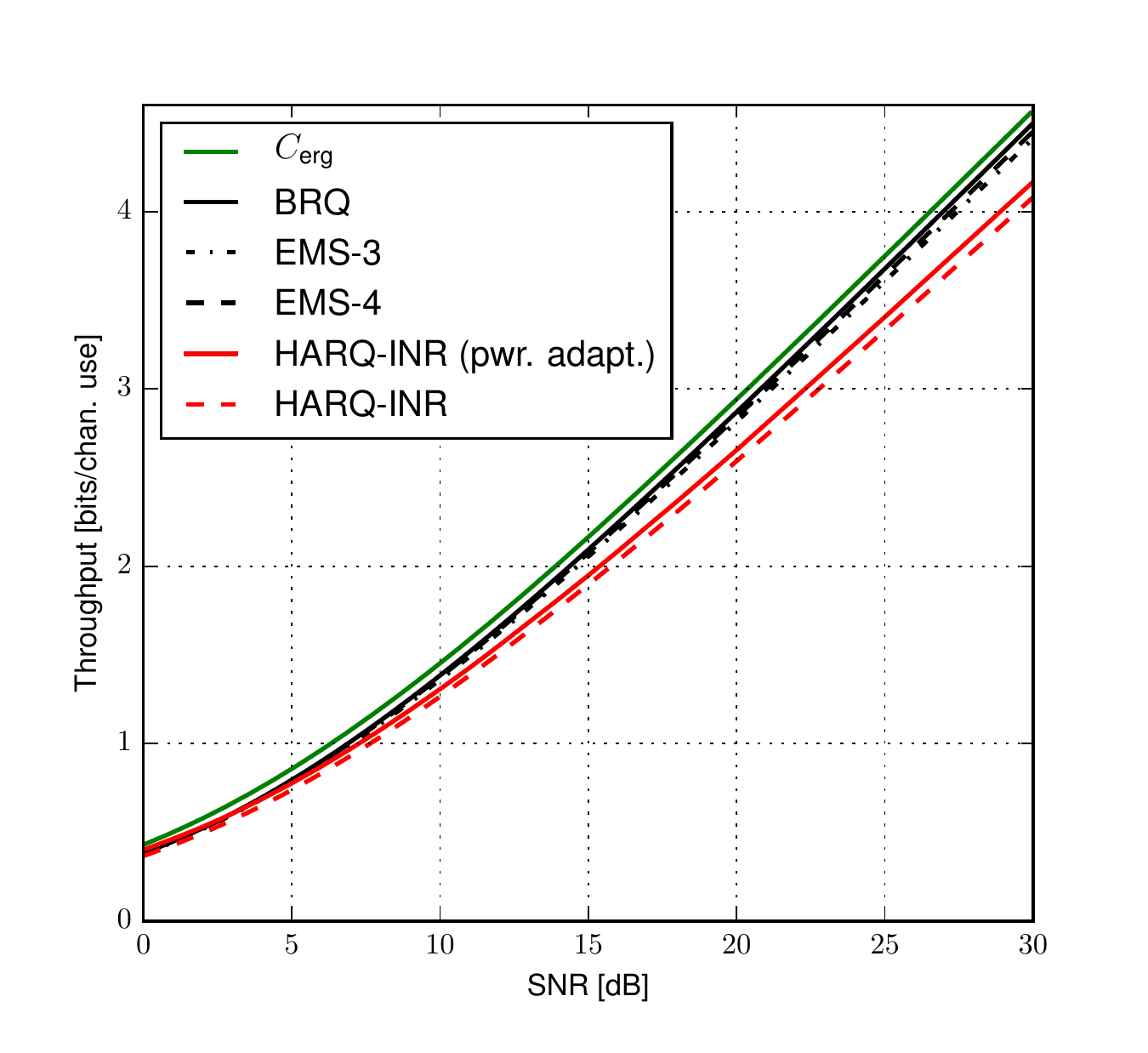}
\label{fig:results_throughput_snr_30}
}
\caption{Throughput versus SNR for the investigated protocols. The throughputs of HARQ-INR and HARQ-INR with power adaptation are computed using \eqref{eq:eta_HARQ} and \eqref{eq:harq_power}, respectively. The throughput of BRQ is computed using \eqref{eq:AverageRateFullCSIT} and for the EMS protocols, we use \eqref{eq:brq_one_bit} and \eqref{eq:brq_one_bit_T}.}
\label{fig:results_throughput_snr} 
\end{figure}

In Fig.~\ref{fig:results_throughput_snr}, the throughput is plotted in terms of SNR for fixed average decoding time $\E{\tau}$. Observe that the back-off from the ergodic capacity of BRQ is approximately constant throughout the range of SNR values while the penalty of the remaining protocols increases for larger SNR.

\section{Discussion and Conclusions}
\label{sec:Discussion}

The objective of this paper was to generalize and extend the BRQ protocol, proposed in \cite{backtrack}, to a broader class of communication strategies termed EMS protocols. EMS protocols are useful when the CSI is only available after the transmission has taken place. The main novelty of EMS protocols is the possibility of appending new information bits before previously transmitted data has been resolved. EMS protocols thereby provides a way to design communication protocols that approach the ergodic capacity with low average decoding time. In contrast to BRQ, EMS protocols in general also benefit from limited feedback. Specifically, it has been shown that even ternary feedback is sufficient to achieve throughput close to that of BRQ. 
This suggests that the main reason for the superior throughput of BRQ and EMS protocols is that, compared to HARQ-type protocols with/without power adaptation, they only terminate a transmission when the CSI is sufficiently good, whereas HARQ-INR terminates a transmission as soon as a sufficient amount of information is accumulated. As a result, HARQ-INR protocol often collects a wasteful amount of mutual information which far surpasses the amount of unresolved information, leading to waste of resources.

Unlike most works in the field of HARQ, we have presented results for systems with an average decoding time constraint as opposed to a strict decoding time constraint. Strict decoding time constraints lead to protocols with a maximum transmission length. Such constraints are motivated by applications like streaming of multimedia data, where data become useless after a certain amount of time. Despite this, there are many applications where data is retransmitted at a packet level upon outage. In other words, a new transmission is initiated with the same data -- perhaps concatenated with data from new data arrivals. For such applications, a constraint on the average decoding time is more applicable. 
Although strict decoding time constraints have not been considered, they are not ruled by the definition of EMS protocols.  An optimal EMS protocol with full delayed CSIT and a constraint on outage probability instead of average decoding time can be computed numerically using dynamic programming.

  We have not treated the impact of the accuracy of the delayed CSI in our throughput comparisons. In the conventional HARQ-INR protocol that rely on, possibly quantized, prior CSI to perform rate and/or power adaptation, the accuracy of CSI has a significant impact on the throughput \cite[pp.~209--213]{Tse}. The main reason for this is that the channel gains change from the time the CSI is estimated to the time the channel is used, which can take a duration that spans multiple slots. This inaccuracy is largely eliminated by relying only on delayed CSI. This follows because the receiver can make a much more precise estimate of the CSI after having observed a time slot. 
For the EMS protocols, however, inaccurate delayed CSI implies that the transmitter cannot precisely append the optimal amount of new information in each step. Our results for the EMS protocols with finite feedback cost show that the precise amount of new information appended in each slot does not significantly alter the achievable throughput. Therefore, we do not expect that the throughput of EMS protocols to suffer significantly if the CSI is inaccurate.

Finally, we note that HARQ-INR has led to several composite protocols that use HARQ-INR as building block. As previously discussed, two examples which are of relevance to this paper are \cite{Szczecinski13} and \cite{multilayer_broadcast}. One can design similar composite protocols using the EMS protocols as building blocks. For example, the broadcast approach to HARQ-INR proposed in \cite{multilayer_broadcast} provides an approach combine multiple HARQ-INR instances that run in parallel in multiple superposition coded layers. We can combine multilayered transmission and EMS protocols similarly. One feasible approach is to divide each transmission into two layers: one with IR for the previous slots and one with new information bits. One can then optimize over the distribution of power in the two layers. In this way the decoder does not need to decode both the IR for previous slot and the new information bits simultaneously. Hence, such protocols might lead to higher throughputs than the present paper report. One can also follow the approach taken in \cite{Szczecinski13} and instantiate several instances of EMS protocols which run in parallel in a TDM fashion.

\ifthenelse{\boolean{thesis}}{}{
\section*{Acknowledgement}
This work has been in part supported by the European Research Council
(ERC Consolidator Grant nr. 648382 WILLOW) within the Horizon 2020
Program.}


\appendices


\section{Proof of Lemma~\ref{lem:strong_converse} (converse)}\label{app:converse_proof}
Fix an EMS protocol defined by  $\{\tau_n\}$, $\{\mathbb{r}^{(n)}_t\}$, $\{\mathbb{v}_t\}$, $\{\mathbb{f}_t^{(n)}\}$, and $\{\mathbb{g}_t^{(n)}\}$. The EMS protocol induces a probability distribution on $(\vect{X}^{\tau_n}, \vect{Y}^{\tau_n}, H^{\tau_n})$ given by $P_{\vect{Y}^{\tau_n},\vect{X}^{\tau_n},H^{\tau_n}}$. 
 To simplify notation, we condition on $H^\infty=h^\infty$ throughout the proof and define the probability distribution $\mathbb{\bar P}$ on $(\vect{X}^{\tau_n},\vect{Y}^{\tau_n})$ by
\begin{IEEEeqnarray}{rCl}
  \ppr{\cdot} &\triangleq& \pr{ \cdot | H^\infty = h^\infty}.
\end{IEEEeqnarray}
Since the stopping time and rate selection functions depend only on the channel realizations, conditioning on $H^\infty=h^\infty$ implies that $\{\tau_n\}$ and $\{R^{(n)}_t\}$ are deterministic sequences. The probability distribution of the channel outputs in the $t$th slot is
\begin{IEEEeqnarray}{rCl}
\pprr_{\vect{Y}_t|\vect{X}_t}(\vect{y}|\vect{x}) \triangleq \prod_{i=1}^n \frac{1}{\sqrt{2\pi}} \e{-\frac{1}{2}(y_i - \sqrt{h_t}x_i)^2}.
\end{IEEEeqnarray}

Since $\tau < \infty$ by assumption, the limit $\lim_{n\rightarrow \infty} \tau_n = \tau$ exists and implies that there exist positive integers $N$ and $n_0$  such that $\tau_n \leq N$ for all $n\geq n_0$. Therefore, we have\footnote{We use the convention that $\sum_{i=j}^{j-1}a_i = 0$ for all $a_i$ and for all integers $j$.}
\begin{IEEEeqnarray}{rCl}
\IEEEeqnarraymulticol{3}{l}{\lim_{n\rightarrow \infty}\max_{k\in\range{1}{\tau_n+1}} \sum_{i=k}^{\tau_n} (R_i^{(n)} - C(h_i))}\nonumber\\
 &=& \lim_{n\rightarrow \infty}\max_{k\in\range{1}{N+1}} \sum_{i=k}^{N} \indi{i \leq \tau_n}(R_i^{(n)} - C(h_i))\label{eq:strong_converse_lim1}\\
&=& \max_{k\in\range{1}{N+1}} \sum_{i=k}^{N} \indi{i \leq \tau}(R_i - C(h_i))\\
&=&  \max_{k\in\range{1}{\tau+1}} \mathbb{u}_{k,\tau}(h^{\tau})\\
&>&0.\label{eq:brq_converse_prop1}
\end{IEEEeqnarray}
The last inequality follows from the condition $\sup_{k\in\range{1}{\tau}} \mathbb{u}_{k,\tau}(h^\tau) >0$. Eq.~\eqref{eq:brq_converse_prop1} implies that there exist a positive integer $n_1$, a positive constant $\gamma$, and a sequence of integers $\{\scseq\}_{n=n_1}^\infty$ with $\scseq\in\range{1}{\tau_n}$ such that 
\begin{IEEEeqnarray}{rCl}
 \sum_{k=\scseq}^{\tau_n} (R_k^{(n)} - C(h_k)) \geq 2\gamma\label{eq:Rin_minus_Ch_geq_2gamma}
\end{IEEEeqnarray}
for all $n\geq n_1$.

To proceed, we prove a variation of the Verdú-Han converse \cite{Ageneralformulaforchannelcapacity}. To state the result, we shall define the information density for $t\in\range{1}{\tau_n}$ as follows
\begin{IEEEeqnarray}{rCl}
  i\farg{\vect{x}_t^{\tau_n}; \vect{y}_t^{\tau_n} | \vect{x}^{t-1}} &=&  \log_2 \frac{\prod_{i=t}^{\tau_n}\pprr_{\vect{Y}_{i}|\vect{X}_{i}}(\vect{y}_i |\vect{x}_i)}{\pprr_{\vect{Y}^{\tau_n}_t|\vect{X}^{t-1}}(\vect{y}_t^{\tau_n} | \vect{x}^{t-1})}\label{eq:strong_converse_inf_dens}
\end{IEEEeqnarray}
where  $\vect{x}^{\tau_n}, \vect{y}^{\tau_n}\in \mathbb{R}^{n\tau_n}$.
\begin{lemma}
Under the above definitions, the following holds for every $n$\ifthenelse{\boolean{thesis}}{
\begin{IEEEeqnarray}{rCl}
 \ppr{\mathcal{E}_n}&\geq&  \max_{t\in\range{1}{\tau_n}} \ppr{\frac{1}{n}i\farg{\vect{X}_t^{\tau_n};\vect{Y}_t^{\tau_n}\Big| \vect{X}^{t-1}} \leq  \sum_{k=t}^{\tau_n} R_k^{(n)} - \gamma }- 2^{-n\gamma}\label{eq:verdu_han_converse}
\end{IEEEeqnarray}}{
\begin{IEEEeqnarray}{rCl}
 \ppr{\mathcal{E}_n}&\geq&  \max_{t\in\range{1}{\tau_n}} \ppr{\frac{1}{n}i\farg{\vect{X}_t^{\tau_n};\vect{Y}_t^{\tau_n}\Big| \vect{X}^{t-1}} \leq  \sum_{k=t}^{\tau_n} R_k^{(n)} - \gamma } \nonumber\\
 &&{}- 2^{-n\gamma}\label{eq:verdu_han_converse}
\end{IEEEeqnarray}}
where $\gamma>0$ is an arbitrary constant.
  \label{lem:verdu_han_converse}
\end{lemma}
\begin{IEEEproof}
 The proof closely follows those found in \cite[Th.~4]{Ageneralformulaforchannelcapacity} or \cite[Lemma ~3.2.2]{Han}. The encoding functions $(\mathbb{f}^{(n)}_{1},\cdots,\mathbb{f}^{(n)}_{\tau_n})$ generates $M_n \triangleq 2^{\lceil n\cumR^{(n)}_{\tau_n}\rceil}$ codewords which we denote by $\{\vect{u}(i)\}_{i=1}^{M_n}$, where $\vect{u}(i) \in\mathbb{R}^{n \tau_n}$. Note that $\pprr_{\vect{X}^t}(\vect{u}^t(i)) = 2^{-n \cumR^{(n)}_t}$ for $i \in\range{1}{M_n}$ and $t\in\range{0}{\tau_n}$ (recall that $\cumR_0^{(n)} = 0$), where $ \vect{u}^t(i)$ denotes the first $nt$ entries of $\vect{u}(i)$. The decoding function $\mathbb{g}_{\tau_n}^{(n)}(\cdot)$  defines disjoint decoding regions $\{\mathcal{D}_i\}_{i=1}^{M_n}$ such that $\mathcal{D}_i \subseteq\mathbb{R}^{n\tau_n}$ and $\bigcup_{i=1}^{M_n}\mathcal{D}_i = \mathbb{R}^{n\tau_n}$.
  Set $\beta \triangleq 2^{-n\gamma}$ and note that
  \ifthenelse{\boolean{thesis}}{
\begin{IEEEeqnarray}{rCl}
 \frac{1}{n}i\farg{\vect{x}_t^{\tau_n};\vect{y}_t^{\tau_n}|\vect{x}^{t-1}} &=& \frac{1}{n}\log_2 \frac{\pprr_{\vect{X}_t^{\tau_n} | \vect{Y}_t^{\tau_n},\vect{X}^{t-1}}(\vect{x}_t^{\tau_n}|\vect{y}_t^{\tau_n},\vect{x}^{t-1})}{\pprr_{\vect{X}_t^{\tau_n} |\vect{X}^{t-1}}(\vect{x}_t^{\tau_n}|\vect{x}^{t-1})}\nonumber\\
  & =& \sum_{k=t}^{\tau_n} R_k^{(n)} + \frac{1}{n} \log_2 \pprr_{\vect{X}_t^{\tau_n}|\vect{Y}_t^{\tau_n},\vect{X}^{t-1}}(\vect{x}_t^{\tau_n}|\vect{y}_t^{\tau_n},\vect{x}^{t-1}).\IEEEeqnarraynumspace
\end{IEEEeqnarray}}{
\begin{IEEEeqnarray}{rCl}
\IEEEeqnarraymulticol{3}{l}{  \frac{1}{n}i\farg{\vect{x}_t^{\tau_n};\vect{y}_t^{\tau_n}|\vect{x}^{t-1}}}\nonumber\\ &=& \frac{1}{n}\log_2 \frac{\pprr_{\vect{X}_t^{\tau_n} | \vect{Y}_t^{\tau_n},\vect{X}^{t-1}}(\vect{x}_t^{\tau_n}|\vect{y}_t^{\tau_n},\vect{x}^{t-1})}{\pprr_{\vect{X}_t^{\tau_n} |\vect{X}^{t-1}}(\vect{x}_t^{\tau_n}|\vect{x}^{t-1})}\nonumber\\
  & =& \sum_{k=t}^{\tau_n} R_k^{(n)} + \frac{1}{n} \log_2 \pprr_{\vect{X}_t^{\tau_n}|\vect{Y}_t^{\tau_n},\vect{X}^{t-1}}(\vect{x}_t^{\tau_n}|\vect{y}_t^{\tau_n},\vect{x}^{t-1}).\IEEEeqnarraynumspace
\end{IEEEeqnarray}}
The last equality follows because \ifthenelse{\boolean{thesis}}{
\begin{IEEEeqnarray}{rCl}
\log_2 \pprr_{\vect{X}_t^{\tau_n} |\vect{X}^{t-1}}(\vect{x}_t^{\tau_n}|\vect{x}^{t-1})
 &=& \log_2 \pprr_{\vect{X}^{\tau_n}}(\vect{x}^{\tau_n}) - \log_2 \pprr_{\vect{X}^{t-1}}(\vect{x}^{t-1})\\
&=& -n\cumR_{\tau_n}^{(n)} + n\cumR_{t-1}^{(n)} \\
&=& -n\sum_{k=t}^{\tau_n} R^{(n)}_k.
\end{IEEEeqnarray}}{
\begin{IEEEeqnarray}{rCl}
\IEEEeqnarraymulticol{3}{l}{\log_2 \pprr_{\vect{X}_t^{\tau_n} |\vect{X}^{t-1}}(\vect{x}_t^{\tau_n}|\vect{x}^{t-1})}\nonumber\\
 &=& \log_2 \pprr_{\vect{X}^{\tau_n}}(\vect{x}^{\tau_n}) - \log_2 \pprr_{\vect{X}^{t-1}}(\vect{x}^{t-1})\\
&=& -n\cumR_{\tau_n}^{(n)} + n\cumR_{t-1}^{(n)} \\
&=& -n\sum_{k=t}^{\tau_n} R^{(n)}_k.
\end{IEEEeqnarray}}
Consequently, we obtain
\begin{IEEEeqnarray}{rCl}
 \IEEEeqnarraymulticol{3}{l}{\ppr{\frac{1}{n}i\farg{\vect{X}_t^{\tau_n};\vect{Y}_t^{\tau_n} \Big| \vect{X}^{t-1}} \leq  \sum_{k=t}^{\tau_n} R_k^{(n)} - \gamma}}\nonumber\\
 \qquad &=& \ppr{\pprr_{\vect{X}_t^{\tau_n}|\vect{Y}_t^{\tau_n},\vect{X}^{t-1}}(\vect{X}_t^{\tau_n}|\vect{Y}_t^{\tau_n}, \vect{X}^{t-1}) \leq \beta }.\IEEEeqnarraynumspace\label{eq:strong_conv_beta_lower_bound}
\end{IEEEeqnarray}
Define\ifthenelse{\boolean{thesis}}{
\begin{IEEEeqnarray}{rCl}
  \mathcal{B}_i &=& \Big\{\vect{y}^{\tau_n}\in\mathbb{R}^{\tau_n n} : \pprr_{\vect{X}_t^{\tau_n}|\vect{Y}_t^{\tau_n}, \vect{X}^{t-1}}\farg{\vect{u}_{t}^{\tau_n}(i)|\vect{y}_t^{\tau_n}, \vect{u}^{t-1}(i)} \leq \beta\Big\}.\label{eq:strong_converse_Bi}\IEEEeqnarraynumspace
\end{IEEEeqnarray}}{
\begin{IEEEeqnarray}{rCl}
  \mathcal{B}_i &=& \Big\{\vect{y}^{\tau_n}\in\mathbb{R}^{\tau_n n} :\nonumber\\
  &&\qquad{}  \pprr_{\vect{X}_t^{\tau_n}|\vect{Y}_t^{\tau_n}, \vect{X}^{t-1}}\farg{\vect{u}_{t}^{\tau_n}(i)|\vect{y}_t^{\tau_n}, \vect{u}^{t-1}(i)} \leq \beta\Big\}.\label{eq:strong_converse_Bi}\IEEEeqnarraynumspace
\end{IEEEeqnarray}}
We obtain a lower bound on $\ppr{\mathcal{E}_n}$ through the following chain of inequalities\ifthenelse{\boolean{thesis}}{
\begin{IEEEeqnarray}{rCl}
\IEEEeqnarraymulticol{3}{l}{\ppr{\frac{1}{n}i\farg{\vect{X}_t^{\tau_n};\vect{Y}_t^{\tau_n}\Big| \vect{X}^{t-1}} \leq \sum_{k=t}^{\tau_n} R_k^{(n)}-\gamma }}\nonumber\\
 &=& \sum_{i=1}^{M_n} \pprr_{\vect{X}^{\tau_n}, \vect{Y}^{\tau_n}}\mathopen{}\left[\vect{u}(i), \mathcal{B}_i\right]\label{eq:strong_converse_0}\\
&=& \sum_{i=1}^{M_n}  \pprr_{\vect{X}^{\tau_n}, \vect{Y}^{\tau_n}}\mathopen{}\left[\vect{u}(i), \mathcal{B}_i\cap \mathcal{D}_i^\complement\right]+\sum_{i=1}^{M_n}  \pprr_{\vect{X}^{\tau_n}, \vect{Y}^{\tau_n}}\left[\vect{u}(i), \mathcal{B}_i\cap \mathcal{D}_i\right]\\
&\leq& \frac{1}{M_n}\sum_{i=1}^{M_n} \pprr_{\vect{Y}^{\tau_n}|\vect{X}^{\tau_n}}\farg{\mathcal{D}_i^\complement| \vect{u}(i)}\nonumber\\
&&+\sum_{i=1}^{M_n} \int_{\mathcal{B}_i\cap \mathcal{D}_i}\pprr_{\vect{Y}^{\tau_n},\vect{X}^{t-1}}\farg{\vect{y}^{\tau_n} , \vect{u}^{t-1}(i)}\nonumber\\
&&{}\qquad\qquad\qquad\qquad\times \pprr_{\vect{X}^{\tau_n}_t|\vect{Y}_t^{\tau_n},\vect{X}^{t-1}}\farg{\vect{u}_{t}^{\tau_n}(i)|\vect{y}_t^{\tau_n},\vect{u}^{t-1}(i)}\dd\vect{y}^{\tau_n}\label{eq:strong_converse_first_ineq}\IEEEeqnarraynumspace\\
&\leq& \ppr{\mathcal{E}_n}+\beta  \sum_{i=1}^{M_n}   \pprr_{\vect{Y}^{\tau_n},\vect{X}^{t-1}}\farg{\mathcal{B}_i\cap \mathcal{D}_i , \vect{u}^{t-1}(i)}\label{eq:strong_converse_used_Bi1}\\
&\leq& \ppr{\mathcal{E}_n}+\beta\sum_{i=1}^{M_n}   \pprr_{\vect{Y}^{\tau_n},\vect{X}^{t-1}}\farg{ \mathcal{D}_i, \vect{u}^{t-1}(i)}\label{eq:strong_converse_used_Bi2}\\
&\leq& \ppr{\mathcal{E}_n}+\beta.\label{eq:strong_converse_final_bound}
\end{IEEEeqnarray}}{
\begin{IEEEeqnarray}{rCl}
\IEEEeqnarraymulticol{3}{l}{\ppr{\frac{1}{n}i\farg{\vect{X}_t^{\tau_n};\vect{Y}_t^{\tau_n}\Big| \vect{X}^{t-1}} \leq \sum_{k=t}^{\tau_n} R_k^{(n)}-\gamma }}\nonumber\\
 &=& \sum_{i=1}^{M_n} \pprr_{\vect{X}^{\tau_n}, \vect{Y}^{\tau_n}}\mathopen{}\left[\vect{u}(i), \mathcal{B}_i\right]\label{eq:strong_converse_0}\\
&=& \sum_{i=1}^{M_n}  \pprr_{\vect{X}^{\tau_n}, \vect{Y}^{\tau_n}}\mathopen{}\left[\vect{u}(i), \mathcal{B}_i\cap \mathcal{D}_i^\complement\right]\nonumber\\
&&{}+\sum_{i=1}^{M_n}  \pprr_{\vect{X}^{\tau_n}, \vect{Y}^{\tau_n}}\left[\vect{u}(i), \mathcal{B}_i\cap \mathcal{D}_i\right]\\
&\leq& \frac{1}{M_n}\sum_{i=1}^{M_n} \pprr_{\vect{Y}^{\tau_n}|\vect{X}^{\tau_n}}\farg{\mathcal{D}_i^\complement| \vect{u}(i)}\nonumber\\
&&+\sum_{i=1}^{M_n} \int_{\mathcal{B}_i\cap \mathcal{D}_i}\pprr_{\vect{Y}^{\tau_n},\vect{X}^{t-1}}\farg{\vect{y}^{\tau_n} , \vect{u}^{t-1}(i)}\nonumber\\
&&{}\qquad\times \pprr_{\vect{X}^{\tau_n}_t|\vect{Y}_t^{\tau_n},\vect{X}^{t-1}}\farg{\vect{u}_{t}^{\tau_n}(i)|\vect{y}_t^{\tau_n},\vect{u}^{t-1}(i)}\dd\vect{y}^{\tau_n}\label{eq:strong_converse_first_ineq}\IEEEeqnarraynumspace\\
&\leq& \ppr{\mathcal{E}_n}+\beta  \sum_{i=1}^{M_n}   \pprr_{\vect{Y}^{\tau_n},\vect{X}^{t-1}}\farg{\mathcal{B}_i\cap \mathcal{D}_i , \vect{u}^{t-1}(i)}\label{eq:strong_converse_used_Bi1}\\
&\leq& \ppr{\mathcal{E}_n}+\beta\sum_{i=1}^{M_n}   \pprr_{\vect{Y}^{\tau_n},\vect{X}^{t-1}}\farg{ \mathcal{D}_i, \vect{u}^{t-1}(i)}\label{eq:strong_converse_used_Bi2}\\
&\leq& \ppr{\mathcal{E}_n}+\beta.\label{eq:strong_converse_final_bound}
\end{IEEEeqnarray}}
Here, \eqref{eq:strong_converse_0} follows from \eqref{eq:strong_conv_beta_lower_bound} and \eqref{eq:strong_converse_Bi}; \eqref{eq:strong_converse_first_ineq} follows because $\mathcal{B}_i \cap \mathcal{D}_i^\complement \subseteq \mathcal{D}_i^\complement$ and because $\pprr_{\vect{X}^{\tau_n}, \vect{Y}^{\tau_n}}$ can be factorized as $\pprr_{\vect{Y}^{\tau_n},\vect{X}^{t-1}}\pprr_{\vect{X}_t^{\tau_n}|\vect{Y}_t^{\tau_n},\vect{X}^{t-1}}$; \eqref{eq:strong_converse_used_Bi1} follows from \eqref{eq:strong_converse_Bi}; and finally, \eqref{eq:strong_converse_final_bound} follows because $\{ \mathcal{D}_i\}_{i=1}^{M_n}$ are disjoint sets. Since \eqref{eq:strong_converse_final_bound} holds for $t\in\range{1}{\tau_n}$, we have established \eqref{eq:verdu_han_converse}.
\end{IEEEproof}
 By Lemma~\ref{lem:verdu_han_converse} and \eqref{eq:Rin_minus_Ch_geq_2gamma}, we have for all $n\geq n_1$\ifthenelse{\boolean{thesis}}{
\begin{IEEEeqnarray}{rCl}
\ppr{\mathcal{E}_n} &\geq&  \ppr{\frac{1}{n}i\farg{\vect{X}_{\scseq}^{\tau_n};\vect{Y}_{\scseq}^\tau\Big| \vect{X}^{\scseq-1}} \leq  \sum_{k=\scseq}^{\tau_n} R_k^{(n)} - \gamma } - 2^{-n\gamma}\IEEEeqnarraynumspace\\
 &\geq& \ppr{\frac{1}{n}i\farg{\vect{X}_{\scseq}^{\tau_n};\vect{Y}_{\scseq}^{\tau_n}\Big| \vect{X}^{\scseq-1}} \leq  \sum_{k=\scseq}^{\tau_n} C(h_k) + \gamma }- 2^{-n\gamma}.\label{eq:strong_converse_first_term}
\end{IEEEeqnarray}}{
\begin{IEEEeqnarray}{rCl}
\ppr{\mathcal{E}_n} &\geq&  \ppr{\frac{1}{n}i\farg{\vect{X}_{\scseq}^{\tau_n};\vect{Y}_{\scseq}^\tau\Big| \vect{X}^{\scseq-1}} \leq  \sum_{k=\scseq}^{\tau_n} R_k^{(n)} - \gamma } \nonumber\\
&&{}- 2^{-n\gamma}\IEEEeqnarraynumspace\\
 &\geq& \ppr{\frac{1}{n}i\farg{\vect{X}_{\scseq}^{\tau_n};\vect{Y}_{\scseq}^{\tau_n}\Big| \vect{X}^{\scseq-1}} \leq  \sum_{k=\scseq}^{\tau_n} C(h_k) + \gamma }\nonumber\\
 &&{} - 2^{-n\gamma}.\label{eq:strong_converse_first_term}
\end{IEEEeqnarray}}
Next, by using the argument in the proof of \cite[Th.~3.7.4]{Han} to analyze the first term in \eqref{eq:strong_converse_first_term}, we find that
\begin{IEEEeqnarray}{rCl}
  \lim_{n\rightarrow \infty}\ppr{\frac{1}{n}i\farg{\vect{X}_{\scseq}^{\tau_n}; \vect{Y}_{\scseq}^{\tau_n} \Big| \vect{X}^{\scseq-1}} \leq \sum_{k=\scseq}^{\tau_n} C(h_k) + \gamma} = 1.\label{eq:strong_converse_final}\IEEEeqnarraynumspace
\end{IEEEeqnarray}
Using \eqref{eq:strong_converse_final} in \eqref{eq:strong_converse_first_term}, we obtain $\lim_{n\rightarrow \infty}\ppr{\mathcal{E}_n} =1$ as desired.

\section{Proof of Lemma~\ref{lem:achievability} (achievability)}\label{app:achiev_proof}
Define the random variable $U_n\in\mathcal{U}_n\triangleq \mathbb{R}^n \times \mathbb{R}^n\times \mathbb{R}^n \times \cdots$ by the probability distribution 
\begin{align}
P_{U_n} \triangleq \mathcal{P}_n\times \mathcal{P}_n \times \mathcal{P}_n \times \cdots
\end{align}
where $\mathcal{P}_n$ denotes probability density of $\sqrt{n} \vect{\tilde X}/\vectornormbig{\vect{\tilde X}}_2$. Here, $\vect{\tilde X}\sim \mathcal{N}(\vect{0},\vect{I}_n)$ and $\vectornorm{\cdot}_2$ denotes the Euclidean distance. Hence, $\mathcal{P}_n$ denotes the uniform distribution on the $n$-dimensional sphere with radius $\sqrt{n}$. We use one realization of $U_n$ to generate the encoder and decoding functions. Then, we show that the conditional probability of error averaged over $U_n$, $\{\mathbf{Z}_t\}_{t=1}^\infty$, and $H^\infty$ given that $\max_{k\in\range{1}{\tau_n}}\mathbb{u}^{(n)}_{k,\tau_n}(H^\infty)\leq -c_n$ tends to zero. Invoking the random coding argument then enables us to show that there must be at least one realization of $U_n$ for each $n$ such that the probability of error tends to zero as $n\rightarrow \infty$. Let the $i$th entry of $u\in\mathcal{U}_n$ be denoted by $u(i)\in\mathbb{R}^n$. By countability of $\mathbb{N}^2$, there exists a bijection between $\mathbb{N}^2$ and $\mathbb{N}$ defined by the mapping  $\imath: \mathbb{N}^2 \mapsto \mathbb{N}$. The encoding functions $\mathbb{f}_{n,t}^{(r)}$, for $r\in \mathbb{R}_+$, are then defined in terms of $u\in\mathcal{U}_n$ as follows
\begin{align}
  \mathbb{f}_{t}^{(n)}(u,\mathbf{b}) &= u\mathopen{}\Bigg(\imath\mathopen{}\Bigg(t,1+\sum_{i=1}^{ \len{\vect{b}}} b_i 2^{i-1}\Bigg)\Bigg)\label{eq:encoder_def}
\end{align}
for every  $\mathbf{b} \in \mathfrak{B}$, where $b_i$ is the $i$th entry of $\mathbf{b}$ and $\len{\cdot}$ denotes the length of a vector. The inner sum in \eqref{eq:encoder_def} is a binary-to-integer conversion that converts the information bit vector $\mathbf{b}$ into an integer-valued index in the range $\range{1}{2^{\len{\vect{b}}}}$. 
Based on the above construction of the encoder, we have that (recall that $B_1^{\lceil n \cumR_t^{(n)} \rceil} = (B_1,\cdots,B_{\lceil n \cumR_t^{(n)} \rceil})$)
\begin{IEEEeqnarray}{rCl}
  \vect{X}_t &=& \mathbb{f}_{t}^{(n)}\fargbig{U_n, B_1^{\lceil n \cumR_t^{(n)} \rceil}}.
\end{IEEEeqnarray}
In order to keep notation simple, we define for $\vect{b}\in\{0,1\}^{\lceil n \cumR_t^{(n)}\rceil }$ and $j,t\in\mathbb{N}$, $j\leq t$,\ifthenelse{\boolean{thesis}}{
\begin{IEEEeqnarray}{rCl}
  \vect{\bar X}^{(n)}_{j:t}(u,\vect{b})
  & \triangleq& \Big[\mathbb{f}_{j}^{(n)}\fargbig{u,b_1^{\lceil n \cumR_j^{(n)} \rceil}}, \cdots,\mathbb{f}_{t}^{(n)}\fargbig{u,b_1^{\lceil n \cumR_t^{(n)} \rceil}}\Big].
\end{IEEEeqnarray}}{
\begin{IEEEeqnarray}{rCl}
  \IEEEeqnarraymulticol{3}{l}{\vect{\bar X}^{(n)}_{j:t}(u,\vect{b})}\nonumber\\
  & \triangleq& \Big[\mathbb{f}_{j}^{(n)}\fargbig{u,b_1^{\lceil n \cumR_j^{(n)} \rceil}}, \cdots,\mathbb{f}_{t}^{(n)}\fargbig{u,b_1^{\lceil n \cumR_t^{(n)} \rceil}}\Big].
\end{IEEEeqnarray}}

Let $\zeta_n\triangleq c_n/2$. For every $\vect{y}^t\in\mathbb{R}^{nt}$, we define the threshold-based decoding functions as follows:\ifthenelse{\boolean{thesis}}{
  \begin{IEEEeqnarray}{rCl}
    g_{t}^{(n)}(u, \mathbf{y}^t, H^t)&\triangleq& \left\{\begin{array}{ll}
\mathbf{b} & \text{if } \exists!\mathbf{b}\in\{0,1\}^{\lceil n \cumR_t^{(n)} \rceil} \text{ s.t. } \forall j\in\range{1}{t}:\\
&  i\fargbig{\vect{\bar X}^{(n)}_{j:t}(u,\mathbf{b}); \vect{y}^t|H_1^t} \geq n \sum_{k=j}^t R^{(n)}_k + n\zeta_n\\
{[]},  & \text{otherwise}.
\end{array}
\right.\IEEEeqnarraynumspace\label{eq:gnt_def}
  \end{IEEEeqnarray}}{
  \begin{IEEEeqnarray}{rCl}
    \IEEEeqnarraymulticol{3}{l}{g_{t}^{(n)}(u, \mathbf{y}^t, H^t)\triangleq}\nonumber\\
    && \left\{\begin{array}{ll}
\mathbf{b} & \text{if } \exists!\mathbf{b}\in\{0,1\}^{\lceil n \cumR_t^{(n)} \rceil} \text{ s.t. } \forall j\in\range{1}{t}:\\
&  i\fargbig{\vect{\bar X}^{(n)}_{j:t}(u,\mathbf{b}); \vect{y}^t|H_1^t} \geq n \sum_{k=j}^t R^{(n)}_k + n\zeta_n\\
{[]},  & \text{otherwise}.
\end{array}
\right.\IEEEeqnarraynumspace\label{eq:gnt_def}
  \end{IEEEeqnarray}}
Here, $[]$ is the vector of length zero which indicates an error, and we have defined the mismatched information density as follows \cite{DispersionNonGaussian}
\begin{IEEEeqnarray}{rCl}
  i(\vect{x}; \vect{y}| h)  &\triangleq& n C(h) + \frac{\vectornorm{\vect{y}}_2^2}{2(h + 1) \log(2)} - \frac{\vectornormbig{\vect{y}-\sqrt{h}\vect{x}}_2^2}{2 \log(2)}.\IEEEeqnarraynumspace
\end{IEEEeqnarray}
We note that $\E{i(\vect{X}_t; \vect{Y}_t | H_t)|H_t=h} = n C(h)$ and by using the same arguments as in \cite{DispersionNonGaussian}, we find that
\begin{IEEEeqnarray}{rCl}
  V_n(h) &\triangleq&  \frac{1}{n}\text{Var}\mathopen{}\left[ i(\vect{X}_t; \vect{Y}_t|H_t) | H_1=h\right]\label{eq:Vdef}\nonumber\\
  &\stackrel{n\rightarrow \infty}{\rightarrow}& \frac{ h(h + 2) }{2(h+1)^2\log^2(2)} \\
  &\leq& \frac{1}{2\log^2(2)}\\
  &\triangleq& V.\label{eq:Vdef3}
\end{IEEEeqnarray}
Thus, for sufficiently large $n$, we have that $V_n(h) \leq 2V$ for all $h\in\mathbb{R}_+$. 

It remains to analyze the probability of error. To do so, we rely on the technique used to prove Shannon's achievability bound in 
\cite[Th.~17.1]{PolyanskiyIT}.
Assume, without loss of generality, that $B_i = 0$ for $i\in\mathbb{N}$. 
We define, for $j\in\range{1}{\tau_n}$, the ``outage'' events as follows\ifthenelse{\boolean{thesis}}{
\begin{IEEEeqnarray}{rCl}
  \mathcal{A}_j
   &\triangleq \bigg\{ i(\vect{\bar X}^{(n)}_{j:\tau_n}(U_n,\vect{0}); \vect{Y}^{\tau_n}_j | H_j^{\tau_n}) < n \sum_{k=j}^{\tau_n} R_k^{(n)} + n\zeta_n  \bigg\}.\label{eq:Aj_def}\IEEEeqnarraynumspace
  \end{IEEEeqnarray}}{
\begin{IEEEeqnarray}{rCl}
  \IEEEeqnarraymulticol{3}{l}{\mathcal{A}_j}\nonumber\\
   &\triangleq \bigg\{ i(\vect{\bar X}^{(n)}_{j:\tau_n}(U_n,\vect{0}); \vect{Y}^{\tau_n}_j | H_j^{\tau_n}) < n \sum_{k=j}^{\tau_n} R_k^{(n)} + n\zeta_n  \bigg\}.\label{eq:Aj_def}\IEEEeqnarraynumspace
  \end{IEEEeqnarray}}
  Here, $\mathbf{0}$ denotes the all-zero vector (we omit specifying the length to keep notation simple). The ``confusion'' events are similarly defined by\ifthenelse{\boolean{thesis}}{
  \begin{IEEEeqnarray}{rCl}
  \mathcal{B}(\mathbf{b}) &\triangleq& \bigcap_{j\in\range{1}{\tau_n}}\bigg\{i(\vect{\bar X}^{(n)}_{j:\tau_n}(U_n,\vect{b}); \vect{Y}_j^{\tau_n} | H_j^{\tau_n})  \geq n \sum_{k=j}^{\tau_n} R_k^{(n)} + n\zeta_n  \bigg\}.\label{eq:confusion_def}
\end{IEEEeqnarray}}{
  \begin{IEEEeqnarray}{rCl}
  \mathcal{B}(\mathbf{b}) &\triangleq& \bigcap_{j\in\range{1}{\tau_n}}\bigg\{i(\vect{\bar X}^{(n)}_{j:\tau_n}(U_n,\vect{b}); \vect{Y}_j^{\tau_n} | H_j^{\tau_n}) \nonumber\\
  &&\qquad\qquad\qquad\qquad{} \geq n \sum_{k=j}^{\tau_n} R_k^{(n)} + n\zeta_n  \bigg\}.\label{eq:confusion_def}
\end{IEEEeqnarray}}
where $\mathbf{b} \in \{0,1\}^{\lceil n \cumR_{\tau_n}^{(n)}\rceil}$.
Here, $\mathcal{A}_j$ is the event that the information density of the correct codeword does not exceed the threshold, while $\mathcal{B}(\cdot)$ is the event that the information density of an incorrect codeword does exceed the threshold. 
Define the (random) set of information bit vectors for $k\in\range{1}{\tau_n}$
\ifthenelse{\boolean{thesis}}{}{
 \begin{figure*}[!t]
\normalsize
\setcounter{MYtempeqncnt}{\value{equation}}
\setcounter{equation}{150}
\begin{IEEEeqnarray}{rCl}
\IEEEeqnarraymulticol{3}{l}{\pr{\bigcup_{\substack{\vect{\bar b}\in \mathbb{B}_1 }}\mathcal{B}(\vect{\bar b}) \Bigg| B^\infty = \vect{0}, \mathcal{\bar H}_n}}\nonumber\\
&=& \E{\pr{\bigcup_{\substack{\vect{\bar b}\in \mathbb{B}_1 }}\mathcal{B}(\vect{\bar b}) \Bigg| B^\infty = \vect{0}, H^\infty}\Bigg| \mathcal{\bar H}_n}\label{eq:confusion_total_expect}\\
&=& \EBigg{\prBigg{\bigcup_{j=1}^{\tau_n} \bigcup_{\mathbf{\bar b}\in \mathbb{B}_j\setminus \mathbb{B}_{j+1} }\bigcap_{q \in\range{1}{t}}\Bigg\{ i\farg{\vect{\bar X}_{q:\tau_n}^{(n)}(U,\mathbf{\bar b}); \vect{Y}^{\tau_n}_q} \geq n \sum_{k=q}^{\tau_n} R_k^{(n)} + n\zeta_n \Bigg\} \Bigg| B^\infty = \vect{0}, H^\infty}\Bigg| \mathcal{\bar H}_n}\label{eq:confusion_used_def}\IEEEeqnarraynumspace\\
&\leq& \EBigg{\prBigg{\bigcup_{j=1}^{\tau_n} \bigcup_{\mathbf{\bar b}\in\mathbb{B}_j \setminus \mathbb{B}_{j+1}}\Bigg\{i\farg{\vect{\bar X}_{j:\tau_n}^{(n)}(U,\mathbf{\bar b}); \vect{Y}^{\tau_n}_j} \geq n \sum_{k=j}^{\tau_n} R_k^{(n)} + n\zeta_n \Bigg\} \Bigg| B^\infty = \vect{0}, H^\infty}\Bigg| \mathcal{\bar H}_n}\label{eq:confusion_union_bound}\IEEEeqnarraynumspace\\
&\leq& \EBigg{\sum_{j=1}^{\tau_n}  2^{\lceil n \sum_{k=j}^{\tau_n} R_k^{(n)}\rceil}\prBigg{ i(\vect{\bar X}^{\tau_n}_j ; \vect{Y}^{\tau_n}_j) \geq n \sum_{k=j}^{\tau_n} R_k^{(n)} +n \zeta_n   \Bigg| H^\infty }\Bigg|\mathcal{\bar H}_n}\label{eq:achiev_barX}\\
&\leq& \EBigg{\sum_{j=1}^{\tau_n}  2^{\lceil n \sum_{k=j}^{\tau_n} R_k^{(n)}\rceil} 2^{ - \left( n\sum_{k=j}^{\tau_n} R_k^{(n)} + n\zeta_n\right)}} \label{eq:yury_lemma}\\
&\leq& \EBigg{\sum_{j=1}^{\tau_n}  2^{ - n\zeta_n+1}}\label{eq:used_mR_def}\\
&=& \taumax 2^{-n c_n/2+1}.\label{eq:achiev_final}
\end{IEEEeqnarray}

\setcounter{equation}{\value{MYtempeqncnt}}
\hrulefill
\vspace*{4pt}
\end{figure*}}\ifthenelse{\boolean{thesis}}{
\begin{IEEEeqnarray}{rCl}
  \mathbb{B}_k
   &\triangleq& \Big\{ \mathbf{b}\in\{0,1\}^{\lceil n \cumR_{\tau_n}^{(n)}\rceil } : b_1^{\lceil n \cumR^{(n)}_{k-1}\rceil} = \mathbf{0}, b_{{\lceil n \cumR^{(n)}_{k-1}\rceil}+1}^{\lceil n \cumR^{(n)}_{\tau_n}\rceil} \not= \mathbf{0} \Big\}.\label{eq:set_of_bit_vectors}\IEEEeqnarraynumspace
\end{IEEEeqnarray}}{
\begin{IEEEeqnarray}{rCl}
  \IEEEeqnarraymulticol{3}{l}{\mathbb{B}_k}\nonumber\\
   &\triangleq& \Big\{ \mathbf{b}\in\{0,1\}^{\lceil n \cumR_{\tau_n}^{(n)}\rceil } : b_1^{\lceil n \cumR^{(n)}_{k-1}\rceil} = \mathbf{0}, b_{{\lceil n \cumR^{(n)}_{k-1}\rceil}+1}^{\lceil n \cumR^{(n)}_{\tau_n}\rceil} \not= \mathbf{0} \Big\}.\label{eq:set_of_bit_vectors}\IEEEeqnarraynumspace
\end{IEEEeqnarray}}
We also define $\mathbb{B}_{\tau_n+1} \triangleq \emptyset$.
Here, we let $\cumR_0^{(n)} = 0$ such that $\mathbb{B}_1$ is the set of all binary vectors of length $\lceil n \cumR^{(n)}_{\tau_n} \rceil$ except the all-zero vector $\vect{0}$. 
Note that $|\mathbb{B}_k|=  2^{\lceil n(R_k + \cdots + R_{\tau_n})\rceil}-1$ and that 
$\vect{\bar X}^{(n)}_{k:\tau_n}(U_n, \mathbf{b})$ 
and $\mathbf{Y}_{k}^{\tau_n}$ are conditionally independent for every $\mathbf{b} \in\mathbb{B}_k$ given $B_1^\infty=\mathbf{0}$ and $H^\infty$. 
Define the error event
\begin{IEEEeqnarray}{rCl}
\mathcal{E}_n(U_n) \triangleq \Big\{\mathbb{g}^{(n)}_{\tau_n}(U_n,\mathbf{Y}_1^{\tau_n}, H^{\tau_n}) \not= B_1^{\lceil n \cumR_{\tau_n} \rceil}\Big\}.
\end{IEEEeqnarray}
Then, we obtain the following probability of error 
\ifthenelse{\boolean{thesis}}{
\begin{IEEEeqnarray}{rCl}
\IEEEeqnarraymulticol{3}{l}{\pr{\mathcal{E}_n(U_n)\Big| \mathcal{\bar H}_n}}\nonumber\\
&=& \pr{\mathcal{E}_n(U_n)\Big|B_1^\infty=\mathbf{0}, \mathcal{\bar H}_n}\label{eq:achiev_err_bound2}\\
   &=& \pr{ \bigcup_{k=1}^{\tau_n} \mathcal{A}_k  \cup \bigcup_{\substack{\vect{\bar b}\in \mathbb{B}_1}}\mathcal{B}(\vect{\bar b})\Bigg| B^\infty = \vect{0},\mathcal{\bar H}_n}\label{eq:achiev_err_bound3}\\
   &\leq& \pr{ \bigcup_{k=1}^{\tau_n}\mathcal{A}_k\Bigg| B^\infty = \vect{0},\mathcal{\bar H}_n}+\pr{\bigcup_{\substack{\vect{\bar b}\in \mathbb{B}_1}}\mathcal{B}(\vect{\bar b}) \Bigg| B^\infty = \vect{0},\mathcal{\bar H}_n}.\label{eq:achiev_three_sums}
   \end{IEEEeqnarray}}{
\begin{IEEEeqnarray}{rCl}
\IEEEeqnarraymulticol{3}{l}{\pr{\mathcal{E}_n(U_n)\Big| \mathcal{\bar H}_n}}\nonumber\\
&=& \pr{\mathcal{E}_n(U_n)\Big|B_1^\infty=\mathbf{0}, \mathcal{\bar H}_n}\label{eq:achiev_err_bound2}\\
   &=& \pr{ \bigcup_{k=1}^{\tau_n} \mathcal{A}_k  \cup \bigcup_{\substack{\vect{\bar b}\in \mathbb{B}_1}}\mathcal{B}(\vect{\bar b})\Bigg| B^\infty = \vect{0},\mathcal{\bar H}_n}\label{eq:achiev_err_bound3}\\
   &\leq& \pr{ \bigcup_{k=1}^{\tau_n}\mathcal{A}_k\Bigg| B^\infty = \vect{0},\mathcal{\bar H}_n}\nonumber\\
   &&\qquad\qquad\qquad{}+\pr{\bigcup_{\substack{\vect{\bar b}\in \mathbb{B}_1}}\mathcal{B}(\vect{\bar b}) \Bigg| B^\infty = \vect{0},\mathcal{\bar H}_n}.\label{eq:achiev_three_sums}
   \end{IEEEeqnarray}}
   Here,  \eqref{eq:achiev_err_bound2} follows from symmetry, \eqref{eq:achiev_err_bound3} follows from \eqref{eq:gnt_def}, \eqref{eq:Aj_def}, and \eqref{eq:confusion_def}; and \eqref{eq:achiev_three_sums} follows from the union bound.
   Next, we upper-bound each of the two terms in \eqref{eq:achiev_three_sums} separately. For the first term, we use the law of total expectation and the union bound to obtain\ifthenelse{\boolean{thesis}}{
\begin{IEEEeqnarray}{rCl}
  \IEEEeqnarraymulticol{3}{l}{\pr{ \bigcup_{k=1}^{\tau_n}\mathcal{A}_k\Bigg| B^\infty = \vect{0},\mathcal{\bar H}_n}}\nonumber\\
   &=& \E{\pr{ \bigcup_{k=1}^{\tau_n}\mathcal{A}_k\Bigg| B^\infty = \vect{0}, H^\infty}\Bigg| \mathcal{\bar H}_n}\\
  &=& \EBigg{ \sum_{k=1}^{\tau_n} \prBigg{ \frac{1}{n}i(\vect{X}^{\tau_n}_k; \vect{Y}^{\tau_n}_k) < \sum_{j=k}^{\tau_n} R_j^{(n)} +\zeta_n\Bigg| H^\infty}\Bigg|\mathcal{\bar H}_n}.\label{eq:achiev_first_term0}\IEEEeqnarraynumspace
\end{IEEEeqnarray}}{
\begin{IEEEeqnarray}{rCl}
  \IEEEeqnarraymulticol{3}{l}{\pr{ \bigcup_{k=1}^{\tau_n}\mathcal{A}_k\Bigg| B^\infty = \vect{0},\mathcal{\bar H}_n}}\nonumber\\
   &=& \E{\pr{ \bigcup_{k=1}^{\tau_n}\mathcal{A}_k\Bigg| B^\infty = \vect{0}, H^\infty}\Bigg| \mathcal{\bar H}_n}\\
  &=& \EBigg{ \sum_{k=1}^{\tau_n} \prBigg{ \frac{1}{n}i(\vect{X}^{\tau_n}_k; \vect{Y}^{\tau_n}_k) \nonumber\\
  &&{} \qquad\qquad\qquad\qquad< \sum_{j=k}^{\tau_n} R_j^{(n)} +\zeta_n\Bigg| H^\infty}\Bigg|\mathcal{\bar H}_n}.\label{eq:achiev_first_term0}\IEEEeqnarraynumspace
\end{IEEEeqnarray}}
For all $h^\infty$ such that $\max_{k\in\range{1}{\tau_n}}\mathbb{u}_{k,\tau_n}^{(n)}(h^{\tau_n}) \leq -c_n$, we upper-bound the inner probability in \eqref{eq:achiev_first_term0} for sufficiently large $n$ using Chebyshev's inequality as follows
\begin{IEEEeqnarray}{rCl}
\IEEEeqnarraymulticol{3}{l}{\pr{ \frac{1}{n}i(\vect{X}^{\tau_n}_k; \vect{Y}^{\tau_n}_k | H_k^{\tau_n}) <  \sum_{j=k}^{\tau_n} R_j^{(n)} + \zeta_n\Bigg| H^\infty = h^\infty}}\nonumber\\
 &\leq& \E{\frac{2V(\tau_n - k + 1)  }{n(\sum_{j=k}^{\tau_n} [C(H_j)-R_j^{(n)}] -\zeta_n)^2}\Bigg|  H^\infty = h^\infty} \IEEEeqnarraynumspace\label{eq:cheb0}\\
&\leq& \E{\frac{2V(\tau_n - k+1) }{n(  c_n - \zeta_n )^2}\Bigg|  H^\infty = h^\infty}\label{eq:cheb1}\\
&=& \E{\frac{8V(\tau_n - k+1) }{nc_n^2}\Bigg|  H^\infty = h^\infty}\label{eq:zeta_choice1}\\
&\leq& \frac{8V \taumax }{n c_n^2}.\label{eq:zeta_choice}
\end{IEEEeqnarray}
Here, \eqref{eq:cheb0} follows from Chebyshev's inequality, from $\E{i(\vect{X}_t; \vect{Y}_t|H_t)|H_t} = n C(H_t)$, and from \eqref{eq:Vdef3}; \eqref{eq:cheb1} follows from $\max_{k\in\range{1}{\tau_n}}\mathbb{u}_{k,\tau_n}^{(n)}(h^{\tau_n}) \leq -c_n$; and \eqref{eq:zeta_choice} follows from $\tau_n \leq \taumax$. As a result of \eqref{eq:achiev_first_term0} and \eqref{eq:zeta_choice}, we have\ifthenelse{\boolean{thesis}}{
\begin{IEEEeqnarray}{rCl}
\pr{ \bigcup_{k=1}^{\tau_n}\mathcal{A}_k\Bigg| B^\infty = \vect{0}, \mathcal{\bar H}_n}&\leq& \E{\sum_{k=1}^{\tau_n}  \frac{8 V\taumax }{n c_n^2}\Bigg| \mathcal{\bar H}_n} \leq \frac{8 \taumax^2 V}{n c_n^2}.\IEEEeqnarraynumspace
\end{IEEEeqnarray}}{
\begin{IEEEeqnarray}{rCl}
\IEEEeqnarraymulticol{3}{l}{\pr{ \bigcup_{k=1}^{\tau_n}\mathcal{A}_k\Bigg| B^\infty = \vect{0}, \mathcal{\bar H}_n}}\nonumber\\
 \qquad\qquad\qquad&\leq& \E{\sum_{k=1}^{\tau_n}  \frac{8 V\taumax }{n c_n^2}\Bigg| \mathcal{\bar H}_n} \leq \frac{8 \taumax^2 V}{n c_n^2}.\IEEEeqnarraynumspace
\end{IEEEeqnarray}}
Next, the second term in \eqref{eq:achiev_three_sums} is upper-bounded as follows \ifthenelse{\boolean{thesis}}
{\begin{IEEEeqnarray}{rCl}
\IEEEeqnarraymulticol{3}{l}{\pr{\bigcup_{\substack{\vect{\bar b}\in \mathbb{B}_1 }}\mathcal{B}(\vect{\bar b}) \Bigg| B^\infty = \vect{0}, \mathcal{\bar H}_n}}\nonumber\\
&=& \E{\pr{\bigcup_{\substack{\vect{\bar b}\in \mathbb{B}_1 }}\mathcal{B}(\vect{\bar b}) \Bigg| B^\infty = \vect{0}, H^\infty}\Bigg| \mathcal{\bar H}_n}\label{eq:confusion_total_expect}\\
&=& \EBigg{\prBigg{\bigcup_{j=1}^{\tau_n} \bigcup_{\mathbf{\bar b}\in \mathbb{B}_j\setminus \mathbb{B}_{j+1} }\bigcap_{q \in\range{1}{t}}\Bigg\{ i\farg{\vect{\bar X}_{q:\tau_n}^{(n)}(U,\mathbf{\bar b}); \vect{Y}^{\tau_n}_q} \nonumber\\
&&{} \qquad\qquad\qquad\qquad\qquad \qquad\geq n \sum_{k=q}^{\tau_n} R_k^{(n)} + n\zeta_n \Bigg\} \Bigg| B^\infty = \vect{0}, H^\infty}\Bigg| \mathcal{\bar H}_n}\label{eq:confusion_used_def}\IEEEeqnarraynumspace\\
&\leq& \EBigg{\prBigg{\bigcup_{j=1}^{\tau_n} \bigcup_{\mathbf{\bar b}\in\mathbb{B}_j \setminus \mathbb{B}_{j+1}}\Bigg\{i\farg{\vect{\bar X}_{j:\tau_n}^{(n)}(U,\mathbf{\bar b}); \vect{Y}^{\tau_n}_j}\nonumber\\
&&\qquad\qquad\qquad\qquad\qquad{} \geq n \sum_{k=j}^{\tau_n} R_k^{(n)} + n\zeta_n \Bigg\} \Bigg| B^\infty = \vect{0}, H^\infty}\Bigg| \mathcal{\bar H}_n}\label{eq:confusion_union_bound}\IEEEeqnarraynumspace\\
&\leq& \EBigg{\sum_{j=1}^{\tau_n}  2^{\lceil n \sum_{k=j}^{\tau_n} R_k^{(n)}\rceil}\prBigg{ i(\vect{\bar X}^{\tau_n}_j ; \vect{Y}^{\tau_n}_j) \geq n \sum_{k=j}^{\tau_n} R_k^{(n)} +n \zeta_n   \Bigg| H^\infty }\Bigg|\mathcal{\bar H}_n}\label{eq:achiev_barX}\\
&\leq& \EBigg{\sum_{j=1}^{\tau_n}  2^{\lceil n \sum_{k=j}^{\tau_n} R_k^{(n)}\rceil} 2^{ - \left( n\sum_{k=j}^{\tau_n} R_k^{(n)} + n\zeta_n\right)}} \label{eq:yury_lemma}\\
&\leq& \EBigg{\sum_{j=1}^{\tau_n}  2^{ - n\zeta_n+1}}\label{eq:used_mR_def}\\
&=& \taumax 2^{-n c_n/2+1}.\label{eq:achiev_final}
\end{IEEEeqnarray}}{[see~\eqref{eq:confusion_total_expect}--\eqref{eq:achiev_final}, shown in the top of the next page]. }
Here, \eqref{eq:confusion_total_expect} follows from the law of total expectation; \eqref{eq:confusion_used_def} follows from \eqref{eq:confusion_def} and \eqref{eq:set_of_bit_vectors}; \eqref{eq:confusion_union_bound} follows from the union bound and because $\left|\mathbb{B}_j \right| = (2^{\lceil n \sum_{k=j}^{\tau_n} R_k\rceil} - 1)$; \eqref{eq:achiev_barX} follows  by defining the random variables $\{\vect{\bar X}_t\}_{t=1}^\infty$ independently according to the probability distribution $\mathcal{P}_n$ such that they are independent of $\{\vect{X}_t\}_{t=1}^\infty$ and $\{\vect{Z}_t\}_{t=1}^\infty$; finally, \eqref{eq:yury_lemma} follows from \cite[Cor.~17.1]{PolyanskiyIT}. 
\addtocounter{equation}{7}Consequently, we have shown that 
\begin{IEEEeqnarray}{rCl}
\pr{\mathcal{E}_n(U_n) \Big| \mathcal{\bar H}_n} \leq \taumax 2^{-n c_n /2+1} + \frac{8 \taumax^2 V}{n c_n^2}
\end{IEEEeqnarray}
for all sufficiently large $n$. As a result, there exists a deterministic sequence $\{u_n^*\}_{n=1}^\infty$ such that 
\begin{IEEEeqnarray}{rCl}
\pr{\mathcal{E}_n(u_n^*) \Big| \mathcal{\bar H}_n} \leq \taumax 2^{-n c_n /2+1} + \frac{8 \taumax^2 V}{n c_n^2}.
\end{IEEEeqnarray}
Define 
\begin{IEEEeqnarray}{rCl}
p_{\text{min},n} \triangleq \min_{\substack{t\in\range{1}{\taumax}: \\\pr{\tau_n=t|\mathcal{\bar H}_n} > 0}} \pr{ \tau_n = t|\mathcal{\bar H}_n}.
\end{IEEEeqnarray}
The condition in \eqref{eq:min_prob_cond} implies that $p_{\text{min},n} \geq g_n$ for all sufficiently large $n$ and therefore we have\ifthenelse{\boolean{thesis}}{
\begin{IEEEeqnarray}{rCl}
\IEEEeqnarraymulticol{3}{l}{\lim_{n\rightarrow \infty}\max_{\substack{t\in \range{1}{\tau_n}:\\ \pr{\tau_n=t|\mathcal{\bar H}_n}>0}}\pr{\mathcal{E}_n(u_n^*)\Big|  \mathcal{\bar H}_n, \tau_n = t}} \nonumber\\
 &\leq& \lim_{n\rightarrow\infty}\frac{1}{p_{\text{min},n}}\sum_{\substack{t \in\range{1}{\taumax}:\\ \pr{\tau_n = t|\mathcal{\bar H}_n} >0 }} \pr{\tau_n = t|\mathcal{\bar H}_n}\pr{\mathcal{E}_n(u_n^*)\Big|  \mathcal{\bar H}_n,\tau_n = t} \\
   &\leq& \lim_{n\rightarrow\infty} \frac{1}{g_n}\pr{\mathcal{E}_n(u_n^*) \Big|  \mathcal{\bar H}_n }\label{eq:law_of_total_prob}\\
  &\leq& \lim_{n\rightarrow\infty} \frac{1}{g_n} \left(\taumax 2^{-n c_n /2+1} + \frac{8 \taumax^2 V}{n c_n^2}\right)\\
   &=& 0.\label{eq:brq_last_eq2}
\end{IEEEeqnarray}}{
\begin{IEEEeqnarray}{rCl}
\IEEEeqnarraymulticol{3}{l}{\lim_{n\rightarrow \infty}\max_{\substack{t\in \range{1}{\tau_n}:\\ \pr{\tau_n=t|\mathcal{\bar H}_n}>0}}\pr{\mathcal{E}_n(u_n^*)\Big|  \mathcal{\bar H}_n, \tau_n = t}} \nonumber\\
 &\leq& \lim_{n\rightarrow\infty}\frac{1}{p_{\text{min},n}}\sum_{\substack{t \in\range{1}{\taumax}:\\ \pr{\tau_n = t|\mathcal{\bar H}_n} >0 }} \Big(\pr{\tau_n = t|\mathcal{\bar H}_n}\nonumber\\
 &&{}\qquad\qquad\qquad\quad\times\pr{\mathcal{E}_n(u_n^*)\Big|  \mathcal{\bar H}_n,\tau_n = t}\Big) \\
   &\leq& \lim_{n\rightarrow\infty} \frac{1}{g_n}\pr{\mathcal{E}_n(u_n^*) \Big|  \mathcal{\bar H}_n }\label{eq:law_of_total_prob}\\
  &\leq& \lim_{n\rightarrow\infty} \frac{1}{g_n} \left(\taumax 2^{-n c_n /2+1} + \frac{8 \taumax^2 V}{n c_n^2}\right)\\
   &=& 0.\label{eq:brq_last_eq2}
\end{IEEEeqnarray}}
Here, \eqref{eq:law_of_total_prob} follows from \eqref{eq:min_prob_cond} and the law of total probability. Moreover, \eqref{eq:brq_last_eq2} follows from \eqref{eq:achievability_conv_cond}, from the upper bound $2^{x} \leq 2/x^2$ that holds for $x\leq 0$, and because $\taumax$ is a nondecreasing sequence:
\begin{IEEEeqnarray}{rCl}
  \frac{\taumax 2^{-n c_n /2+1}}{g_n} &\leq& \frac{16 \bar \tau_n}{g_n n^2c_n^2} \leq \frac{o(1)}{\bar \tau_n n } = o(1).
\end{IEEEeqnarray}

\section{Proof of Theorem~\ref{thm:brq_opt} (upper bound)}\label{app:converse_zero_outage}

We shall prove that $\eta_{\text{opt}}(T) \leq \eta_{\text{BRQ}}(T)$ for $T > 1$. We do this by applying the converse result in Lemma~\ref{lem:strong_converse}, which implies that a zero outage EMS protocol must satisfy $\sup_{k\in\range{1}{\tau}}\mathbb{u}_{k,\tau}(H^\tau) \leq 0$ almost surely. To see this, note first that we must have $\tau < \infty$ almost surely. Otherwise, $\sup_n \E{\tau_n} = \infty$. Additionally, suppose that a zero outage EMS protocol satisfies $\pr{\sup_{k\in\range{1}{\tau}}\mathbb{u}_{k,\tau}(H^\tau) > 0} > 0$. Then,\ifthenelse{\boolean{thesis}}{
\begin{IEEEeqnarray}{rCl}
 \liminf_{n\rightarrow \infty}\pr{\mathcal{E}_n}  &\geq& \prBig{\sup_{k\in\range{1}{\tau}}\mathbb{u}_{k,\tau}(H^{\tau}) > 0,\tau<\infty}\nonumber\\
  && {}\qquad\times\liminf_{n\rightarrow \infty}\EBig{\pr{\mathcal{E}_n|H^\infty} \Big| \sup_{k\in\range{1}{\tau}}\mathbb{u}_{k,\tau}(H^{\tau}) > 0,\tau<\infty}\\
  &\geq& \prBig{\sup_{k\in\range{1}{\tau}}\mathbb{u}_{k,\tau}(H^{\tau}) > 0}\nonumber\\
  &&{}\qquad\times\EBig{\liminf_{n\rightarrow \infty}\pr{\mathcal{E}_n|H^\infty} \Big| \sup_{k\in\range{1}{\tau}}\mathbb{u}_{k,\tau}(H^{\tau}) > 0,\tau<\infty}\label{eq:fatou}\IEEEeqnarraynumspace\\
  &=& \prBig{\sup_{k\in\range{1}{\tau}}\mathbb{u}_{k,\tau}(H^{\tau}) > 0}\label{eq:strong_converse_used}\\
  &>& 0.
\end{IEEEeqnarray}}{
\begin{IEEEeqnarray}{rCl}
 \IEEEeqnarraymulticol{3}{l}{\liminf_{n\rightarrow \infty}\pr{\mathcal{E}_n}}\nonumber\\
  &&\geq \prBig{\sup_{k\in\range{1}{\tau}}\mathbb{u}_{k,\tau}(H^{\tau}) > 0,\tau<\infty}\nonumber\\
  && {}\times\liminf_{n\rightarrow \infty}\EBig{\pr{\mathcal{E}_n|H^\infty} \Big| \sup_{k\in\range{1}{\tau}}\mathbb{u}_{k,\tau}(H^{\tau}) > 0,\tau<\infty}\\
  &&\geq \prBig{\sup_{k\in\range{1}{\tau}}\mathbb{u}_{k,\tau}(H^{\tau}) > 0}\nonumber\\
  &&{}\times\EBig{\liminf_{n\rightarrow \infty}\pr{\mathcal{E}_n|H^\infty} \Big| \sup_{k\in\range{1}{\tau}}\mathbb{u}_{k,\tau}(H^{\tau}) > 0,\tau<\infty}\label{eq:fatou}\IEEEeqnarraynumspace\\
  &&= \prBig{\sup_{k\in\range{1}{\tau}}\mathbb{u}_{k,\tau}(H^{\tau}) > 0}\label{eq:strong_converse_used}\\
  &&> 0.
\end{IEEEeqnarray}}
Here, \eqref{eq:fatou} follows from Fatou's lemma \cite[Th.~16.3]{billingsley} and because $\tau<\infty$ almost surely, and \eqref{eq:strong_converse_used} follows from Lemma~\ref{lem:strong_converse}.
Therefore, we can find $\tau$, $\{\mathbb{r}_t\}$, and $\{\mathbb{v}_t\}$ of an optimal zero outage EMS protocol satisfying the constraint $\E{\tau}\leq T$ by solving the optimization problem:
\begin{subequations}
\begin{IEEEeqnarray}{rCl}
\zeta_1(T)\triangleq&\sup_{\{\mathbb{r}_t\},\{\mathbb{v}_t\},\tau} & T \E{\cumR_{\tau}}/\E{\tau} \\
&\text{s.t.} &  \E{\tau}\leq T\\
& & \prBig{\sup_{k\in\range{1}{\tau}} \mathbb{u}_{k,\tau}(H_1^\tau)\leq 0} = 1.\IEEEeqnarraynumspace
\end{IEEEeqnarray}
\label{thm:opt_EMS_optproblem0}%
\end{subequations}
Here, $T\geq 1$ and $\cumR_\tau \triangleq \sum_{t=1}^{\tau} \mathbb{r}_t(\mathbb{v}_{t-1}(H^{t-1}))$.
It turns out that it is convenient to scale the objective function in \eqref{thm:opt_EMS_optproblem0} by $T$. We shall prove that the solution to \eqref{thm:opt_EMS_optproblem0} coincides with the BRQ-EMS protocol, i.e., we find that $\zeta_1(T) = T \eta_{\text{BRQ}}(T)$ under the condition in \eqref{eq:opt_cond}.

Since the transmitter has full delayed CSIT, it is sufficient to maximize over all composite rate selection-feedback functions $\mathbb{r}\mathbb{v}_t: \mathbb{R}_+^{t-1} \mapsto \mathbb{R}_+$ such that $\mathbb{r}_t(\mathbb{v}_{t-1}(H_1^{t-1})) = \mathbb{rv}_t(H_1^{t-1})$. Moreover, we also define the optimization problem
\begin{subequations}
\begin{IEEEeqnarray}{rCl}
\zeta(T)\triangleq&\sup_{\{\mathbb{r}\mathbb{v}_t\},\tau} & \E{\cumR_\tau} \\
&\text{s.t.} &  \E{\tau}\leq T\\
& & \prBig{\sup_{k\in\range{1}{\tau}} \mathbb{u}_{k,\tau}(H_1^\tau)\leq 0} = 1.
\end{IEEEeqnarray}
\label{thm:opt_EMS_optproblem}%
\end{subequations}
It clearly follows from the constraint $\E{\tau}\leq T$ that $\zeta(T) \leq \zeta_1(T)$, but if it can be shown that $\zeta(\cdot)$ is an increasing function, then $\zeta_1(T) = \zeta(T)$. Indeed, suppose that $\zeta(\cdot)$ is an increasing function and that there exists $\tilde T>1$ such that $\zeta(\tilde T) < \zeta_1(\tilde T)$. Let $\tau^*$ be the solution of \eqref{thm:opt_EMS_optproblem0} for $T = \tilde T$. Then, we must have that $\zeta(\E{\tau^*}) = \zeta_1(\E{\tau^*})$ and that $\E{\tau^*} < \tilde T$. But since $\zeta(T)$ is an increasing function, we also have that $\zeta(\tilde T) > \zeta(\E{\tau^*}) = \zeta_1(\E{\tau^*}) = \zeta_1(\tilde T)$ which cannot be true since $\zeta(T) \leq \zeta_1(T)$ for all $T>1$. We shall later use this fact to prove equality between $\zeta_1(T)$ and $\zeta(T)$ under the condition in \eqref{eq:opt_cond}.

To solve the optimization problem in \eqref{thm:opt_EMS_optproblem}, we first relate $\zeta(T)$ to the decoding time $\tau_{\text{opt}}$ (recall that $\tau_{\text{opt}}$ depends on $\{\mathbb{rv}_t\}$) defined in \eqref{eq:opt_EMS_tau} as follows  \ifthenelse{\boolean{thesis}}{
\begin{IEEEeqnarray}{rCl}
\zeta(T)
&=& \sup_{\substack{\{\mathbb{r}\mathbb{v}_t\}:\\ \E{\tau_{\text{opt}}}\leq T}} \mathopen{}\Bigg\{\EBigg{\sum_{t=1}^{\tau_{\text{opt}}} \mathbb{r}\mathbb{v}_t(H^{t-1}) }+ \max_{\tau\geq \tau_{\text{opt}}  } \EBigg{\sum_{t=\tau_{\text{opt}+1}}^{\tau} \mathbb{r}\mathbb{v}_t(H^{t-1})}\Bigg\}.\label{eq:rid_of_tau}\IEEEeqnarraynumspace
\end{IEEEeqnarray} }{
\begin{IEEEeqnarray}{rCl}
\IEEEeqnarraymulticol{3}{l}{\zeta(T)}\nonumber\\
&=& \sup_{\substack{\{\mathbb{r}\mathbb{v}_t\}:\\ \E{\tau_{\text{opt}}}\leq T}} \mathopen{}\Bigg\{\EBigg{\sum_{t=1}^{\tau_{\text{opt}}} \mathbb{r}\mathbb{v}_t(H^{t-1}) }\nonumber\\
&&\qquad\qquad{}+ \max_{\tau\geq \tau_{\text{opt}}  } \EBigg{\sum_{t=\tau_{\text{opt}+1}}^{\tau} \mathbb{r}\mathbb{v}_t(H^{t-1})}\Bigg\}.\label{eq:rid_of_tau}\IEEEeqnarraynumspace
\end{IEEEeqnarray}}
Here, the inner maximization is subject to the constraints $\E{\tau}\leq T$ and $\pr{\sup_{k\in\range{\tau_{\text{opt}}+1}{\tau}} \mathbb{u}_{k,\tau}(H^\tau)\leq 0}=1$, and we have used that any feasible (in the sense defined by the constraints in \eqref{thm:opt_EMS_optproblem}) decoding time $\tau$ must satisfy $\tau\geq \tau_{\text{opt}}$ almost surely and that the constraints $\mathbb{u}_{k,\tau}(H_1^\tau)\leq 0$, by the definition of $\tau_{\text{opt}}$, are automatically satisfied for $k\in\range{1}{\tau_{\text{opt}}}$ when $\pr{\sup_{k\in\range{\tau_{\text{opt}}+1}{\tau}} \mathbb{u}_{k,\tau}(H_1^\tau)\leq 0}=1$ because\ifthenelse{\boolean{thesis}}{
\begin{IEEEeqnarray}{rCl}
  \mathbb{u}_{k,\tau}(H_1^\tau)
   &=& \mathbb{u}_{1,\tau_{\text{opt}}}(H_1^{\tau_{\text{opt}}}) - \mathbb{u}_{1,k-1}(H_1^{k-1}) + \mathbb{u}_{\tau_{\text{opt}+1},\tau}(H_{1}^\tau)\IEEEeqnarraynumspace\\
  &\leq& 0.
\end{IEEEeqnarray}}{
\begin{IEEEeqnarray}{rCl}
  \IEEEeqnarraymulticol{3}{l}{\mathbb{u}_{k,\tau}(H_1^\tau)}\nonumber\\
   &=& \mathbb{u}_{1,\tau_{\text{opt}}}(H_1^{\tau_{\text{opt}}}) - \mathbb{u}_{1,k-1}(H_1^{k-1}) + \mathbb{u}_{\tau_{\text{opt}+1},\tau}(H_{1}^\tau)\IEEEeqnarraynumspace\\
  &\leq& 0.
\end{IEEEeqnarray}}
It follows that the inner maximization in \eqref{eq:rid_of_tau} is upper-bounded by $(T-\E{\tau_{\text{opt}}})C_{\text{erg}}$, which implies that\ifthenelse{\boolean{thesis}}{
\begin{IEEEeqnarray}{rCl}
  \zeta(T)
  &\leq& \sup_{\substack{\{\mathbb{r}\mathbb{v}_t\}:\\\E{\tau_{\text{opt}}}\leq T} }\mathopen{} \Bigg\{\E{\sum_{t=1}^{\tau_{\text{opt}}} \mathbb{r}\mathbb{v}_t(H^{t-1}) } + (T-\E{\tau_{\text{opt}}})C_{\text{erg}} \Bigg\}\IEEEeqnarraynumspace\\
  &=& \sup_{1< T_1 \leq T}\mathopen{}\left\{ \zeta_{\text{opt}}(T_1) + (T-T_1)C_{\text{erg}}\right\}\label{eq:decompose_tau}
\end{IEEEeqnarray}}{
\begin{IEEEeqnarray}{rCl}
 \IEEEeqnarraymulticol{3}{l}{ \zeta(T)}\nonumber\\
  &\leq& \sup_{\substack{\{\mathbb{r}\mathbb{v}_t\}:\\\E{\tau_{\text{opt}}}\leq T} }\mathopen{} \Bigg\{\E{\sum_{t=1}^{\tau_{\text{opt}}} \mathbb{r}\mathbb{v}_t(H^{t-1}) } + (T-\E{\tau_{\text{opt}}})C_{\text{erg}} \Bigg\}\IEEEeqnarraynumspace\\
  &=& \sup_{1< T_1 \leq T}\mathopen{}\left\{ \zeta_{\text{opt}}(T_1) + (T-T_1)C_{\text{erg}}\right\}\label{eq:decompose_tau}
\end{IEEEeqnarray}}
where we have defined 
\begin{align}
\zeta_{\text{opt}}(T) &\triangleq \sup_{\substack{\{\mathbb{r}\mathbb{v}_t\}:\\\E{\tau_{\text{opt}}}\leq T} }\E{\sum_{t=1}^{\tau_{\text{opt}}} \mathbb{r}\mathbb{v}_t(H^{t-1}) }\label{eq:Ropt_def}
\end{align} 
for $T\geq 1$.
We can upper bound $\zeta_{\text{opt}}(\cdot)$ using weak duality as follows\ifthenelse{\boolean{thesis}}{
  \begin{IEEEeqnarray}{rCl}
     \zeta_{\text{opt}}(T)     &\leq& \min_{\lambda>0 }\mathopen{}\Bigg\{\lambda (T-1) + \sup_{\{\mathbb{r}\mathbb{v}_t\}}\mathopen{}\Bigg\{ \E{\sum_{t=1}^{\tau_{\text{opt}}} \mathbb{r}\mathbb{v}_t(H^{t-1}) } - \lambda (\E{\tau_{\text{opt}}}-1) \Bigg\}\Bigg\}.\IEEEeqnarraynumspace \label{eq:dyn_opt1}
  \end{IEEEeqnarray}}{
  \begin{IEEEeqnarray}{rCl}
     \zeta_{\text{opt}}(T)     &\leq& \min_{\lambda>0 }\mathopen{}\Bigg\{\lambda (T-1) + \sup_{\{\mathbb{r}\mathbb{v}_t\}}\mathopen{}\Bigg\{ \E{\sum_{t=1}^{\tau_{\text{opt}}} \mathbb{r}\mathbb{v}_t(H^{t-1}) } \nonumber\\
     &&\quad\qquad\qquad\qquad\qquad\quad{}- \lambda (\E{\tau_{\text{opt}}}-1) \Bigg\}\Bigg\}.\IEEEeqnarraynumspace \label{eq:dyn_opt1}
  \end{IEEEeqnarray}}
  We solve the inner maximization in \eqref{eq:dyn_opt1} using dynamic programming. For given $\lambda>0$, let $\{\mathbb{rv}^*_i\}$ be the solution to the inner maximization problem in \eqref{eq:dyn_opt1}. Then, observe that $\mathbb{r v}^*_t$ depends only on $H_1^{t-1}$ through $\mathbb{u}_1^{t-1}(H_1^{t-1})$. Intuitively, this means that the rate selection depends only on the amount of unresolved information up to time $t$. We define functions $\mathbb{\overline{rv}}_t: \mathbb{R} \mapsto \mathbb{R}_+$ and let $\mathbb{\overline{rv}}_t(\mathbb{u}_{1,t-1}(h_1^{t-1}))\triangleq \mathbb{rv}_t(h_1^{t-1})$ for $t\in\mathbb{N}$ and $h_1^{t-1}\in\mathbb{R}_+^{t-1}$.
  
Now, define the value function (see e.g.~\cite{Ross1995})
  \begin{align}
    V_t(u)\triangleq\max_{\{\mathbb{\overline{r v}}_i\}}\E{\sum_{i=t}^{\tau_t(u)} \mathbb{\overline{r v}}_i( \mathbb{\bar u}_{t,i-1}(u,H_t^{i-1}) )  - \lambda (\tau_t(u)-t) } \label{eq:dyn_opt2}
  \end{align}
  where 
  \begin{align}
    \tau_t(u) &\triangleq \min\mathopen{}\big\{ \bar t\geq t: \mathbb{\bar u}_{t,\bar t}(u,H_t^{\bar t}) < 0 \big\}
    \end{align}
    for $t\in\mathbb{N}$ and
    \begin{align}
    \mathbb{\bar u}_{k,t}(u, h_k^t) &\triangleq u + \sum_{i=k}^t [\mathbb{\overline{rv}}_i(\mathbb{\bar u}_{k,i-1}(u, h_k^{i-1}))- C(h_i)]
  \end{align}
  for $t,k\in\mathbb{N}$.
  Using these definitions, the inner maximization in \eqref{eq:dyn_opt1} can be expressed in terms of $V_t(\cdot)$ in \eqref{eq:dyn_opt2}:
  \begin{align}
  V_1(0)&=  \max_{\{\mathbb{r}\mathbb{v}_t\}} \E{\sum_{i=1}^{\tau_{\text{opt}}} \mathbb{r}\mathbb{v}_i(H_1^{i-1})  - \lambda (\tau_{\text{opt}}-1) } \label{eq:V10}
  \end{align}
To apply dynamic programming, the value function $V_t$ in \eqref{eq:dyn_opt2} is expressed in a recursive form as follows\ifthenelse{\boolean{thesis}}{
\begin{IEEEeqnarray}{rCl}
    V_t( u)
    &=&   \max_{r\geq 0 } \Big\{r+ \EBig{\indi{C(H) \leq u+r}(V_{t+1}(u+r- C(H))- \lambda) }\Big\}.\IEEEeqnarraynumspace\label{eq:opt_over_tau2}
\end{IEEEeqnarray}}{
\begin{IEEEeqnarray}{rCl}
    V_t( u)
    &=&   \max_{r\geq 0 } \Big\{r+ \EBig{\indi{C(H) \leq u+r}\nonumber\\
    &&{}\quad\qquad\qquad \times(V_{t+1}(u+r- C(H))- \lambda) }\Big\}.\IEEEeqnarraynumspace\label{eq:opt_over_tau2}
\end{IEEEeqnarray}}
  Here, we have defined $F_C(r) \triangleq \pr{\C{H} \leq r}$ and let $P_C(\cdot)$ be the probability density of $C(H)$.
  The problem is thereby formulated as a standard infinite horizon dynamic programming problem \cite{Ross1995}. Consequently, the value function $V_t$ is time-invariant such that $V_t(u) =  V_{t+1}(u)$ for all $t\in\mathbb{N}$ and $u\in\mathbb{R}$. We denote the time-invariant value function by $V(u)\triangleq V_1(u)$. As a result, we obtain\ifthenelse{\boolean{thesis}}{
  \begin{IEEEeqnarray}{rCl}
     V(u)
    &=&\max_{r\geq 0 } \Big\{r+ \EBig{\indi{C(H) \leq u+r}(V(u+r- C(H))- \lambda) }\Big\}.\IEEEeqnarraynumspace\label{eq:bellman}
  \end{IEEEeqnarray}}{
  \begin{IEEEeqnarray}{rCl}
     V(u)
    &=&\max_{r\geq 0 } \Big\{r+ \EBig{\indi{C(H) \leq u+r}\nonumber\\
    &&{}\quad\quad\qquad\qquad \times(V(u+r- C(H))- \lambda) }\Big\}.\IEEEeqnarraynumspace\label{eq:bellman}
  \end{IEEEeqnarray}}
  It remains to guess $ V(u)$ satisfying \eqref{eq:bellman}. 
  We claim that the value function has the form $ V(u)=A-u$ for $u\in [0,r_A]$, where 
    \begin{align}
  A \triangleq \max_{r\geq 0} \left\{ r+\frac{\overline{C}(r)-F_C(r) \lambda}{1-F_C(r)}\right\}\label{eq:equation_for_A}
  \end{align}
  and $r_A$ is the maximizer in \eqref{eq:equation_for_A}. Here, $\overline{C}(r) \triangleq \int_{0}^r x P_C(x)\dd x$.
 Indeed, by substituting $V(u)=A-u$ into \eqref{eq:bellman}, we have \ifthenelse{\boolean{thesis}}{
  \begin{IEEEeqnarray}{rCl}
    V(u) &=& \max_{r \geq 0} \Big\{r+ \EBig{\indi{C(H) \leq u + r }(A - u- r+ C(H)- \lambda )}\Big\}\IEEEeqnarraynumspace\\
    &=&  \max_{\bar r\geq u} \left\{ \bar r(1-F_C(\bar r))+ \overline{C}(\bar r)+ F_C(\bar r)\left(A- \lambda \right)\right\}-u\label{eq:Vu_second_last_step}\\
    &=& A - u\label{eq:Vu_last_step}
    \end{IEEEeqnarray}}{
  \begin{IEEEeqnarray}{rCl}
    V(u) &=& \max_{r \geq 0} \Big\{r+ \EBig{\indi{C(H) \leq u + r }\nonumber\\
    &&{} \quad\quad\qquad\qquad\times(A - u- r+ C(H)- \lambda )}\Big\}\IEEEeqnarraynumspace\\
    &=&  \max_{\bar r\geq u} \left\{ \bar r(1-F_C(\bar r))+ \overline{C}(\bar r)+ F_C(\bar r)\left(A- \lambda \right)\right\}\nonumber\\
    &&{}-u\label{eq:Vu_second_last_step}\\
    &=& A - u\label{eq:Vu_last_step}
    \end{IEEEeqnarray}}
  for every $u\in[0,r_A]$. Here, \eqref{eq:Vu_second_last_step} follows by the substitution $\bar r = u+r$ and \eqref{eq:Vu_last_step} follows from \eqref{eq:equation_for_A} because
\begin{align}
  0 &= \max_{r\geq 0} \left\{ r-A+\frac{\overline{C}(r)-F_C(r) \lambda}{1-F_C(r)}\right\} \\
  &= \max_{r\geq 0} \left\{ (r-A)(1-F_C(r))+\overline{C}(r)-F_C(r) \lambda\right\} \\
 &= \max_{r\geq 0} \left\{ r(1-F_C(r))+\overline{C}(r)+F_C(r)(A- \lambda)\right\} - A.\label{eq:Aopt_eq}
\end{align}
Since $r_A$ is a maximizer of the RHS of \eqref{eq:equation_for_A}, it is also a maximizer of \eqref{eq:Aopt_eq}, and thus also of the optimization problem in \eqref{eq:Vu_second_last_step}. This proves that $V(u) = A- u$ for $u\in[0,r_A]$.

We shall shortly prove that \eqref{eq:opt_cond} implies that
\begin{IEEEeqnarray}{rCl}
T \eta_{\text{BRQ}}(T) = F_C^{-1}\farg{1-\frac{1}{T}} +T \overline{C}\mathopen{}\left(F_C^{-1}\farg{1-\frac{1}{T}}\right)\label{eq:concave_cond}\IEEEeqnarraynumspace
\end{IEEEeqnarray}
is concave in $T$. 
  Consequently, we have shown the following\ifthenelse{\boolean{thesis}}{
  \begin{IEEEeqnarray}{rCl}
\zeta_{\text{opt}}(T)&\leq& \min_{\lambda > 0}\mathopen{}\bigg\{\lambda (T-1)+ \max_{r \geq 0} \mathopen{}\left\{ r+\frac{\overline{C}(r)-\lambda F_C(r) }{1-F_C(r)}\right\}\bigg\}\label{eq:A_subt}\IEEEeqnarraynumspace\\
 &=& \min_{\lambda > 0}\mathopen{}\bigg\{\lambda (T-1)+ \max_{\nu\geq 1} \mathopen{}\bigg\{  F_C^{-1}\farg{1-\frac{1}{\nu}}\nonumber\\
 &&{}\qquad\qquad\qquad\qquad\qquad+\nu \overline{C}\farg{ F_C^{-1}\farg{1-\frac{1}{\nu}}} - \lambda(\nu-1)\bigg\}\bigg\}\IEEEeqnarraynumspace\label{eq:nu_subt}\\
&=& \max_{\nu \in [1,T]} \mathopen{}\left\{ F_C^{-1}\farg{1-\frac{1}{\nu}} + \nu \overline{C}\farg{F_C^{-1}\farg{1-\frac{1}{\nu}}} \right\}\label{eq:slater}\\
&=&  F_C^{-1}\farg{1-\frac{1}{T}} + T \overline{C}\farg{F_C^{-1}\farg{1-\frac{1}{T}}}\label{eq:last_step}\\
&=& T\eta_{\text{BRQ}}(T).\label{eq:zeta_opt_eq_Tetabrq}
  \end{IEEEeqnarray}}{
  \begin{IEEEeqnarray}{rCl}
\IEEEeqnarraymulticol{3}{l}{\zeta_{\text{opt}}(T)}\nonumber\\
 \quad&\leq& \min_{\lambda > 0}\mathopen{}\bigg\{\lambda (T-1)+ \max_{r \geq 0} \mathopen{}\left\{ r+\frac{\overline{C}(r)-\lambda F_C(r) }{1-F_C(r)}\right\}\bigg\}\label{eq:A_subt}\IEEEeqnarraynumspace\\
 &=& \min_{\lambda > 0}\mathopen{}\bigg\{\lambda (T-1)+ \max_{\nu\geq 1} \mathopen{}\bigg\{  F_C^{-1}\farg{1-\frac{1}{\nu}}\nonumber\\
 &&{}\qquad\quad+\nu \overline{C}\farg{ F_C^{-1}\farg{1-\frac{1}{\nu}}} - \lambda(\nu-1)\bigg\}\bigg\}\label{eq:nu_subt}\\
&=& \max_{\nu \in [1,T]} \mathopen{}\left\{ F_C^{-1}\farg{1-\frac{1}{\nu}} + \nu \overline{C}\farg{F_C^{-1}\farg{1-\frac{1}{\nu}}} \right\}\label{eq:slater}\\
&=&  F_C^{-1}\farg{1-\frac{1}{T}} + T \overline{C}\farg{F_C^{-1}\farg{1-\frac{1}{T}}}\label{eq:last_step}\\
&=& T\eta_{\text{BRQ}}(T).\label{eq:zeta_opt_eq_Tetabrq}
  \end{IEEEeqnarray}}
  Here, \eqref{eq:A_subt} follows \eqref{eq:dyn_opt1}, \eqref{eq:V10}, and from $V(0) = A$; \eqref{eq:nu_subt} follows from the substitution $r = F_C^{-1}(1-1/\nu)$; \eqref{eq:slater} follows because \eqref{eq:concave_cond} is concave in $T$ and by Slater's condition \cite[pp.~226--227]{Boyd2004}; and   \eqref{eq:last_step} holds since the objective function in \eqref{eq:slater} is increasing in $\nu$. Since we have already shown that $\eta_{\text{opt}}(T) \geq \eta_{\text{BRQ}}(T)$, it follows that \eqref{eq:zeta_opt_eq_Tetabrq} implies $\zeta_{\text{opt}}(T) = T \eta_{\text{BRQ}}$.

Next, we need to show that $\zeta_1(T)=\zeta_{\text{opt}}(T)$. Because of the concavity of $\zeta_{\text{opt}}(\cdot)$, the upper bound $\zeta_{\text{opt}}(T) \leq TC_{\text{erg}}$, and $\zeta'_{\text{opt}}(T) > C_{\text{erg}}$ for $T > 1$, \eqref{eq:decompose_tau} implies that $\zeta(T) = \zeta_{\text{opt}}(T) = T \eta_{\text{BRQ}}(T)$. Here, $\zeta'_{\text{opt}}(T)$ denote the derivative of $\zeta_{\text{opt}}(T)$. Moreover, since $\eta_{\text{BRQ}}(\cdot)$ is an increasing function, it follows as previously argued that $\zeta_1(T) = T \eta_{\text{BRQ}}(T)$ as desired.

It remains to establish the claim in \eqref{eq:concave_cond} that $T \eta_{\text{BRQ}}(T)$ is concave in $T$. To do so, we show that the second derivative of $T \eta_{\text{BRQ}}(T)$ with respect to $T$ is negative. 
It turns out that the second derivative of $T \eta_{\text{BRQ}}(T)$ with respect to $T$ is given by
\ifthenelse{\boolean{thesis}}{
\begin{IEEEeqnarray}{rCl}
  \log(2)\frac{\partial^2 (T\eta_{\text{BRQ}})}{\partial T^2}
  &=& -\frac{1}{(1+F_H^{-1}(1-1/T)) T^3 P_H(F^{-1}_H(1-1/T))}\nonumber\\
  && {} - \frac{1}{(1+F_H^{-1}(1-1/T))^2 T^4 P_H(F^{-1}_H(1-1/T))^2}\nonumber\\
  && {} - \frac{P_H'(F_H^{-1}(1-1/T))}{(1+F_H^{-1}(1-1/T)) T^4 P_H(F^{-1}_H(1-1/T))^3}.\label{eq:second_derivative}\IEEEeqnarraynumspace
\end{IEEEeqnarray}}{
\begin{IEEEeqnarray}{rCl}
  \IEEEeqnarraymulticol{3}{l}{\log(2)\frac{\partial^2 (T\eta_{\text{BRQ}})}{\partial T^2}}\nonumber\\ 
  &=& -\frac{1}{(1+F_H^{-1}(1-1/T)) T^3 P_H(F^{-1}_H(1-1/T))}\nonumber\\
  && {} - \frac{1}{(1+F_H^{-1}(1-1/T))^2 T^4 P_H(F^{-1}_H(1-1/T))^2}\nonumber\\
  && {} - \frac{P_H'(F_H^{-1}(1-1/T))}{(1+F_H^{-1}(1-1/T)) T^4 P_H(F^{-1}_H(1-1/T))^3}.\label{eq:second_derivative}\IEEEeqnarraynumspace
\end{IEEEeqnarray}}
Here, $P_H'(\cdot)$ denotes the derivative of $P_H(\cdot)$.
By multiplying the RHS of \eqref{eq:second_derivative} by the positive term $(1+F_H^{-1}(1-1/T))^2 T^4$ and by using the substitution $T = 1/(1-F_H(h))$, we find that $\frac{\partial^2 (T\eta_{\text{BRQ}})}{\partial T^2}\leq 0$ is equivalent to the condition in \eqref{eq:opt_cond}, hence establishing the desired result.

%

\ifCLASSOPTIONcaptionsoff
  \newpage
\fi



\bibliographystyle{IEEEtran}

\bibliography{references}
%
%

%

\begin{IEEEbiographynophoto}{Kasper Fløe Trillingsgaard}
 (S'12) received his B.Sc. degree in electrical engineering and M.Sc. degree in wireless communications from Aalborg University, Denmark, in 2011 and 2013, respectively. He is currently pursuing a Ph.D. degree in electrical engineering at the same institution. He was a visiting student at New Jersey Institute of Technology, NJ, USA, in 2012 and at Chalmers University of Technology, Sweden, in 2014. His research interests are in the areas of information and communication theory. 
\end{IEEEbiographynophoto}

\begin{IEEEbiographynophoto}{Petar Popovski}
  (S’97–A’98–M’04–SM’10-F'16) is a Professor in
wireless communications at Aalborg
University, Denmark. He received Dipl.-Ing. in electrical engineering
(1997) and Magister Ing. in communication engineering (2000) from Sts.
Cyril and Methodius University, Skopje, Macedonia, and Ph.D. from
Aalborg University, Denmark, in 2004. He has more than 270
publications in journals, conference proceedings and books and has
more than 30 patents and patent applications. He is a Fellow of IEEE,
a holder of a Consolidator Grant from the European Research Council
and recipient of the Elite Researcher Award (2016) in Denmark. He is currently an Editor for IEEE
Transactions on Communications and Area Editor for IEEE Trans. Wireless Communications. From 2012 to 2014 he served as the Chair of IEEE ComSoc Emerging Technology Committee on Smart Grid
Communications. He is a Steering Committee member for IEEE Internet of
Things Journal, as well as Steering committee member of IEEE
SmartGridComm. His research interests are in the area of wireless
communication, networking, and communication/information theory.
\end{IEEEbiographynophoto}




\end{document}